\documentclass[twocolumn,twocolappendix]{aastex631}
\usepackage{amsmath,amssymb}
\usepackage{newtxtext,newtxmath}
\usepackage[T1]{fontenc}
\usepackage{ae,aecompl}

\usepackage{graphicx}
\usepackage{gensymb}

\usepackage{multirow}

\usepackage{xcolor}

\providecommand{\ionsuper}[2]{\ensuremath{\mathrm{{#1}}^{\mathrm{{#2}}}}}

\newcommand{\cb}{C\&B}
\newcommand{\snn}{S99}

\newcommand{\fvar}{$f_{\mathrm{var}}$}

\newcommand{\civ}{\ion{C}{4}}
\newcommand{\heii}{\ion{He}{2}}

\providecommand{\e}[1]{\ensuremath{\times 10^{#1}}}

\newcommand{\hst}{\textit{HST}}
\newcommand{\hstcos}{\textit{HST}/COS}

\newcommand{\jwst}{\textit{JWST}}

\newcommand{\ott}{\ensuremath{\mathrm{O}_{32}}}

\newcommand{\dynesty}{\textsc{dynesty}}
\newcommand{\cloudy}{\textsc{cloudy}}

\newcommand{\calcos}{\textsc{calcos}}
\newcommand{\pyneb}{\textsc{pyneb}}
\newcommand{\beagle}{\textsc{beagle}}
\newcommand{\parsec}{\textsc{parsec}}

\newcommand{\xid}{\hbox{$\xi_\mathrm{d}$}}

\newcommand{\hii}{\mbox{H\,{\sc ii}}}

\defcitealias{senchynaUltravioletSpectraExtreme2017}{S17}
\defcitealias{senchynaExtremelyMetalpoorGalaxies2019}{S19}
\defcitealias{bergCarbonOxygenAbundances2016}{B16}
\defcitealias{bergChemicalEvolutionCarbon2019}{B19}

\shorttitle{Stellar metallicities in \civ{} emitters}
\shortauthors{Senchyna et al.}

\graphicspath{{./}{figs/}}

\begin{document}

\title{Direct Constraints on the Extremely Metal-Poor Massive Stars Underlying Nebular \ion{C}{4} Emission from Ultra-Deep \hstcos{} Ultraviolet Spectroscopy}

\author[0000-0002-9132-6561]{Peter Senchyna}
\altaffiliation{Carnegie Fellow}
\email{psenchyna@carnegiescience.edu}
\affiliation{The Observatories of the Carnegie Institution for Science, 813 Santa Barbara Street, Pasadena, CA 91101, USA}
\affiliation{Steward Observatory, University of Arizona, 933 N Cherry Ave, Tucson, AZ 85721, USA}

\author{Daniel P.\ Stark}
\affiliation{Steward Observatory, University of Arizona, 933 N Cherry Ave, Tucson, AZ 85721, USA}

\author{St\'{e}phane Charlot}
\affiliation{Sorbonne Universit\'{e}, CNRS, UMR7095, Institut d'Astrophysique de Paris, F-75014, Paris, France}

\author{Adele Plat}
\affiliation{Steward Observatory, University of Arizona, 933 N Cherry Ave, Tucson, AZ 85721, USA}
\affiliation{Sorbonne Universit\'{e}, CNRS, UMR7095, Institut d'Astrophysique de Paris, F-75014, Paris, France}

\author{Jacopo Chevallard}
\affiliation{Sorbonne Universit\'{e}, CNRS, UMR7095, Institut d'Astrophysique de Paris, F-75014, Paris, France}

\author{Zuyi Chen}
\affiliation{Steward Observatory, University of Arizona, 933 N Cherry Ave, Tucson, AZ 85721, USA}

\author{Tucker Jones}
\affiliation{Department of Physics, University of California Davis, 1 Shields Avenue, Davis, CA 95616, USA}

\author{Ryan L. Sanders}
\altaffiliation{Hubble Fellow}
\affiliation{Department of Physics, University of California Davis, 1 Shields Avenue, Davis, CA 95616, USA}

\author{Gwen C. Rudie}
\affiliation{The Observatories of the Carnegie Institution for Science, 813 Santa Barbara Street, Pasadena, CA 91101, USA}

\author{Thomas J. Cooper}
\affiliation{The Observatories of the Carnegie Institution for Science, 813 Santa Barbara Street, Pasadena, CA 91101, USA}

\author{Gustavo Bruzual}
\affiliation{Instituto de Radioastronom{\'i}a y Astrof{\'i}sica, UNAM, Campus Morelia, Michoacan, M{\'e}xico, C.P. 58089, M{\'e}xico}

\begin{abstract} 
\noindent
Metal-poor nearby galaxies hosting massive stars have a fundamental role to play in our understanding of both high-redshift galaxies and low metallicity stellar populations.
But while much attention has been focused on their bright nebular gas emission, the massive stars that power it remain challenging to constrain.
Here we present exceptionally deep \textit{Hubble Space Telescope} ultraviolet spectra targeting six galaxies that power strong nebular \civ{} emission approaching that encountered at $z>6$.
We find that the strength and spectral profile of the nebular \civ{} in these new spectra follow a sequence evocative of resonant scattering models, indicating that the hot circumgalactic medium likely plays a key role in regulating \civ{} escape locally.
We constrain the metallicity of the massive stars in each galaxy by fitting the forest of photospheric absorption lines, reporting measurements driven by iron that lie uniformly below 10\% solar.
Comparison with the gas-phase oxygen abundances reveals evidence for enhancement in O/Fe above solar across the sample, robust to assumptions about the absolute gas-phase metallicity scale.
This supports the idea that these local systems are more chemically-similar to their primordial high-redshift counterparts than to the bulk of nearby galaxies.
Finally, we find significant tension between the strong stellar wind profiles observed and our population synthesis models constrained by the photospheric forest in our highest-quality spectra.
This reinforces the need for caution in interpreting wind lines in isolation at high-redshift, but also suggests a unique path towards validating fundamental massive star physics at extremely low metallicity with integrated ultraviolet spectra.
\end{abstract}


\section{Introduction}

Nearby star-forming dwarf galaxies beyond the Local Group are a crucial testbed for the physics of massive stars and gas at very low metallicities.
The possible nature of these blue, \hii{} region-like galaxies as genuinely primordial systems captured during their first episodes of star formation was emphasized and debated soon after their discovery \citep[e.g.][]{sargentIsolatedExtragalacticII1970,searleInferencesCompositionTwo1972,searleHistoryStarFormation1973}.
A modern view paints these systems as different from true nascent galaxies in important ways, including their absolute mass scale and likely longer-term bursty star formation histories.
While debate continues regarding the precise differences with their high-redshift counterparts, metal-poor local dwarf galaxies remain the most accessible venue in which to empirically confront models for the very low metallicity young stellar populations we expect to encounter at the highest redshifts.

The value of such rare very metal-poor dwarfs is particularly well-illustrated by efforts to understand the strong \ion{C}{4} emission recently uncovered in the reionization era.
In two of the first deep rest-ultraviolet (UV) spectra obtained for Ly$\alpha$-emitters at $z\gtrsim 6$, nebular emission in the resonant \civ{} $\lambda \lambda 1548, 1550$ doublet is detected at implied doublet equivalent widths of 20--40 \AA{} \citep{starkSpectroscopicDetectionIV2015,mainaliEvidenceHardIonizing2017,schmidtGrismLensAmplifiedSurvey2017}.
Strong limits placed on other UV lines in one of the systems support a stellar (over a power-law) ionizing radiation source \citep{mainaliEvidenceHardIonizing2017}; yet stellar models struggle to reproduce such prominent emission, only approaching the observed equivalent widths at extremely low metallicities.
Unfortunately, an empirical basis for this potential connection to very metal-poor stars remains elusive.
Nebular emission in \civ{} is exceedingly rare at any strength in star-forming galaxies at lower redshifts, where this doublet is generally dominated by a broad P-Cygni profile powered in the winds of luminous OB stars \citep[e.g.][]{prinjaHSTUVMeasurementsWind1998,shapleyRestFrameUltravioletSpectra2003,steidelReconcilingStellarNebular2016,rigbyMagellanEvolutionGalaxies2018}.
The canonical picture from observations of star-forming galaxies in the local Universe with the original loadout of Hubble Space Telescope (\hst{}) UV instruments supported this picture of \civ{} as a purely wind and ISM absorption feature \citep[e.g.][]{leithererUltravioletSpectroscopicAtlas2011}.
In this context, the detection of such prominent nebular \civ{} in several pathfinder spectroscopic programs foreshadows potentially significant challenges in interpreting rest-UV spectra at the highest redshifts in the \jwst{} and ELT eras.

Fortunately, the highly-sensitive fourth-generation \hst{} / Cosmic Origins Spectrograph (COS) has unveiled UV spectra of a dramatically different character than those found in previous local surveys.
In particular, several \hstcos{} programs have now identified a population of star-forming galaxies with \civ{} in clear nebular emission alongside nebular \heii{}, at equivalent widths of $\sim 1$--11~\AA{} \citep[][]{bergCarbonOxygenAbundances2016,senchynaUltravioletSpectraExtreme2017,senchynaExtremelyMetalpoorGalaxies2019,bergChemicalEvolutionCarbon2019,bergIntenseIVHe2019}.
These systems confirm that star-forming galaxies can power nebular \civ{} approaching the large equivalent widths observed at $z>6$, but only at both sufficiently young effective ages (or high specific star formation rates; sSFRs) and extremely low metallicities ($12+\log\mathrm{O/H}\lesssim 7.7$; corresponding to $Z/Z_\odot \lesssim 10\%$), a parameter space occupied locally by primarily relatively low-mass galaxies missed by previous UV programs \citep{senchynaExtremelyMetalpoorGalaxies2019}.

The establishment of this local reference sample of \civ{}-emitters represents a unique opportunity to directly link high-ionization nebular emission to the fundamental properties of the underlying stellar populations, which are generally inaccessible at high redshift.
However, several outstanding questions in the interpretation of these galaxies suggest our picture of them is not yet complete.
While it is immediately clear that low gas-phase metallicities and young ages are a necessary condition for nebular \civ{} production in nearby star-forming systems, it is less obvious why galaxies with nearly identical H$\beta$ equivalent widths and gas-phase metallicities would present such different \civ{} equivalent widths \citep{bergChemicalEvolutionCarbon2019,senchynaExtremelyMetalpoorGalaxies2019}.
In addition, the significant gap between the largest \civ{} nebular emission equivalent width observed locally of 11~\AA{} and the 20--40~\AA{} emission measured at $z>6$ is suggestive of potential differences between this reionization era population and these local metal-poor dwarfs of still unclear origin.

There is a crucial quantity as-yet missing in our attempts to understand these findings: the stellar metallicity.
In a situation common to \hii{} regions and star-forming galaxies, strong optical nebular lines readily yield the gas-phase abundances of oxygen and nitrogen (and carbon through the UV \ion{C}{3}] line) as well as neon and argon; all of which except for nitrogen are $\alpha$ elements, synthesized and released mainly in intermediate-mass to massive stars on short timescales.
While these elements play an outsized role in shaping the behavior and appearance of \hii{} regions, the spectra of massive stars are primarily governed by iron and iron-peak elements, whose numerous freed electrons and electronic transitions in the EUV--UV dominate their opacities.
The full promise of nearby star-forming galaxies as laboratories for studying young stellar populations and their impact on their environs cannot be fully realized without constraints on the total stellar metallicity and in particular their iron abundances.

Unfortunately, typical CNO gas-phase abundances tell us little about the total metallicity of the underlying massive stars.
In contrast to the CNO and $\alpha$ elements, iron is released into the ISM for subsequent star formation mainly by Type~Ia supernovae which occur after a significant $\sim$~Gyr delay after a star formation episode.
As a  result, $\alpha$/Fe ratios can be expected to vary dramatically with star formation history, generally decreasing significantly with time after a peak in star formation rate as $\alpha$-rich yields from massive stars are slowly diluted by iron-rich Type~Ia products.
This is observed directly in the abundance patterns of individual stars in the Milky Way, its satellites, and other Local Group galaxies \citep[e.g.][]{tolstoyStarFormationHistoriesAbundances2009,bensbyExploringMilkyWay2014}.
The natural expectation for assembling galaxies in the very early Universe which have just begun to form stars is thus a significant enhancement in $\alpha/$Fe over the solar value, potentially even above the Type~II limit due to Population III enrichment based upon the abundances of the lowest-metallicity Galactic stars \citep[e.g.][]{frebelNearFieldCosmologyExtremely2015}.
An increasing number of reports suggest that such an enhancement is necessary to reconcile the low stellar metallicities relative to the gas-phase oxygen abundance inferred from observations of nebular emission and UV continuum light in massive galaxies at $z\sim 2$ \citep[e.g.][]{steidelReconcilingStellarNebular2016,stromNebularEmissionLine2017,sandersMOSDEFSurveyDirectmethod2020,toppingMOSDEFLRISSurveyConnection2020,cullenNIRVANDELSSurveyRobust2021,stromChemicalAbundanceScaling2021}.

Despite this surge of interest at $z\sim 2$, relatively little progress has been made in assessing the stellar abundances of extreme metal-poor star-forming galaxies nearby like the nebular \civ{} emitters.
The optical absorption lines commonly used to assess stellar metallicities such as the Lick indices \citep[e.g.][]{gallazziAgesMetallicitiesGalaxies2005} are washed-out by light from the hot stars and nebular continuum dominating these high-sSFR systems, and regardless do not directly assess the abundances of the massive stars themselves.
Another approach is to estimate the iron abundance in the gas-phase through collisionally-excited lines, but this is subject to significant systematic uncertainties in the ionization correction factors adopted and the assumed depletion onto dust \citep[e.g.][]{izotovChemicalCompositionMetalpoor2006,kojimaEMPRESSIIHighly2020,bergCharacterizingExtremeEmission2021} and is furthermore still an indirect probe of the massive star abundances.

The most direct path towards stellar abundances in these systems is through their UV continuum light; and in particular, the photospheric iron lines encountered there \citep[e.g.][]{bouretQuantitativeSpectroscopyStars2003,hillierTaleTwoStars2003,rixSpectralModelingStarforming2004}.
These lines probe iron in the atmospheres of the massive stars which dominate these galaxies, and bypass the substantial modeling uncertainties inherent in analysis of the stellar wind lines.
Unfortunately, the generally single-orbit \hstcos{} spectra in which the strong nebular \civ{} emission was initially observed are too shallow to detect these blended absorption features; indeed, the target galaxies are often too low in metallicity for even the usually-strong stellar P-Cygni wind profile in \civ{} to be detected in these spectra.
To remedy this, we obtained new ultra-deep \hstcos{} G160M FUV spectroscopy with exposure times of 5--10~orbits per galaxy for six galaxies powering nebular \civ{} emission during \hst{} Cycles 26 and 27.
These new spectra provide a more detailed view of \civ{} and strong constraints on stellar wind and photospheric lines in the FUV continuum, allowing us to link this high-ionization emission to the fundamental properties of the underlying stellar populations.

This paper focuses on \civ{} and the stellar features in these new ultra-deep \hstcos{} spectra, and is organized as-follows.
In Section~\ref{sec:sample}, we describe the target sample and derive constraints on their gas-phase metallicities and bulk stellar properties from their optical spectra.
Section~\ref{sec:uvobs} presents the new \hstcos{} spectroscopic observations including their reduction and basic analysis.
We describe our fits to the UV continuum with stellar population synthesis models to constrain the stellar metallicity in Section~\ref{sec:fitting}, and the results of these fits are presented in Section~\ref{sec:results}.
Finally, we discuss these results in contexts ranging from local galaxy abundances to high-redshift nebular emission in Section~\ref{sec:discuss}, before summarizing and concluding in Section~\ref{sec:summary}.

Unless otherwise stated, equivalent widths are reported as positive for emission and negative for absorption; and all logarithmic quantities presented herein should be assumed to be in base-10.
When comparing with solar abundances, we adopt the composition assumed by our fiducial stellar population synthesis models from Charlot \& Bruzual \citep[in-prep., \cb{}; see][]{gutkinModellingNebularEmission2016,platConstraintsProductionEscape2019}, which differ only slightly from the compilation adopted by the underlying \parsec{} stellar evolution code as described in \citet{bressanPARSECStellarTracks2012}.
In particular, this sets the present-day solar oxygen abundance at $12+\log\mathrm{O/H}=8.83$\footnote{Slightly larger but within the 1$\sigma$ uncertainties of the \citet{caffauSolarChemicalAbundances2011} measurement of $8.76\pm 0.07$; see \citet{gutkinModellingNebularEmission2016}.} (corresponding to a gas-phase abundance with typical depletion of $8.71$), solar iron at $12+\log\mathrm{Fe/H}=7.52$, and total metallicity at $Z_\odot = 0.01524$ \citep{caffauPhotosphericSolarOxygen2008,caffauSolarChemicalAbundances2011}.

\section{A sample of very metal-poor galaxies with prominent active star formation}
\label{sec:sample}

The population of low-mass dwarf galaxies in the nearby Universe that are undergoing active star formation provide a unique view onto very metal-poor stellar populations and the nebular emission they excite, under conditions potentially approaching those encountered at high redshift.
A subset of these dwarfs at $12+\log\mathrm{O/H}<8.0$ and particularly high sSFR (or young dominant ages) which power nebular emission in the resonant \civ{} $\lambda \lambda 1548,1550$ doublet have been recently uncovered, inviting comparison to the $z>6$ Ly$\alpha$ emitters in which this emission has been detected at unprecedented strength.
However, the shallow \hstcos{} spectra in which this emission was initially identified preclude detailed assessment of the underlying stellar populations.

We selected a sample of six of these local \civ{}-emitters to follow-up with deep FUV spectroscopy targeting both \civ{} and the photospheric signatures of their constituent massive stars.
The target systems were selected from the \hstcos{} G140L samples assembled by \citet[][hereafter \citetalias{bergCarbonOxygenAbundances2016}]{bergCarbonOxygenAbundances2016} and \citet[][\citetalias{bergChemicalEvolutionCarbon2019}]{bergChemicalEvolutionCarbon2019}, and the G160M+G185M samples of \citet[][\citetalias{senchynaUltravioletSpectraExtreme2017}]{senchynaUltravioletSpectraExtreme2017} and \citet[][\citetalias{senchynaExtremelyMetalpoorGalaxies2019}]{senchynaExtremelyMetalpoorGalaxies2019}, which together with \citet{woffordStarsGasMost2021} comprise the entire sample of local star-forming systems with reported nebular \ion{C}{4} detections.
In particular, we selected our sample to include the two highest equivalent width nebular emitters in this doublet \citepalias[J104457 at 11 \AA{} and HS~1442+4250 at 4 \AA{};][]{bergCarbonOxygenAbundances2016,senchynaExtremelyMetalpoorGalaxies2019}; two of the systems with the highest gas-phase metallicities still powering nebular emission in \ion{C}{4} \citepalias[SB~2 and 82, at $12+\log\mathrm{O/H} \simeq 7.8$--$7.9$;][]{senchynaUltravioletSpectraExtreme2017}; and two metal-poor systems with extremely high equivalent width optical line emission yet relatively modest \ion{C}{4} equivalent widths $<4$~\AA{} \citepalias[J082555 and J120202;][]{bergCarbonOxygenAbundances2016,bergChemicalEvolutionCarbon2019}.
These systems were targeted with \hstcos{} G160M/1533 spectroscopy with allocations of 5--10 orbits each over the course of two \hst{} programs (GO:15646 and GO:15881, PI: Stark) which we describe in Section~\ref{sec:uvobs}.
The basic properties of this sample of \civ{} emitters are summarized in Table~\ref{tab:basicprop}, and described in the remainder of this section.

\begin{table*}
\centering
\footnotesize
\caption{Basic properties of the six local star-forming systems targeted for deep G160M spectra.
Included here are the target names; the coordinates of the region targeted with \hstcos{} in the SDSS imaging frame; the redshift measured from the \hstcos{} spectrum; the corresponding luminosity distance estimated from Cosmicflows-3 \citep[CF3;][]{grazianiPeculiarVelocityField2019,kourkchiCosmicflows3TwoDistanceVelocity2020}; and gas-phase measurements of \ott{} ([\ion{O}{3}] $\lambda 4959+\lambda 5007$ / [\ion{O}{2}] $\lambda 3727$ flux ratio, extinction-corrected), the rest-frame H$\beta$ equivalent width, and direct-$T_e$ oxygen abundance.
}
\label{tab:basicprop}

\begin{tabular}{lllccccccr}
\hline
Target & RA & Dec                       & $z$           & Distance      & \ott{}    & H$\beta$  & $\mathrm{E(B-V)}$     & $12+\log_{10}(\mathrm{O/H})$  & Optical    \\
Name   & (J2000) & (J2000)              & (\hstcos{})   & (Mpc, CF3)    &           &  $W_0/$\AA{}         &  int. (Gal.)  &                & Source                       \\
\hline

J082555 & 08:25:55.52 & +35:32:32.0 & 0.0024 & 13 & $16.1 \pm 1.6$ & $264 \pm 21$ & $0.06 (0.04)$ & $7.44 \pm 0.04$ & SDSS+Bok\\
SB~2 & 09:44:01.87 & $-$00:38:32.2 & 0.0049 & 21 & $7.3 \pm 0.8$ & $276 \pm 11$ & $0.15 (0.05)$ & $7.81 \pm 0.05$ & SDSS+MMT \citepalias{senchynaUltravioletSpectraExtreme2017}\\
J104457 & 10:44:57.80 & +03:53:13.1 & 0.0129 & 57 & $18.1 \pm 2.0$ & $264 \pm 21$ & $0.08 (0.04)$ & $7.48 \pm 0.06$ & SDSS+Bok\\
SB~82 & 11:55:28.34 & +57:39:51.9 & 0.0172 & 77 & $9.6 \pm 1.0$ & $175 \pm 3$ & $0.14 (0.03)$ & $7.93 \pm 0.04$ & SDSS+MMT \citepalias{senchynaUltravioletSpectraExtreme2017}\\
J120202 & 12:02:02.49 & +54:15:51.0 & 0.0120 & 58 & $9.2 \pm 0.4$ & $271 \pm 8$ & $0.05 (0.01)$ & $7.50 \pm 0.03$ & SDSS (BOSS)\\
HS1442+4250 & 14:44:11.47 & +42:37:35.5 & 0.0022 & 11 & $10.2 \pm 0.5$ & $113 \pm 4$ & $0.02 (0.01)$ & $7.66 \pm 0.04$ & MMT \citepalias{senchynaExtremelyMetalpoorGalaxies2019}\\

\hline
\end{tabular}

\end{table*}

\begin{figure*}
    \includegraphics[width=1.0\textwidth]{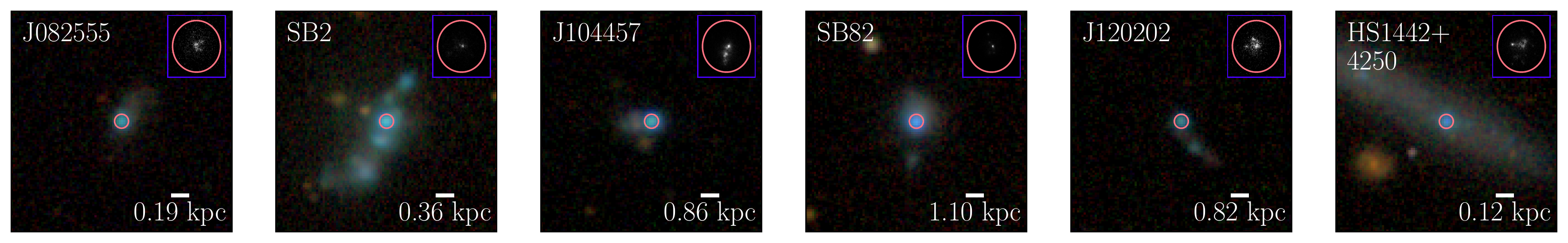}
    \caption{
        Optical $g,r,i$ (RGB) SDSS images centered on our targets, with the 2.5$''$-diameter \hstcos{} aperture indicated by a red circle centered on the target $u$-band centroid.
        An inset panel for each target displays one of the \hstcos{} NUV target acquisition images obtained for each, on a linear scale.
        A 3$''$ scalebar is also shown and labeled with the corresponding distance in comoving Mpc at the estimated distance of each target (Table~\ref{tab:basicprop}).
    }
    \label{fig:sdssmontage}
\end{figure*}

\subsection{Gas-phase conditions and metallicities}
\label{sec:gasmet}

The optical nebular emission lines provide key insight onto the properties of the ionized gas in these galaxies including the abundance of oxygen in the gas-phase.
This is of particular interest as a point of comparison in the stellar abundance analysis which the new ultraviolet data enables.
Leveraging the detections of the auroral [\ion{O}{3}] $\lambda 4363$ line for all six, we derive a self-consistent set of gas-phase oxygen abundances via the direct-$T_e$ method (following the general procedure described also in \citetalias{senchynaUltravioletSpectraExtreme2017} and \citetalias{senchynaExtremelyMetalpoorGalaxies2019}).
For this analysis we rely on the designated \texttt{sciencePrimary} SDSS spectrum for all of the targets except for HS~1442+4250 which does not have an SDSS spectrum; in this case, we rely instead on the MMT Blue Channel spectrum described in \citetalias{senchynaExtremelyMetalpoorGalaxies2019}.
First, we measure the flux and equivalent width of the necessary optical lines by fitting a Gaussian (or pair thereof for close doublets) plus linear local continuum model to each, using \textsc{emcee} \citep{foreman-mackeyEmceeMCMCHammer2013} to compute the uncertainties on the fit parameters.
The fluxes of these lines are corrected for reddening, first Galactic (using the \citealt{schlaflyMeasuringReddeningSloan2011} reddening map and assuming a \citealt{fitzpatrickCorrectingEffectsInterstellar1999} reddening law) and then intrinsic assuming an SMC extinction curve \citep{gordonQuantitativeComparisonSmall2003}.
We use the \pyneb{} library to derive constraints on the gas parameters of interest from these line measurements\footnote{Adopting the \pyneb{} tabulated atomic and collisional data for \ionsuper{O}{+} from \citet{froesefischerBreitPauliEnergyLevels2004,wieseAtomicTransitionProbabilities1996,tayalOscillatorStrengthsElectron2007}; for \ionsuper{O}{2+} from \citet{froesefischerBreitPauliEnergyLevels2004,storeyTheoreticalValuesOIII2000,storeyCollisionStrengthsNebular2014}; and for \ionsuper{S}{+} from \citet{tayalBreitPauliTransitionProbabilities2010,irimiaBreitPauliOscillator2005}.}.
In particular, we adopt a standard two-zone \hii{} region model, computing the $T_e$ and $n_e$ relevant for the \ionsuper{O}{+} and \ionsuper{O}{2+} regions through cross-convergence of the $n_e$-sensitive \ion{S}{2} $\lambda \lambda 6731,6716$ ratio with the $T_e$-sensitive ([\ion{O}{2}] $\lambda\lambda 3726,3729$\footnote{Unresolved in our observations, and hereafter referred to as [\ion{O}{2}] $\lambda 3727$.}, [\ion{O}{2}] $\lambda\lambda 7320,7330$) and ([\ion{O}{3}] $\lambda 4363$, [\ion{O}{3}] $\lambda 5007$) ratios (respectively).
The exception to this procedure is HS~1442+4250 for which [\ion{O}{2}] $\lambda\lambda 7320,7330$ is undetected in our MMT spectrum; in this case, we instead rely on the empirical relationship between $T_e(\mathrm{O II})$ and $T_e(\mathrm{O III})$ derived for metal-poor galaxies by \citet{izotovChemicalCompositionMetalpoor2006} to estimate the temperature of the lower-ionization region.
We estimate uncertainties in the final quantities of interest by bootstrap resampling the Gaussian line flux uncertainty distributions and recalculating all quantities of interest (with line ratio combinations recognized as invalid by \pyneb{} according to our adopted atomic data discarded).

In addition to the primary SDSS and MMT spectra, these gas-phase measurements must also in four cases rely on shallow bluer spectra to obtain measurements of the [\ion{O}{2}] $\lambda3727$ doublet which remains below the 3800~\AA{} blue cutoff of the original SDSS I/II spectrograph at these low redshifts ($z<0.02$)\footnote{Note that J120202 was observed with the BOSS spectrograph as one of the 173 galaxies targeted in the Census of Nearby Galaxies ancillary program (following-up systems identified in Palomar Transient Facility narrowband H$\alpha$ imaging), and thus [\ion{O}{2}] is accessible in the original optical spectrum by virtue of its bluer cutoff.}.
Specifically, SB~2 and 82 have [\ion{O}{2}] constraints from MMT Blue Channel spectra described in \citetalias{senchynaUltravioletSpectraExtreme2017}.
We also rely on new Bok spectra to constrain [\ion{O}{2}] $\lambda 3727$ for J082555 and J104457; both were observed with the Bok B\&C spectrograph and the 300 g/mm grating with 3$\times$15 minutes integrations on 2021-Mar-03.
These data were reduced using standard longslit techniques in \textsc{IRAF}. 
As done also for the supplementary MMT spectra, we translate the measured [\ion{O}{2}] doublet flux measured in the Bok spectra to the SDSS flux scale by applying a correction derived from the ratio of the strong optical lines in common between the SDSS and Bok spectra; and adopt a conservative uncertainty of 10\% in this rescaled doublet flux.

The resulting direct-$T_e$ oxygen abundances (Table~\ref{tab:basicprop}) confirm that these \civ{}-emitters harbor very metal-poor gas.
The values range from $12+\log\mathrm{O/H}=7.44$ in the lowest-metallicity case (J082555) up to 7.93 in the highest (SB~82).
This corresponds to a range of 5--15\% the solar photospheric oxygen value, with all but SB~2 and SB~82 falling below 10\% and thus in the realm of designated extremely metal-poor galaxies (XMPs).
Since these are measured in essentially the same way and with the same data as in \citetalias{senchynaUltravioletSpectraExtreme2017} and \citetalias{senchynaExtremelyMetalpoorGalaxies2019}, the metallicity measurements are by construction in very good agreement with those results.
Our metallicity estimates are slightly different from those presented in \citetalias{bergCarbonOxygenAbundances2016} for J082555 (0.07 dex higher) and J104457 (0.03 dex higher), and from \citetalias{bergChemicalEvolutionCarbon2019} for J120202 (0.13 dex lower).
We attribute these generally small discrepancies to differences in our assumed atomic data and methodology (most significantly, our uniform [\ion{O}{2}] $\lambda 3727$ access and our choice to estimate $T_e$([\ion{O}{2}]) from the [\ion{O}{2}] $\lambda 7325/\lambda 3727$ ratio).

We also measure uniformly-large flux ratios \ott{} ([\ion{O}{3}]$\lambda 4959+ \lambda 5007$ / [\ion{O}{2}] $\lambda 3727$, after extinction correction) of 7--18; and generally very prominent gas emission relative to the stellar continuum with H$\beta$ at (rest-frame) equivalent widths of 113--276 \AA{}.
Together, these measurements paint a picture of systems dominated by recently-formed stars with strong ionizing continua.

\subsection{Bulk stellar properties}
\label{sec:optother}

The optical spectroscopy and imaging for these systems provides quantitative constraints on the age and mass of stars formed in their recent dominant star formation event.
This can act as a valuable prior in the interpretation of the UV spectra we will focus on in the remainder of this paper.
To extract these optical constraints, we first measure 3$''$-diameter aperture photometry (corresponding approximately to the \hstcos{} 2.5$''$ diameter aperture convolved with the median SDSS $r$-band seeing of 1.3$''$) in all five ($u'g'r'i'z'$) SDSS broadband mosaics for each source centered on the spectroscopic target.
We then fit these fluxes with \beagle{} \citep{chevallardModellingInterpretingSpectral2016} alongside the equivalent width of H$\beta$ measured in the optical spectrum for each.
Together with the strong band contamination in $g'$ and $r'$ produced by the [\ion{O}{3}] and Balmer lines at these low redshifts, these measurements provide constraints on both the total stellar mass and SFR of the current star formation event.
The models employed here are the same as those used to fit the UV continuum described below (Section~\ref{sec:spsmods}), but with IMF upper mass cutoff of 300~$M_\odot$ (minimal differences were found between the 300 and 600~$M_\odot$ models in the photospheric metallicity results presented in Section~\ref{sec:results}) and with nebular emission lines included so as to self-consistently account for the strong line contributions in the optical broadbands.
We perform this modeling assuming both a constant star formation history with variable age (or start time) and a simple (single, variable age) stellar population, representing two approximations of a recent `burst' of star formation.
In both cases we adopt uniform priors for the two corresponding star formation history parameters, age (allowed to vary from $1$--$100$ Myr) and stellar mass (up to $10^9$~$\mathrm{M_\odot}$).
We allow the tied stellar and gas-phase metallicity (0.006--1.0~$Z_\odot$), nebular $\log U$ ($-4$ -- $-1$), and dust attenuation (assuming an SMC curve, $\hat{\tau}_V\sim 0$--2) to vary over their respective uniform priors.

The absolute magnitude of the mass and star formation rates derived from this fitting depend upon the assumed luminosity distance to these targets.
While fortunately our analysis does not depend strongly on the exact value of these stellar masses or thus this distance, it is important to note that at very low redshifts $z<0.01$ (corresponding to velocities $<1000$~km/s) peculiar motions begin to approach and potentially exceed the velocity of the Hubble flow.
Without a means of more robust distance measurement, our best option remains comparison of the measured redshifts to a model for the peculiar velocity field in the local Universe.
We update our distance estimates in this analysis to the Cosmicflows-3 local flow model \citep{grazianiPeculiarVelocityField2019,kourkchiCosmicflows3TwoDistanceVelocity2020}, first converting our heliocentric redshifts to the Local Sheet reference frame \citep{tullyOurPeculiarMotion2008,kourkchiCosmicflows3TwoDistanceVelocity2020} and then obtaining the corresponding distance from the smoothed Cosmicflows-3 model (none of our objects were found to have multiple degenerate solutions).
The distances thereby obtained are displayed in Table~\ref{tab:basicprop}.
Differences from previous estimates relying on other local flow models are generally small; with respect to the published distances of \citetalias{senchynaUltravioletSpectraExtreme2017} and \citetalias{senchynaExtremelyMetalpoorGalaxies2019} in particular, we find changes of $\leq 2$ Mpc ($\lesssim 10\%$) for SB~2, SB~82, and HS~1442+4250.

The resulting measurements of the effective young stellar population age and (distance-dependent) stellar mass within the spectroscopic aperture are presented in Table~\ref{tab:sedfits}.
These fits confirm that the broadband data for these systems are consistent with low-mass stellar populations $\sim 10^4$--$10^6$~$\mathrm{M_\odot}$ 
with young effective ages in the range of 3--22~Myr assuming a constant star formation history or 2--3~Myr assuming a single-age stellar population, both of which are able to reasonably match the photometric data and H$\beta$ equivalent widths.
However, a more detailed view of the massive stars present requires deep UV spectra.

\begin{table}
\centering
\begin{tabular}{lcc|cc}
\hline
Target  & \multicolumn{2}{c|}{CSFH} & \multicolumn{2}{c}{SSP}  \\
Name    & $\log (\mathrm{M/M_\odot})$ & $t/\mathrm{Myr}$ & $\log (\mathrm{M/M_\odot})$ & $t/\mathrm{Myr}$ \\

\hline

J082555 & $4.03 \pm 0.03$ & $3.08^{+0.24}_{-0.18}$& $4.25 \pm 0.05$ & $2.11^{+0.04}_{-0.04}$\\ 
SB2 & $5.14 \pm 0.02$ & $3.38^{+0.08}_{-0.09}$& $5.35 \pm 0.01$ & $2.00^{+0.01}_{-0.01}$\\ 
J104457 & $5.53 \pm 0.02$ & $2.59^{+0.09}_{-0.05}$& $5.67 \pm 0.02$ & $1.97^{+0.01}_{-0.01}$\\ 
SB82 & $6.29 \pm 0.02$ & $6.25^{+0.24}_{-0.29}$& $6.69 \pm 0.01$ & $2.01^{+0.01}_{-0.02}$\\ 
J120202 & $5.17 \pm 0.21$ & $4.72^{+0.31}_{-0.45}$& $5.22 \pm 0.04$ & $2.46^{+0.17}_{-0.09}$\\ 
HS1442+4250 & $4.60 \pm 0.02$ & $19.47^{+1.02}_{-1.14}$& $4.61 \pm 0.03$ & $3.47^{+0.05}_{-0.48}$\\ 

\hline

\end{tabular}

\caption{
    Results of \beagle{} SED fits to the broadband SDSS $ugriz$ photometry with nebular line contamination (explicitly including H$\beta$ equivalent widths).
    We include the stellar mass and age assuming both a constant star formation history with variable start and a single-age stellar population model for the young stellar component which dominates the optical and UV for these systems.
    \label{tab:sedfits}
}
\end{table}

\section{Ultra-deep COS spectroscopy}
\label{sec:uvobs}

Constraining the massive stars underlying extreme nebular emission at low metallicities requires deep UV spectroscopy to access their faint metal line features.
These necessary spectra were obtained for the target systems in two GO programs (proposal IDs 15646 and 15881, PI: Stark) selected and executed as part of Cycles 26 and 27.
All of the spectroscopic observations for both programs were made using the new G160M/1533 \AA{} central wavelength setting introduced in Cycle 26 \citep[e.g.][]{foxFluxCalibrationNew2019}.
This setting produces spectra spanning 1339--1710 \AA{} with the FUVA/FUVB detector gap at 1520--1530, which at the low redshifts $<0.02$ of our targets (Table~\ref{tab:basicprop}) allows us to simultaneously access the \ion{C}{4} $\lambda \lambda 1548,1550$ and \ion{He}{2} $\lambda 1640$ nebular and stellar wind complexes alongside the wind and photospheric complexes at $\sim1350$--$1500$ \AA{}.

In this section, we first describe these new FUV spectroscopic observations and their calibration (Section~\ref{sec:uvcal}).
Then, we provide an overview of these new spectra and the strength of nebular and stellar features derived therefrom (Section~\ref{sec:uvmeas}) before proceeding to describe the detailed continuum fitting we perform in the following Section.

\subsection{Observations and Calibration}
\label{sec:uvcal}

The spectroscopic observations were conducted similarly for all objects in both programs.
Given the uncertain nature of the strength of the target continuum features at these low metallicities, we allotted a uniform 10 (GO:15646) or 5 (GO:15881) orbits in total to each system.
These exposure times were designed to ensure a minimum signal-to-noise (S/N) ratio per-resolution element (or `resel'; 6 pixels, or $0.07$ \AA{} for G160M) of 10 (corresponding to S/N $>30$ per-\AA{}).
To maximize the ability of our observations to be scheduled, these allotted orbits were split between two (5+5 orbits) or three (2+2+1 orbits) separate telescope visits (respectively; see Table~\ref{tab:cosobs}).
Since our initial coordinates were based upon centroids of the SDSS $u$-band images which are well-aligned astrometrically, each visit began with a single \texttt{ACQ/IMAGE} exposure with the NUV camera (plus the standard follow-up confirmation image) to acquire and center the target star-forming regions in the primary science aperture (PSA).
All acquisition exposures were conducted with the PSA and \texttt{MIRRORA} using exposure times based upon simple point-source ETC calculations and previous acquisition exposures for the targets (median on-source time of 60s).
As observed on prior acquisition of these targets, all resolve into a dominant compact star-forming region with outlying flux, consisting of some mixture of subdominant compact regions and diffuse flux (Figure~\ref{fig:sdssmontage}).

After target acquisition, the remainder of each visit consisted of spectroscopic exposures in the \texttt{TIME-TAG} mode.
Taking advantage of the faintness of the targets, the entire duration of each orbit was devoted to on-target exposure, with buffer dumps taking place during target occultation.
Internal wavelength calibrations were conducted simultaneously using the recommended \texttt{TAGFLASH} (\texttt{FLASH=YES}) lamp flashes.
To minimize the impact of fixed-pattern noise, the instrument was also shifted to a new \texttt{FP-POS} setting between orbits, cycling through all 4 for approximately-equal exposure time spread over the course of the visit blocks for each target.
The position of the COS FUV channel spectroscopic trace has been moved several times since installation to minimize the impact of gain sag on observations; all of the spectroscopic observations in both of the programs presented here were conducted at Lifetime Position 4 (LP4).

This nominal observing plan was disrupted by a handful of telescope exception events, some of which affected data acquisition and required repeat observations.
We detail these events below in the order that they occurred:
\begin{itemize}
\item On 2019-Oct-09 during Visit 6 of GO:15646, the \hst{} Fine Guidance Sensors (FGS) lost guide star lock during the final exposure for that visit on HS~1442+4250.
As a safety measure, this caused the shutter to close, losing approximately 10 minutes of the planned exposure time.
A single additional orbit was awarded in response to Hubble Observing Problem Report (HOPR) 91647 to re-execute the failed exposure, which proceeded successfully (Visit 56).
\item While initiating Visit 3 of GO:15881 on 2020-Feb-18, the FGS failed to acquire the guide stars before the target acquisition sequence was scheduled to begin.
As a result, following COS safety protocols, the shutter was not allowed to open for the ACQ/IMAGE exposure or the planned spectroscopic exposure.
This failed visit was rescheduled as Visit 53 following HOPR 91715.
\item Full guide star acquisition was unexpectedly delayed during the rescheduled Visit 53 on 2020-Mar-30.
However, the spectroscopic exposure proceeded for the planned duration, and the flux and S/N in the resulting spectrum were in good agreement with the prior observations of J082555 from this program.
Thus, no repeat observations were necessary, and we proceeded with the collected data.
\item Finally, guide star acquisition was again delayed during Visit 8 of GO:15646.
But as for Visit 53, the acquired spectrum closely matched the continuum level and S/N achieved in the previous identical Visit 7 also targeting J104457 which proceeded normally.
We thus used the data collected during this visit as usual.
\end{itemize}
In summary, Visit 3 of GO:15881 and one orbit of Visit 6 of GO:15646 failed and required repeat observations; but the other exceptions were non-critical and did not affect the collected data.

The data for both programs was reduced, calibrated, and combined using \calcos{} (version 3.3.10).
This reduction was performed using the latest reference files (as-of February~2021) from the \hst{} Calibration Reference Data System (CRDS).
Briefly, \calcos{} first corrects and calibrates the raw \texttt{TIME-TAG} photon event lists for each FUV spectroscopic exposure, including corrections for temperature and geometric distortion effects, pulse height filtering for cosmic ray and background rejection, flat-field correction, and wavecal offsets using the contemporaneous \texttt{TAGFLASH} exposures recorded in the same \texttt{TIME-TAG} table.
These corrected photon event tables are then binned to a two-dimensional spectral image, from which a one-dimensional spectrum is extracted.

By default, the extraction is performed using the new \texttt{TWOZONE} algorithm for these FUV spectra. 
In this case, \calcos{} first straightens and centroids the object trace, before summing flux over the object profile (with data quality flags assessed in a second `inner zone' subset of this full profile) and subtracting a per-column background level.
However, since it was optimized for point sources, this algorithm might underestimate the flux of very spatially-extended sources, whose flux might also overlap with regions of more significant gain-sag.
We visually confirmed that the bulk of the FUV light from all six of our sources is reasonably compact in the cross-dispersion direction in the calibrated two-dimensional \texttt{.flt} images, and in particular is well contained within the nominal extraction zone.
To test quantitatively whether deviations from an unresolved source profile might still impact on our final spectra, we also re-extract the spectra using the older \texttt{BOXCAR} algorithm (with \texttt{TRCECORR} and \texttt{ALGNCORR} both set to be omitted).
The resulting one-dimensional spectra are in good agreement with those extracted using the \texttt{TWOZONE} method, with only a small (median 1\%, all $\lesssim 4\%$) increase in the overall continuum flux level.
Thus, we proceed with the preferred default \texttt{TWOZONE} extraction method for our analysis.

After extraction of each spectroscopic exposure, the data from all visits and exposures are joined to produce a final combined one-dimensional spectrum.
The fluxes at each physical wavelength bin in all calibrated sub-spectra without serious data quality flags are combined, weighted by the individual exposure times.
Including data taken at all four \texttt{FP-POS} settings ensures that no large data quality gaps interrupt the final spectrum, and minimizes the residual impact of fixed-pattern features which remain at low levels after flat-fielding.
We also produced \texttt{x1dsum} combined spectra for each visit which contributed to the final combined spectrum.
We do not find any clear signs of either total flux variation or significant changes in spectral features upon comparison of these temporally-separated per-visit and combined spectra.

While \hstcos{} is well-suited for observation of faint UV sources, some difficulties arise in calibrating realistic uncertainties for such observations.
For instance, the uncertainty in the instrumental dark current correction represents a substantial portion of the \calcos{} error budget for faint sources, but this value can be overestimated.
As recommended in the COS Data Handbook \citep{rafelskiCOSDataHandbook2018}, we compare the reported \calcos{} uncertainties to the observed scatter in the flux array to determine whether any correction to the uncertainties is required.
In particular, we compare the standard deviation in the flux to the median-filtered error array at a resolution of $10$ pixels ($\sim 0.12$ \AA{}; a small enough wavelength range for noise to still dominate over real flux variation at the resolution of our spectra).
This comparison reveals that the observed noise in the final spectrum is consistently smaller than predicted by the formal \calcos{} uncertainty array.
We find a fairly constant median value for this ratio for each spectrum, with only a small ($\lesssim 15\%$) shift towards more inflated \calcos{} uncertainties relative to observed scatter in the longer-wavelength Segment~A of the FUV detector compared to Segment~B.
This ratio of observed to formally-expected noise varies from $\simeq0.7$ for the faintest FUV sources in our sample (SB~2, J120202, J082555) up to $0.85$ in the brightest (SB~82), consistent with an origin in over-estimation of instrumental noise sources dominant for faint targets in the \calcos{} pipeline.
We adopt this ratio as-measured for each spectrum as a constant uncertainty correction factor, and multiply the \calcos{} error spectrum for each target accordingly.

We summarize all of the \hstcos{} spectroscopic observations performed for both programs in Table~\ref{tab:cosobs}.
After revising the uncertainties as described, the final combined G160M spectra all exceed the target S/N of 10 per-resel at $1450$ \AA{}, or equivalently $\geq 30$ $\text{\AA{}}^{-1}$: sufficient for a detailed investigation of the stellar continuum in these extreme \civ{} emitters.

\begin{table*}
\centering
\caption{
Details of the \hstcos{} UV spectroscopic observations conducted for each of the target systems.
All executed visits are listed for each object and for the entirety of both proposals (GO:15646 and GO:15881).
Visits with an exception report (some of which did not yield useful data as a result) are indicated with a \textdagger{} and discussed in Section~\ref{sec:uvobs}; the one visit during which usable spectroscopic data was not taken at all is marked with a \textdaggerdbl{}.
The final columns list the S/N per resolution element (6 pixels, or $\sim 0.073$~\AA{}) averaged near 1450 \AA{} both as initially reported by \calcos{} in the reduced one-dimensional spectrum, and after correction based upon the observed scatter in the continuum (see text).
\label{tab:cosobs}
}

\begin{tabular}{lllcccc}

\hline

Target  & Proposal /   & Visits         &  $F_\lambda(1450\mathrm{\AA{}})$ &     Total Exposure     & \multicolumn{2}{c}{S/N(1450~\AA{}) per resel ($0.07$~\AA{})}  \\
Name    & Program ID   &  (UT start)    & $\times 10^{-15} \; \mathrm{erg/s/cm^2/\text{\AA{}}}$   &      Time (s)  & (\calcos{} reported)  & (corrected) \\

\hline
\hline

\multirow{4}*{J082555} & \multirow{4}*{GO:15881 / \texttt{LE4Z}} & 01 (2020-03-01) &\multirow{4}*{1.16} & \multirow{4}*{12869} & \multirow{4}*{8.4} & \multirow{4}*{12.5}\\
 & & 02 (2020-04-24) & & & & \\
 & & 03 (2020-02-18)\textdaggerdbl & & & & \\
 & & 53 (2020-03-30)\textdagger & & & & \\
\hline
\multirow{2}*{SB2} & \multirow{2}*{GO:15646 / \texttt{LDXT}} & 03 (2020-04-05) &\multirow{2}*{0.87} & \multirow{2}*{26925} & \multirow{2}*{10.8} & \multirow{2}*{14.9}\\
 & & 04 (2020-04-06) & & & & \\
\hline
\multirow{2}*{J104457} & \multirow{2}*{GO:15646 / \texttt{LDXT}} & 07 (2020-05-18) &\multirow{2}*{1.55} & \multirow{2}*{26365} & \multirow{2}*{15.7} & \multirow{2}*{21.0}\\
 & & 08 (2020-04-07)\textdagger & & & & \\
\hline
\multirow{2}*{SB82} & \multirow{2}*{GO:15646 / \texttt{LDXT}} & 01 (2019-11-21) &\multirow{2}*{3.51} & \multirow{2}*{29256} & \multirow{2}*{24.0} & \multirow{2}*{28.2}\\
 & & 02 (2019-11-21) & & & & \\
\hline
\multirow{3}*{J120202} & \multirow{3}*{GO:15881 / \texttt{LE4Z}} & 04 (2020-04-26) &\multirow{3}*{1.05} & \multirow{3}*{13498} & \multirow{3}*{7.7} & \multirow{3}*{10.9}\\
 & & 05 (2020-04-26) & & & & \\
 & & 06 (2020-04-12) & & & & \\
\hline
\multirow{3}*{HS1442+4250} & \multirow{3}*{GO:15646 / \texttt{LDXT}} & 05 (2019-10-11) &\multirow{3}*{2.08} & \multirow{3}*{27137} & \multirow{3}*{18.6} & \multirow{3}*{24.7}\\
 & & 06 (2019-10-09)\textdagger & & & & \\
 & & 56 (2019-12-26) & & & & \\
\hline

\end{tabular}

\end{table*}

\subsection{Overview of the New Spectra}
\label{sec:uvmeas}

The deep FUV spectra described above provide a much deeper view onto the ionized gas and massive stars that the targeted galaxies host.
In this section we provide a broad overview of the spectral signatures of both in these new data.

The most striking feature dominating the new G160M/1533 spectra of all six targets is the extraordinarily strong nebular emission from highly-ionized gas species on which these galaxies were selected.
Both components of the semi-forbidden \ion{O}{3}] $\lambda 1661, 1666$ doublet, the \ion{He}{2} $\lambda 1640$ recombination line, and the \ion{C}{4} $\lambda \lambda 1548,1550$ resonant doublet are detected in all six systems.
We fit these lines individually with Gaussians after re-binning the spectra to one resolution element (six pixels).
Comparison with the vacuum wavelengths of these transitions yield the new systemic redshift measurements for our systems presented in Table~\ref{tab:basicprop} and utilized throughout our analysis.
We find small $<10$~km/s ($\lesssim 1\%$) error in this redshift measurement based on the line-to-line variation and the formal fitting uncertainties in the central wavelengths, and thus the uncertainties in these redshifts are likely close to the uncertainty in the medium grating wavelength calibration of 15~km/s \citep{hirschauerCosmicOriginsSpectrograph2021}.
Measurements of these line fluxes are presented in Table~\ref{tab:uvmeas}.

These new deep FUV spectra are especially notable for providing an exquisite view of the multi-component high-ionization \civ{} doublet, illustrating the power of deep UV spectra to probe the detailed physics of star-forming galaxies.
While all of the targets were selected on the identification of nebular \ion{C}{4} emission in archival \hstcos{} spectra, these shallower spectra did not provide a complete picture of this doublet.
In particular, J082555 and J120202 were previously only observed with the low-resolution COS/G140L grating in the FUV which affords nearly an order of magnitude lower spectral resolution than G160M, precluding clear disentanglement of the nebular emission profile from interstellar absorption.
In addition, the broad ($>1000$~km/s) and metallicity-dependent stellar wind contribution to this complex was generally too weak to be confidently detected in the single-orbit spectra obtained previously for this metal-poor galaxy sample \citep[though see][]{senchynaExtremelyMetalpoorGalaxies2019}, leaving us with minimal information about the massive stars underlying this emission.

\begin{figure*}
    \includegraphics[width=1.0\textwidth]{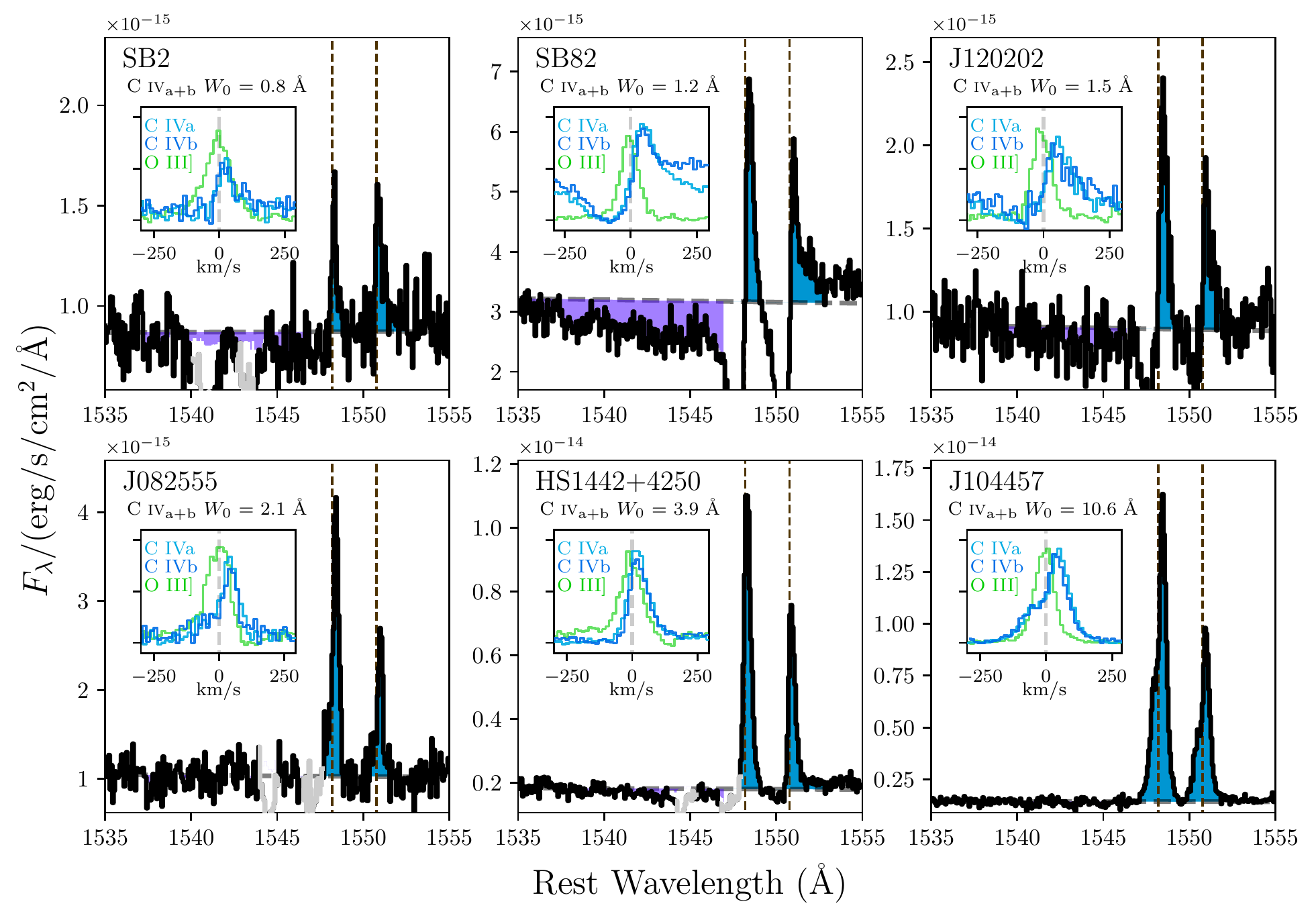}
    \caption{
        The new ultra-deep spectra presented in this paper provide a clear view of the \civ{} $\lambda \lambda 1548,1550$ complex, here ordered by the total equivalent width of nebular emission in this doublet (as indicated) and plotted after binning to 2 resels.
        The rest wavelengths of the two transitions are indicated by vertical dashed lines, and the redshift has been determined using joint fits to the \ion{O}{3}] $\lambda \lambda 1661,1666$ doublet and \heii{} $\lambda 1640$ line in the same spectrum.
        Stellar P-Cygni absorption is clearly detected in several of the targets and indicated by purple shading, while the nebular emission is highlighted in blue; the spectrum is shaded light-grey at wavelengths impacted by intervening MW \civ{} absorption.
        Inset panels show a comparison between the profiles of the two narrow \civ{} components and \ion{O}{3}] $\lambda 1666$ in velocity space.
        All six emit strongly (at 1--11 \AA{} combined equivalent width) in this resonant doublet, indicative of a significant quantity of hot and highly-ionized gas.
        However, a sequence is apparent from purely redshifted emission and blueshifted absorption in the lowest equivalent width emission profiles to broad and sometimes double-peaked emission in the most prominent emitters, likely a signature of resonant scattering.
    }
    \label{fig:civ_overview}
\end{figure*}

We plot the new very high-S/N G160M \civ{} profiles obtained for all six target systems in Figure~\ref{fig:civ_overview}.
All show very clearly-detected emission in both components of the doublet.
We directly integrate this emission for each component separately after first fitting and subtracting-off any visible absorption components of MW and systemic \civ{}, yielding equivalent width measurements ranging from 0.8--10.6 \AA{} (Table~\ref{tab:uvmeas}).
In addition to finding stronger emission than previously reported for J082555 and J120202 \citepalias{senchynaExtremelyMetalpoorGalaxies2019,bergChemicalEvolutionCarbon2019}, we detect clear evidence for non-Gaussianity in the narrow profiles of \civ{} in all six targets.
To illustrate this, an inset panel for each spectrum is included in Figure~\ref{fig:civ_overview} displaying the velocity profile of both \civ{} components and \ion{O}{3}] $\lambda 1666$ for comparison (with redshift determined jointly from the \ion{O}{3}] and \heii{} lines in the same spectra).
In all cases, the \civ{} profiles show structure not evident in the \ion{O}{3}] emission, with peaks of emission uniformly redder than the theoretically-expected wavelengths.

Intriguingly, this velocity structure in the narrow \civ{} profiles appear to correlate with the measured strength of emission in this doublet.
The weakest nebular \civ{} emitters in the sample ($<2$~\AA{}, top row of Figure~\ref{fig:civ_overview}) all show strongly redshifted emission, with either negligible emission or strong absorption blueward of zero velocity.
In contrast, the three strongest emitters (bottom row) show no strong blue-side absorption and include two cases of double-peaked emission profiles \citep[J082555 and J104457; the latter already noted and discussed extensively by][]{bergIntenseIVHe2019}.
The detection of such a sequence, evocative of that followed by Ly$\alpha$ \citep{verhamme3DLyaRadiation2006}, strongly suggests that the \civ{} in these local systems is subject to resonant scattering in its escape from these galaxies \citep[as argued by][]{bergIntenseIVHe2019}.
We discuss the consequences of this realization in more detail in Section~\ref{sec:disc_neb}.

The \civ{} profiles shown in Figure~\ref{fig:civ_overview} also illustrate the power of these spectra to unveil features encoded directly in the stellar continuum.
Strong stellar P-Cygni absorption is detected in SB~82 at $-2.4$ \AA{} equivalent width in depth and in SB~2 and HS~1442+4250 with equivalent widths of $-0.5$ \AA{} (Table~\ref{tab:uvmeas}; integrated from 1530--1550 \AA{} after subtracting Gaussian fits to the ISM/MW and nebular lines and with linear continuum determined in the windows [1490, 1520] and [1560, 1590]).
The other three systems with lower gas-phase metallicities are not formally detected via this full absorption trough integration at their achieved S/N (Table~\ref{tab:uvmeas}), but hints of broad stellar absorption at $<1$~\AA{} EW in depth are still visible in their spectra (see also Fig.~\ref{fig:cbs99_wind}).
This high-velocity P-Cygni feature is produced by outflowing gas in the expanding atmospheres of the massive O stars powering the UV continuum in these galaxies.
At the low metallicities of these systems, these metal line-driven stellar winds and their spectral signatures are usually far too weak to be detected in typical UV spectra.
These spectra thus place some of the strongest and only constraints to date on the winds of massive stars below the metallicity of the Magellanic Clouds where individual stars can be studied en-masse.

In addition to information about uncertain winds, the stellar continuum in these deep UV spectra also provide a rarefied view of photospheric absorption in the underlying OB stars, which provide leverage on the stellar metallicity.
As a first attempt at quantifying this absorption, we measure and report the strength of several relatively broad indices probing stellar features that are commonly measured especially in low-S/N and low-resolution spectra of star-forming galaxies.
In particular, we measure the Bl1425 and Fe1453 indices defined by \citet[][; see also \citealt{vidal-garciaModellingUltravioletlineDiagnostics2017}]{fanelliSpectralSynthesisUltraviolet1992} which are dominated by many stellar photospheric lines of \ion{Fe}{5} and in the former case, also \ion{Fe}{4}, \ion{C}{2} $\lambda 1429$, and \ion{Si}{3} $\lambda 1417$.
In a similar pattern to the \civ{} P-Cygni strengths, we find strong absorption detections of $0.5$--$0.8$ \AA{} EW in depth in SB~82 and HS~1442+4250, weaker absorption in SB~2 of $0.2$--$0.4$ \AA{}, and a range of absorption largely $\lesssim 0.2$ \AA{} in magnitude and often undetected in these broad indices in the remaining three spectra of the lowest-$\mathrm{O/H}$ systems (Table~\ref{tab:uvmeas}).

These initial empirical measurements of the stellar continuum already show clear evidence for a range of stellar absorption strengths, potentially indicative of a significant spread in metallicities.
At first impression, this trend appears broadly in-line with the metallicity trend found in the gas-phase oxygen abundances.
However, quantitative comparison of the new deep UV spectra to stellar models is necessary to extract robust constraints for comparison.
At the high S/N and dispersion achieved in our spectra, our analysis can take advantage of a more detailed continuum comparison than usually possible, which is especially important in the low-metallicity regime where the very broad stellar indices just presented show relatively low dynamic range with variations in metallicity \citep[e.g.][]{vidal-garciaModellingUltravioletlineDiagnostics2017}.
The following section is devoted to describing the model comparison methodology adopted in this work.

\begin{table*}
\footnotesize
\centering
\caption{UV nebular emission line and stellar absorption indices measured in the new G160M spectra. Nebular emission in \ion{C}{4}, \ion{He}{2}, and both \ion{O}{3}] components are detected in all six targets, and constraints on the integrated equivalent width of several key broad stellar absorption indices are also presented.
\label{tab:uvmeas}
}
\begin{tabular}{l|c|cccc|ccc}

\hline
Target & \ion{C}{4} $\lambda 1549$ & \multicolumn{4}{c}{Nebular line fluxes ($10^{-15}$ erg/s/$\mathrm{cm^2}$)} &  \multicolumn{3}{|c}{Stellar index equivalent widths (integrated, \AA{})}  \\
Name & Nebular $W_0$/\AA{} & \ion{C}{4} $\lambda 1549$ & \ion{He}{2} $\lambda 1640$ & \ion{O}{3}] $\lambda 1661$ & \ion{O}{3}] $\lambda 1666$  & \ion{C}{4} $\lambda 1549$ (P-Cygni) & Bl1425 & Fe1453 \\
\hline
\hline

J082555 &$2.14 \pm 0.17$ &$2.19 \pm 0.16$ &$0.15 \pm 0.07$ &$1.35 \pm 0.09$ &$2.83 \pm 0.10$ &$-0.22 \pm 0.18$ &$0.02 \pm 0.09$ &$0.24 \pm 0.13$ \\
SB2 &$0.77 \pm 0.12$ &$0.66 \pm 0.10$ &$1.25 \pm 0.06$ &$1.50 \pm 0.12$ &$3.08 \pm 0.13$ &$-0.53 \pm 0.11$ &$-0.21 \pm 0.08$ &$-0.41 \pm 0.10$ \\
J104457 &$10.60 \pm 0.48$ &$15.57 \pm 0.45$ &$3.12 \pm 0.16$ &$2.92 \pm 0.13$ &$7.13 \pm 0.16$ &$0.37 \pm 0.09$ &$0.24 \pm 0.07$ &$-0.27 \pm 0.07$ \\
SB82 &$1.23 \pm 0.08$ &$3.67 \pm 0.21$ &$1.46 \pm 0.10$ &$2.19 \pm 0.10$ &$5.81 \pm 0.14$ &$-2.37 \pm 0.06$ &$-0.48 \pm 0.05$ &$-0.77 \pm 0.05$ \\
J120202 &$1.46 \pm 0.14$ &$1.26 \pm 0.11$ &$0.67 \pm 0.08$ &$0.84 \pm 0.07$ &$2.02 \pm 0.08$ &$0.01 \pm 0.17$ &$-0.12 \pm 0.12$ &$-0.14 \pm 0.13$ \\
HS1442+4250 &$3.87 \pm 0.11$ &$7.09 \pm 0.18$ &$2.88 \pm 0.07$ &$1.40 \pm 0.10$ &$2.99 \pm 0.10$ &$-0.54 \pm 0.08$ &$-0.53 \pm 0.04$ &$-0.64 \pm 0.06$ \\

\hline

\end{tabular}
\end{table*}

\section{Ultraviolet continuum fitting}
\label{sec:fitting}

The ultraviolet continuum of star-forming galaxies is dominated by light which has been imprinted-on by the photospheres and expanding atmospheres of massive OB stars.
The resulting pattern of emission and absorption from various atomic species provides direct insight onto the fundamental properties of the composite stellar population responsible for this light, including crucially the stellar iron content --- both directly via the strength of photospheric absorption lines of ionized iron, and indirectly through the strength of wind features produced by the iron line-driven winds.
In order to extract constraints on this metallicity and other stellar population properties, it is necessary to compare these patterns to composites of model stellar atmospheres.
In this section, we describe our approach to this model comparison; first by describing the stellar population model framework we utilize, followed by our description of the fitting procedure we follow to constrain the parameters describing this composite population.

\subsection{Stellar Population Synthesis Models}
\label{sec:spsmods}

First we must choose a set of stellar population synthesis models to which our deep UV continuum constraints can be compared.
We choose to focus on the updated \citet{bruzualStellarPopulationSynthesis2003} models (Charlot \& Bruzual in-preparation, hereafter \cb{}), which have been iteratively improved since described by \citet{gutkinModellingNebularEmission2016}.
We refer the reader to \citet{platConstraintsProductionEscape2019} for a more thorough description of the current state of these models, choosing here to highlight the details most pertinent to our continuum fitting analysis.
While we focus on \cb{}, we also repeat our analysis with minor modifications for Starburst99 \citep[hereafter \snn{};][]{leithererEffectsStellarRotation2014} as well; these fits and results are described in Appendix~\ref{app:othermods}.

\subsubsection{Stars}

These evolutionary tracks followed by massive stars and the atmosphere models assigned to them on this journey are the most crucial ingredients shaping the emergent stellar features.
The \cb{} models are underpinned by evolutionary tracks from the PAdova and TRieste Stellar Evolution Code \citep[\parsec{};][]{bressanPARSECStellarTracks2012,chenPARSECEvolutionaryTracks2015} which are used to predict the evolution of stars up to initial masses of $600~\mathrm{M_\odot}$ from $Z=0.0001$ up to $0.040$ (corresponding to $Z/Z_\odot = 0.0066$--$2.6$).
These models include modern treatments of mass loss and the Wolf-Rayet (WR) phase appropriate for such massive stars, extending to much lower metallicities than have previously been employed in such calculations.
As described in \citet[][their Appendix~A]{platConstraintsProductionEscape2019}, these predictions have accordingly been coupled with updated high-resolution fully-theoretical WR atmosphere models from the Potsdam Wolf-Rayet group \citep[PoWR; e.g.][and references therein]{todtPotsdamWolfRayetModel2015}.
This library provides line-blanketed, non-LTE, spherically-expanding atmosphere predictions extensible to WNE, WNL, WC, and WO spectral types down to sub-SMC metallicities.

Next these stellar tracks and atmospheres are used to construct composite SSP spectral models.
The \cb{} models are provided on a uniform moderate-resolution wavelength grid of 0.5~\AA{} steps over the FUV and at 135 ages from $10^4$ to $10^8$ years, which we retain for comparison to the observed spectra.
We adopt a \citet{chabrierGalacticStellarSubstellar2003} IMF and consider upper mass cutoffs of 300 and 600 $\mathrm{M_\odot}$.

Finally, an assumption about the star formation history must be made to produce a composite spectrum.
We consider both a constant star formation history (CSFH) starting at variable time $t$ and combinations of single-age stellar populations (SSPs) to represent the young stellar populations underlying the FUV continuum.
We explore the results of both choices in more detail below, but consider the CSFH results our fiducial products.

\subsubsection{Nebular Emission}

At the young ages of these star-forming complexes, the contribution of nebular continuum to the FUV is non-negligible, reaching up to $\sim 50\%$ of the total continuum at 1450 \AA{} at the lowest metallicities $Z\lesssim0.001$ and youngest ages $\lesssim 2$~Myr.
This additional radiation dilutes the stellar continuum, and must be taken into account when analyzing the strength of stellar absorption features in such observed spectra.
In the range of the considered FUV wavelengths, the nebular continuum emission is dominated by two-photon decay from the first excited state of hydrogen along the forbidden $2s\to 1s$ pathway, with a much smaller contribution from free-bound recombination to the first excited state \citep{drainePhysicsInterstellarIntergalactic2011}.
To account for this addition to the UV light, we utilize the nebular continuum predictions of the \cloudy{} models \citep{ferland2017ReleaseCloudy2017} computed self-consistently alongside the \cb{} stellar models.
We use the standard ionization-bounded predictions and gas-phase abundances described by \citet{gutkinModellingNebularEmission2016} and \citet{platConstraintsProductionEscape2019}.
We fix the gas density $n_H$ to $10^2$ $\mathrm{cm^{-3}}$ (close to the median of 190 $\mathrm{cm^{-3}}$ we find from the optical \ion{S}{2} doublet in metallicity calculation); dust-to-metal mass ratio \xid{} to $0.3$ \citep[which is close to the solar value:][]{gutkinModellingNebularEmission2016}; and volume-average ionization parameter $\log U$ to $-2.5$.
This \cloudy{} calculation also includes dust attenuation from the dust internal to the \hii{} regions computed in a physically-consistent way which we apply to the stellar continuum.
We effectively assume throughout that the stellar and nebular continua are both subject to the same dust column (i.e.\ we assume no differential reddening of the stellar continuum).
This is likely appropriate for the very young stellar populations considered herein, which are unlikely to have dispersed significantly and are predominantly co-spatial with the nebular emission they have excited \citep[e.g.][]{charlotNebularEmissionStarforming2001,senchynaUltravioletSpectraExtreme2021}.

For the purposes of this analysis, these nebular parameters are fixed to the above values and the nebular continuum with no emission lines is simply added to the stellar continuum for each SSP spectrum.
However, we briefly consider the potential impact of varying these quantities on the predicted nebular continuum contribution and on the derived stellar properties.
First, we find that increasing the ionization parameter from the fiducial value of $\log U = -2.5$ has the effect of decreasing the relative contribution of the nebular continuum in the FUV.
This is a consequence of the dust included in the spherical \cloudy{} models used to generate the nebular continuum; an increase in the ionization parameter corresponds to an increase in the filling factor of gas and dust in these models at fixed $Q$ and $n_H$ \citep{platConstraintsProductionEscape2019}, so a larger proportion of the fixed number of ionizing photons are absorbed by dust rather than by the gas, reducing the total recombination rate and thus the UV nebular continuum.
As a result, increasing $\log U$ will decrease the total continuum and increase the predicted equivalent width of the stellar absorption features at fixed other parameters, leading to a lower inferred metallicity for a given observed stellar continuum.
However, we find that this effect is generally small; leading to downward offsets in the derived $Z$ of $<0.05$ dex up to $\log U=-1.5$ and $0.1$--$0.2$ dex in the very worst case of $\log U=-1.0$ \citep[likely far larger than is appropriate even the most extreme systems in our sample; e.g.][]{bergIntenseIVHe2019}.

We also examine the impact of the gas density $n_\mathrm{H}$ on the nebular continuum and our results.
The two-photon emission channel which dominates the nebular continuum in this FUV range can be suppressed by collisional deexcitation at sufficiently high electron densities.
Thus an increase in the assumed $n_\mathrm{H}$ would qualitatively produce a similar effect to an increase in the model $\log U$; diminishing the level of the nebular continuum and decreasing the inferred metallicity.
However, we compute models with $n_\mathrm{H}=500 \; \mathrm{cm}^{-3}$ (much larger than inferred in any of our galaxies) to test the magnitude of this effect, and find that the decrease in the nebular continuum relative to our primary $100 \; \mathrm{cm}^{-3}$ models amounts to $<10\%$ over the considered wavelength range, leading to a much smaller effect than that induced by $\log U$.
To summarize, we estimate that uncertainties in $\log U$ and $n_\mathrm{H}$ have only a small impact on our conclusions, and would produce a minor systematic error towards inferred metallicities $\lesssim 0.1$ dex higher than reality if either $\log U$ or $n_\mathrm{H}$ are underestimated by our fiducial model assumptions.

\subsection{Spectra Comparison and Fitting}
\label{sec:fittingdetails}

With stellar models assembled, we proceed to comparison with the observed FUV spectra.
In this paper we are primarily interested in leveraging the information encoded in the forest of FUV photospheric absorption lines, which directly access ionized species including iron in the atmospheres of the massive stars present.
We accomplish this through direct comparison of the measured spectra to the models, taking full advantage of the resolution and wavelength coverage of the ultra-deep spectra.
To minimize the confounding effects of extinction (potentially a source of significant systematic uncertainty in the FUV even at the modest $\mathrm{E(B-V)}$ values of our sample; Table~\ref{tab:basicprop}), we perform all of this model comparison in continuum-normalized space.
In this subsection we describe the processing both the observed and model spectra undergo, and the Bayesian model comparison we then perform to extract constraints on the key model parameters of interest.
We present and discuss the results of this fitting in the following sections.

First, we must ensure that the data and model comparison only includes features that we wish to compare.
Most crucially, the data are impacted by nebular line emission (Section~\ref{sec:uvmeas}), as well as narrow gas absorption lines both from surrounding or outflowing gas at the galaxy systemic redshift as well as from the MW halo (notably strong lines of \ion{Al}{2}, \ion{C}{2}, \ion{C}{4}, \ion{Si}{2}, \ion{Si}{4}, and \ion{Fe}{2}) which must be excluded if we are to compare the stellar continuum cleanly.
We conservatively mask $\pm 1.5$ \AA{} around the expected wavelength of each of these transitions, with wavelengths taken from NIST and the literature where appropriate, which we find is well beyond their typical measured width and is sufficient to remove them from consideration.
We also mask the 1--2.5 \AA{} closest to the FUV chip edges, which tend to be noisier and suffer from some residual calibration artifacts.

Our primary focus is on the photospheric lines, but the deep constraints on the winds lines provide important information about the massive stars present as well.
In model comparison we choose to always mask the stellar wind lines \ion{O}{5} $\lambda 1371$ and \ion{Si}{4} $\lambda \lambda 1393, 1402$; \ion{O}{5} has generally poor morphological agreement between the data and models, and \ion{Si}{4} is strongly impacted by difficult-to-clean interstellar absorption.
We generally also mask generously around the \civ{} $\lambda \lambda 1548,1550$ and \heii{} $\lambda 1640$ wind lines as well, but the wings of the \civ{} and \heii{} wind profiles beyond the central regions impacted by nebular emission are in contrast straightforward to compare (and we include the uncontaminated portions of the \civ{} profile in some fits as described below).

Next, both the observed and model spectra must be continuum-normalized in a self-consistent way.
In both cases, we fit a cubic spline to the wavelength region from 1310--1710 \AA{} (encompassing the rest-wavelength span of all of the G160M/1533 spectra presented here) after masking the strong wind lines (and other gas lines in the case of the observed spectra),
with intermediate knots placed at in the relatively clean continuum at rest-frame 1450 and 1590~\AA{}.
This common procedure ensures that data comparison occurs in a common normalized flux frame; but we also include a nuisance parameter in our fitting to account for a potential residual constant offset.

The actual model comparison proceeds as-follows.
First, the normalized model spectrum is assembled from the specified stellar metallicity and age (or other star formation history parameters).
We linearly interpolate the flux in continuum-normalized space between the gridpoints so that the predicted flux is continuous in $Z$ and $t$.
Then, the raw continuum-normalized spectral data is corrected to the restframe by redshift $z_\mathrm{adopt} = f_z \, z$, where $z$ is the systemic redshift estimated from the UV nebular emission lines in the same spectrum (Table~\ref{tab:basicprop}) and $f_z$ is a free model parameter accounting for potential uncertainties in this correction.
To account for the different spectral resolutions of the data and model spectra, the data are then smoothed by a Gaussian filter whose size $\sigma_s$ in COS spectral pixels is allowed to vary as another free parameter.
We then median-bin the data onto the model wavelength grid over the region of interest, with masked regions ignored.
The uncertainties on the binned data are computed in the usual way, but with variance adjusted by a fractional amount \fvar{}:
\begin{equation}
\sigma_{F,\mathrm{adopt}}^2 = \sigma_{F,\mathrm{COS}}^2 + (f_{\mathrm{var}} \cdot F_\mathrm{\lambda,COS} )^2 .
\end{equation}
This variable \fvar{} thus accounts for the possibility that our (already revised downwards; see Sec.~\ref{sec:uvcal}) COS spectral uncertainties might underestimate the real error.
Our Bayesian likelihood function $\mathscr{L}$ assuming uncorrelated Gaussian uncertainties is then given as (modulo a constant)
\begin{equation}
\ln \mathscr{L} = -\frac{1}{2}\sum_i\left[ \frac{(F_{\mathrm{i,COS}}-F_\mathrm{i,mod}-\delta y)^2}{\sigma_{F_i,\mathrm{adopt}}^2} + \ln \sigma_{F_i,\mathrm{adopt}}^2 \right] + C
\end{equation}
where $\delta y$ is the freely-varying constant adjustment to the model continuum normalization; and all fluxes $F$ and uncertainties thereon $\sigma_F$ are continuum-normalized quantities.

In total, we now have a model for our continuum-normalized spectrum consisting of $\geq6$ parameters $\Theta$:
\begin{itemize}
    \item First, the $\geq 1$ parameters describing the star formation history. 
    This is simply the effective stellar population age $\log (t/\mathrm{yr})$ in the case of a CSFH or SSP alone.
    We either allow this quantity to vary over a uniform prior spanning the model age grid; or (as discussed below) assume a narrow Gaussian prior ($\sigma=0.1$~dex.) fixed to the values derived for each object from the optical by \beagle{} (Table~\ref{tab:sedfits}).
    \item We also consider combinations of this base SFH with an additional SSP.
    This adds two parameters; the age of this second SSP $\log(t_{\mathrm{+SSP}})$ and its fractional contribution to the continuum at 1450 \AA{} $f_{\mathrm{+SSP}}$.
    \item The stellar metallicity, $\log Z$.
    We always allow $\log Z$ to vary over a uniform prior from the minimum metallicity of the grid up to $0.5 Z_\odot$, well above the metallicities preferred by these spectra.
    \item The multiplicative adjustment to the measured redshift $f_z$, which we allow to vary by $\pm 3$\%.
    \item The linear offset in the model normalization $\delta y$, which we allow to vary over $(-0.1, 0.1)$.
    \item The Gaussian smoothing kernel size $\sigma_s$. 
    For the \cb{} models, we allow this to vary from 30--70 pixels, corresponding to $0.4$--$0.8$ \AA{}.
\end{itemize}
As usual, we are now interested in estimating the posterior distribution $\mathscr{P}(\Theta|\mathbf{D},M)$; the likelihood as a function of these model parameters given this model $M$ and our data $\mathbf{D}$.

To explore this parameter space and obtain constraints on the parameters of interest, we utilize \dynesty{} \citep{speagleDYNESTYDynamicNested2020}.
This code implements both static and dynamic nested sampling \citep{skillingNestedSampling2004,skillingNestedSamplingGeneral2006,higsonDynamicNestedSampling2019}, which is a method for estimating simultaneously both the evidence $P(\mathbf{D}|M)$ and the posterior by integration in shells of constant likelihood.
Though this method has more stringent requirements for prior specification than for instance Markov Chain Monte Carlo, this also forces it to fully-explore the specified prior volume, making it far less likely to miss regions of high likelihood.
We use the static nested sampler, adopting the default configuration of uniform sampling with multi-ellipsoidal decomposition bounding \citep{ferozMULTINESTEfficientRobust2009}.
We extract confidence intervals on the key parameters of interest from the weighted sample distributions and draw samples from the posterior using the helper functions built-into \dynesty{} for precisely these functions \citep[see also][]{holResamplingAlgorithmsParticle2006}.

\begin{figure}
\plotone{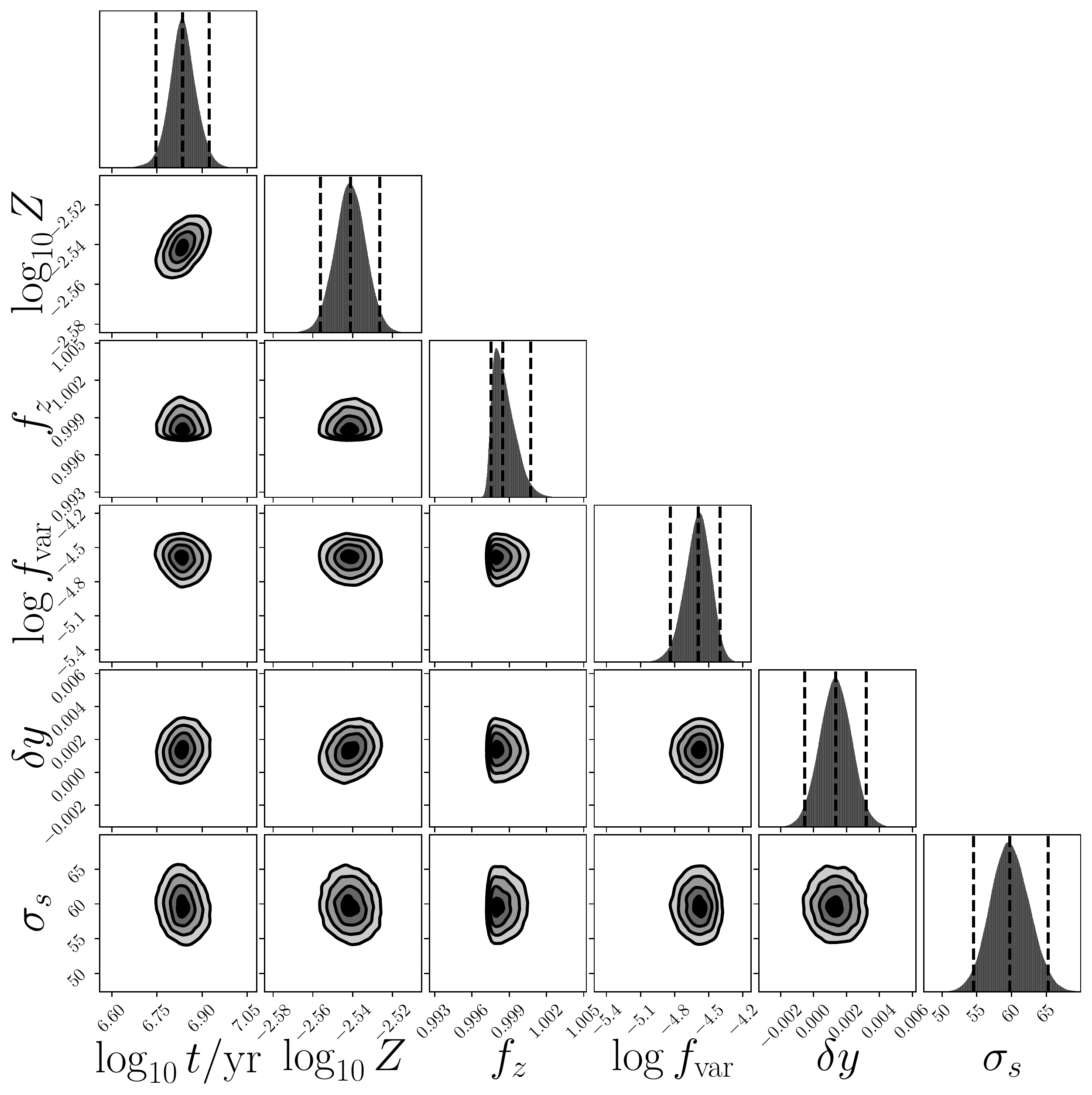}
\caption{
\label{fig:corner}
Corner plot for the \dynesty{} fit to the full SB~82 UV spectrum, with all wind features but \civ{} masked and assuming a variable-age CSFH and flat priors.
All of the parameters including the `nuisance' quantities are very well-constrained, with near-Gaussian distributions with minimal covariance found for the key physical parameters.
}
\end{figure}

For illustration, we highlight the resulting posterior distribution in a corner plot presented in Figure~\ref{fig:corner} for the particularly well-constrained case of SB~82.
In general, the fit agreement to the photospheric continuum alone with wind lines masked is reasonable, with reduced-$\chi^2$ of-order unity and excellent morphological agreement with the highest S/N spectra (Figure~\ref{fig:photo1450comp}).
Our photospheric fits with the \cb{} models uniformly support very small values for the multiplicative uncertainty factor \fvar{}$< 0.01$, supporting the notion that our revised spectral uncertainties are not significantly underestimated.
We discuss the fit results and their implied constraints on the stellar population properties of interest in the following section.

\section{Results}
\label{sec:results}

The UV continuum fitting described in the previous section yields Bayesian constraints on the basic properties of the massive star populations in the context of the assumed stellar models.
We now examine these fitting results (with important derived quantities compiled in Table~\ref{tab:metresults}) and consider them in the context of the other measurements of these systems.

\subsection{Metallicities from the photospheric continuum}
\label{sec:photcomp}

\subsubsection{Reproducing the photospheric lines}

\begin{figure*}
\plotone{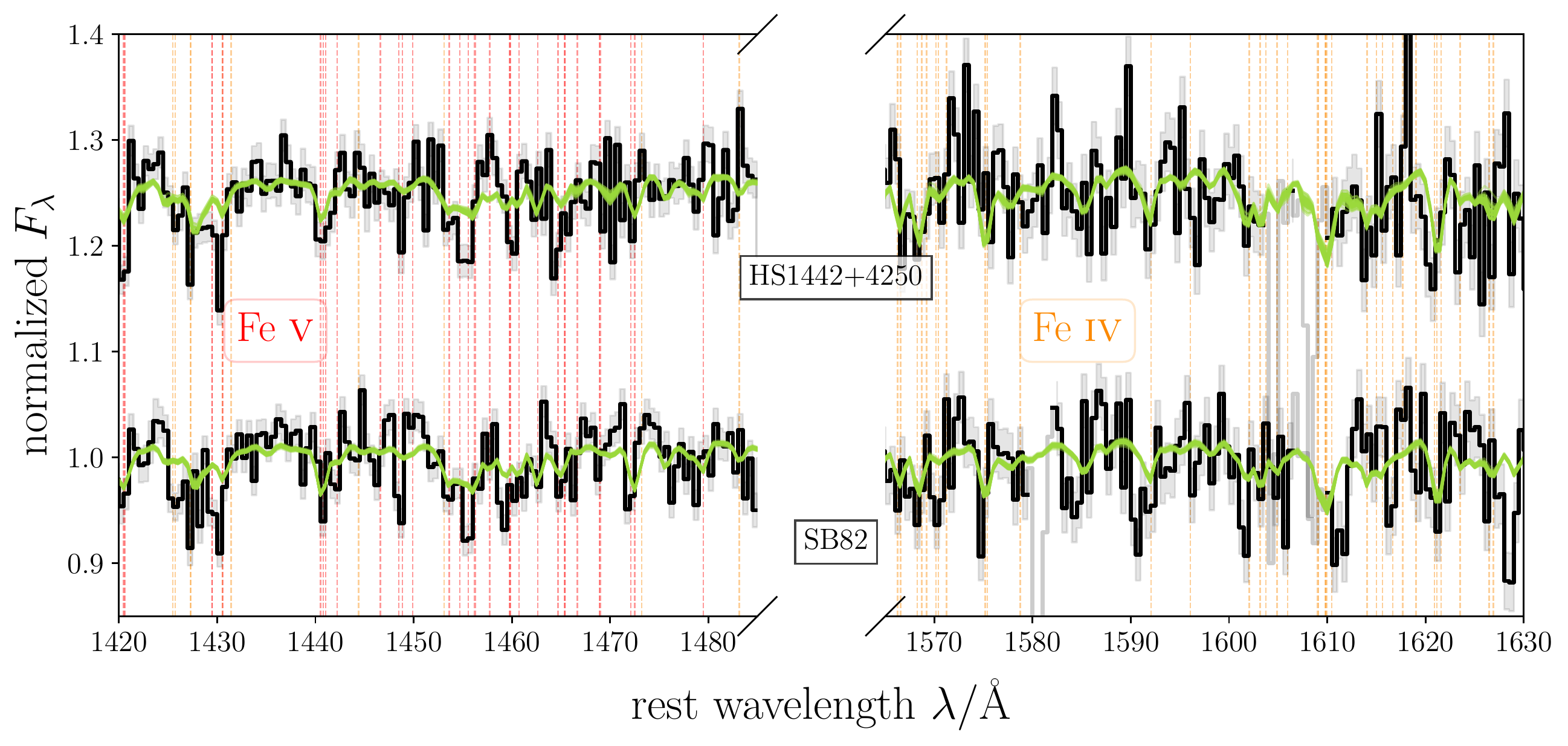}
\caption{
    Zoom-in on stellar continuum fits at the photospheric complex near 1450\AA{} for the two highest-S/N spectra in our sample: SB~82 and HS~1442+4250.
    The observed continuum-normalized data binned to model resolution is overlaid by \cb{} model spectra drawn from the posterior found by \dynesty{}, for the full cleaned G160M spectrum without stellar wind lines included.
    Many blended iron absorption features are detected at model resolution and well-reproduced by the \cb{} model spectra.
    \label{fig:photo1450comp}
}
\end{figure*}

First, we examine how well the model spectra reproduce the forest of iron photospheric lines that provide direct leverage on the metallicity of the massive stars.
In Figure~\ref{fig:photo1450comp} we compare model spectra drawn from the posterior distribution of the \cb{} fits against the observed binned spectra for the two highest-S/N spectra in our sample, SB~82 and HS~1442+4250, focusing on the regions 1420--1480 and 1570--1630~\AA{} (the \cb{} and \snn{} fits for the other galaxies are included in Figure~\ref{fig:cbs99_photo}).
These two regions of the spectrum of hot OB stars are expected to be dominated by a forest of \ion{Fe}{5} and \ion{Fe}{4} lines (respectively), with minimal contamination from lines impacted by ISM absorption or stellar winds.
By focusing on lines with minimal wind impact, we bypass much of the significant additional modeling uncertainties inherent to reproducing the structure of and radiative transfer through the expanding outer atmosphere of a wide variety of hot stars whose wide-ranging spectra must then be combined in the correct proportion (see Section~\ref{sec:windcomp} below).
We indicate the expected vacuum wavelengths of the transitions predicted to be most prominent in hot stellar atmospheres for each of these iron species as reported by the Vienna Atomic Line Database \citep[VALD3;][and references therein]{kuruczRobertKuruczOnline2006,ryabchikovaMajorUpgradeVALD2015}.

The comparison in Figure~\ref{fig:photo1450comp} reveals remarkably good agreement between the model and observed spectra in this region within the data uncertainties.
In this wavelength range, a majority of the fluctuations in the observed binned continuum are due to real photospheric metal line absorption, detected cleanly at the resolution of the \cb{} SPS models.
The fidelity with which the \cb{} model spectra are able to match most of the `wiggles' evident in the observed binned continuum in these systems attests both to the reality of these features in the data and to the precision of these model spectra.
The lower-S/N spectra of the other targets (most notably J082555 and J120202) unsurprisingly reveal less striking agreement with the models (Figure~\ref{fig:cbs99_photo}).
However, this is appropriately reflected in correspondingly large uncertainties in the derived metallicities.

\begin{figure}
\plotone{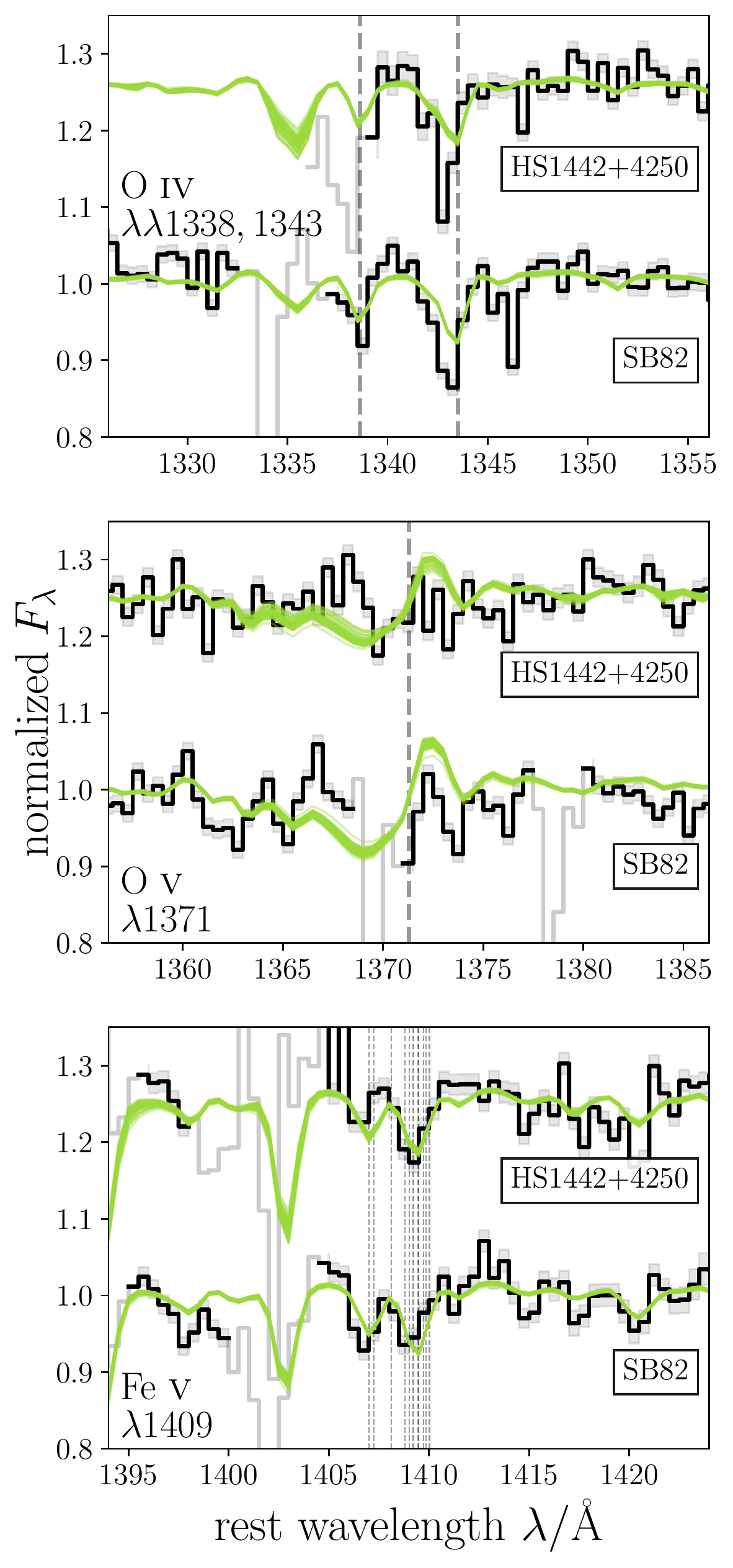}
\caption{
    Zoom-in on \cb{} stellar continuum fits at several key lines of oxygen for the two highest-S/N spectra (HS~1442+4250 and SB~82): \ion{O}{4}~$\lambda \lambda 1338,1343$, \ion{O}{5}~$\lambda 1371$, and for comparison the strong blend of iron transitions near 1409~\AA{}.
    The models which are predominantly constrained to match the numerous lines of Fe in the continuum appear to systematically underestimate the strength of \ion{O}{4} in these systems; qualitatively consistent with an overabundance of these elements relative to a solar chemistry.
    The stellar wind-dominated model profile of \ion{O}{5} is strongly inconsistent with the data, indicative of deficiencies in the treatment of massive star winds in the underlying model atmospheres (Section~\ref{sec:windcomp}).
    \label{fig:otherphotocomp}
}
\end{figure}

We note that the wavelength region spanned by our spectra also provide access to several other strong photospheric lines of non-iron peak species.
Most prominent and least likely to be contaminated by ISM absorption among these in our systems are \ion{O}{4} $\lambda\lambda 1338,1343$, \ion{O}{5} $\lambda 1371$, and \ion{S}{5} $\lambda 1502$.
We highlight the oxygen line complexes in comparison to a prominent blend of iron transitions for our two highest-S/N spectra in Figure~\ref{fig:otherphotocomp} (the constraints on these lines for the other targets and the \snn{} fits are visible in Figures~\ref{fig:cbs99_photo} and \ref{fig:otherwindcomp}).
Our fits are driven primarily by the far more numerous lines of iron in the continuum; and the model spectra (both \cb{} and \snn{}) which best-match this iron line forest appear to systematically underestimate the strength of these lines of oxygen and other excited $\alpha$-elements.
This comparison is complicated by the fact that stellar winds can contaminate several of these transitions especially \ion{O}{5} $\lambda 1371$ whose strong model mismatch we discuss further in Section~\ref{sec:windcomp}, though we note that only a small velocity asymmetry as would be expected from a wind contribution is detected in \ion{O}{4} $\lambda\lambda 1338,1343$.
Modulo the uncertain wind contribution, this systematic inability to match the strength of these photospheric lines of $\alpha$ elements is qualitatively consistent with an overabundance of these elements with respect to iron in the observed galaxies relative to the solar chemistry assumed by the models employed here.
We discuss the consequences of this possibility further in Section~\ref{sec:disc_stellar}.

\subsubsection{Stellar metallicity constraints}

The photospheric continuum fits provide Bayesian constraints on the stellar metallicity of these systems driven by the iron line forest.
We present these constraints (medians and 68\% confidence intervals) in Table~\ref{tab:metresults} for several different sets of model assumptions.
Here, we focus on the metallicity posterior derived assuming a constant star formation history with a Gaussian prior on age set by the $ugriz$ SED fits to each system (Table~\ref{tab:sedfits}, with Gaussian $\sigma$ of 0.05~dex).
We adopt this measurement as our fiducial metallicity for several reasons: it involves a minimal number of free parameters for the star formation history, and recovery of low metallicities near the edge of our metallicity prior is broadly successful even at S/N well below that reached by our sample with such an age prior (Appendix~\ref{app:snretrieval}).
We note that minimal changes in the derived metallicity are observed when a conservative flat prior is adopted in age instead beyond a broadening of the posterior constraints, with the largest shift being a $\sim 1.9\sigma$ increase in $Z$ for J104457 (Table~\ref{tab:metresults}).
We discuss the other star formation history results in Section~\ref{sec:windcomp} and Appendix~\ref{app:sfh} below, and the comparison with \snn{} in Appendix~\ref{app:othermods}.

The metallicities derived from the UV photospheric line fits are uniformly low.
The UV spectra of these systems prefer metallicities below 10\% solar ($\log Z = \log (0.1 \; Z_\odot) = \log (0.001524) = -2.82$), with all but SB~2 confidently below this metallicity at $>1$-$\sigma$.
This confirms that these systems are almost uniformly host to extremely metal-poor massive stars, despite the fact that two have $T_e$-derived oxygen abundances greater than the commonly-adopted 10\% solar cutoff at $12+\log\mathrm{O/H}=7.7$ (Table~\ref{tab:basicprop}).
However, the results for our sample as a whole are also confidently not in the realm of nearly metal-free stars.
All six have at least some photospheric metal line detection evident (Fig.~\ref{fig:cbs99_photo}); and only the two lowest-metallicity and lowest-S/N spectra (J082555 and J120202) have posterior distributions approaching the bottom of our model metallicity range (and thus prior) at $\log Z = -4$.
The relatively confidently-measured metallicities for the remaining four objects range from $-3.2$ to $-2.9$ in $\log Z$, corresponding to 4--8\% solar.

The highest photospheric metallicities we infer in our sample occur in SB~2 ($\log Z = -2.93^{+0.19}_{-0.15}$), HS~1442+4250 ($-2.94^{+0.10}_{-0.15}$), and SB~82 ($-3.00^{+0.05}_{-0.04}$).
These systems reveal the clearest detections of photospheric absorption lines (Fig.~\ref{fig:photo1450comp},\ref{fig:cbs99_photo}), with inferred model `wiggles' notably more prominent than those in the three other systems.
These systems also reveal the three highest abundances of gas-phase O/H; with the other three lower-O/H galaxies all constrained to confidently (at the 84\% confidence level) $\log Z<-3.0$ ($<6\%$ solar).
Further detailed comparison of these three lowest-metallicity systems is challenging due to both weak continuum features and low-S/N in the spectra for J082555 and J120202.
However, the apparent first-order correspondence between the gas and stellar abundances is less clear when the three higher metallicity systems are examined more closely.
In the gas phase, these three highest-$Z$ systems vary in O/H by 0.27~dex, with HS~1442+4250 at $12+\log\mathrm{O/H}=7.66\pm 0.04$ and SB~82 at $12+\log\mathrm{O/H}=7.93\pm 0.04$.
However, their stellar metallicities are far more tightly clustered over $\sim 0.1$~dex, with HS~1442+4250 and SB~82 in particular highly inconsistent in gas-phase O/H but with fairly tightly overlapping stellar metallicity distributions.
The similarity in their inferred stellar $Z$ is consistent with the qualitatively-similar strength of photospheric absorption in their spectra (Fig.~\ref{fig:photo1450comp}) and empirically-measured absorption indices (Table~\ref{tab:uvmeas}).

This analysis of the UV continuum features in these galaxies has revealed a surprising fact: sorting these galaxies by gas-phase O/H yields a different rank-ordering and a different spread than implied by the stellar photospheric lines.
This immediately has important consequences for any work where only gas-phase abundances are known, which we discuss further in Section~\ref{sec:disc_neb}.
It also strongly suggests that another variable is at-play in setting the gas-phase oxygen and stellar iron abundances.

\begin{table*}
\centering

\caption{
    Stellar metallicities inferred from the population synthesis fits described in Section~\ref{sec:fitting}.
    The fiducial stellar metallicity used in this paper is first presented, inferred from fits to the photospheric line-dominated normalized UV stellar continuum assuming a CSFH with age prior set by SED fitting.
    We also include stellar metallicities and ages inferred from fits with a flat prior to the same masked and normalized UV continuum assuming a both a CSFH and a pair of SSPs with variable age below or above 5~Myr.
    Finally, we include metallicities derived given these two SFH assumptions when the broad stellar wind components of the \civ{} profile are fit directly alongside the photospheric lines.
\label{tab:metresults}
}
\begin{tabular}{l|c|cc|cccc|cc}

\hline
Target  & Photospheric & \multicolumn{2}{c|}{CSFH, no age prior} & \multicolumn{4}{c|}{2 SSPs} & \multicolumn{2}{c}{$\log Z$ with stellar \civ{}}   \\
Name &  $\log Z$ (fiducial) &  $\log Z$ & $\log t/\mathrm{yr}$ &   $\log Z$ & $\log t_1/\mathrm{yr}$ & $\log t_2/\mathrm{yr}$ & $f_\mathrm{SSP_2}/f_\mathrm{SSP_1}$ & CSFH & 2 SSPs \\
\hline
\hline

J082555 & $-3.43^{+0.34}_{-0.49}$ & $-3.47^{+0.32}_{-0.40}$ & $7.44^{+0.28}_{-0.24}$ & $-3.46^{+0.22}_{-0.32}$ & $5.30^{+0.60}_{-0.68}$ & $6.97^{+0.10}_{-0.11}$ & $0.46^{+0.14}_{-0.13}$ & $-3.51^{+0.12}_{-0.17}$ & $-3.57^{+0.26}_{-0.24}$\\ 
SB2 & $-2.93^{+0.19}_{-0.15}$ & $-2.91^{+0.18}_{-0.21}$ & $6.68^{+0.49}_{-0.24}$ & $-3.03^{+0.16}_{-0.13}$ & $6.20^{+0.02}_{-0.04}$ & $7.22^{+0.12}_{-0.20}$ & $0.22^{+0.09}_{-0.08}$ & $-2.69^{+0.02}_{-0.07}$ & $-2.99^{+0.05}_{-0.05}$\\ 
J104457 & $-3.14^{+0.13}_{-0.11}$ & $-2.78^{+0.07}_{-0.16}$ & $5.32^{+0.64}_{-0.54}$ & $-2.79^{+0.08}_{-0.14}$ & $4.81^{+0.48}_{-0.40}$ & $6.78^{+0.16}_{-0.06}$ & $0.10^{+0.08}_{-0.07}$ & $-2.96^{+0.09}_{-0.08}$ & $-3.14^{+0.16}_{-0.17}$\\ 
SB82 & $-3.00^{+0.05}_{-0.04}$ & $-3.01^{+0.06}_{-0.07}$ & $7.01^{+0.10}_{-0.08}$ & $-3.07^{+0.07}_{-0.06}$ & $6.24^{+0.02}_{-0.10}$ & $6.82^{+0.09}_{-0.07}$ & $0.28^{+0.16}_{-0.08}$ & $-2.54^{+0.01}_{-0.01}$ & $-2.62^{+0.01}_{-0.01}$\\ 
J120202 & $-3.35^{+0.23}_{-0.43}$ & $-3.32^{+0.21}_{-0.47}$ & $6.75^{+0.69}_{-1.02}$ & $-3.22^{+0.18}_{-0.20}$ & $5.26^{+0.55}_{-0.57}$ & $6.86^{+0.25}_{-0.12}$ & $0.24^{+0.16}_{-0.12}$ & $-3.30^{+0.47}_{-0.67}$ & $-3.61^{+0.13}_{-0.13}$\\ 
HS1442+4250 & $-2.94^{+0.10}_{-0.15}$ & $-2.96^{+0.11}_{-0.13}$ & $7.21^{+0.14}_{-0.16}$ & $-3.26^{+0.06}_{-0.04}$ & $6.34^{+0.00}_{-0.00}$ & $7.07^{+0.06}_{-0.07}$ & $0.30^{+0.06}_{-0.06}$ & $-2.65^{+0.01}_{-0.01}$ & $-2.80^{+0.02}_{-0.02}$\\ 

\hline

\end{tabular}
\end{table*}

\subsection{The O/Fe abundance ratio}
\label{sec:abundcomp}

\begin{figure}
\plotone{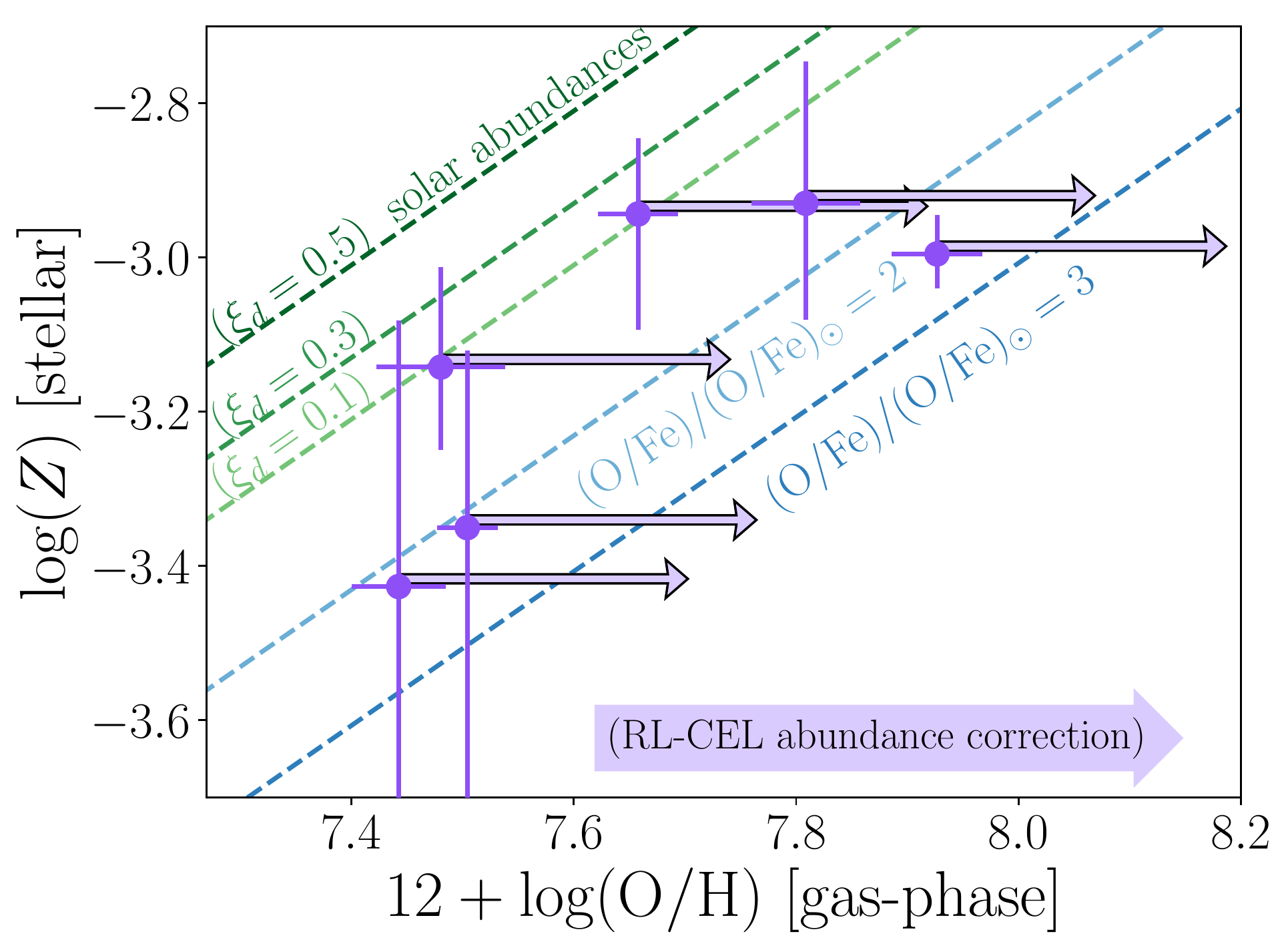}
\caption{
    Comparing the gas-phase oxygen abundances derived from the direct-$T_e$ method and stellar metallicity constraints found from ultraviolet photospheric continuum fits for the galaxies in our sample.
    We overplot green dashed lines corresponding to an assumed solar abundance scale with varying dust depletion factors \xid{} according to \citet[][]{gutkinModellingNebularEmission2016}, and two blue dashed lines indicating the enhancement expected from an O/Fe overabundance relative to solar of 2 and 3 times (with \xid{}$=0.3$).
    We also highlight with arrows the shift introduced to the gas-phase abundances when accounting for an offset between the recombination line and collisionally-excited line abundance scales (see Section~\ref{sec:abundcomp}).
    Even without any abundance scale shift, the measured stellar metallicities lie systematically below those expected given the gas-phase oxygen abundances under the assumption of solar abundances.
    \label{fig:abundances_abs}
}
\end{figure}

Direct comparison of the iron-driven stellar photospheric metallicities and gas-phase oxygen measurements should yield an estimate of their relative abundance, and thereby insight into the $\alpha$/Fe ratio in the star-forming ISM of these galaxies.
We plot our two abundance measurements in Figure~\ref{fig:abundances_abs}, which immediately makes clear the lack of a direct proportionality between these quantities as discussed in the previous subsection and thus the possibility of a varying O/Fe ratio.
However, comparing these abundances requires converting the ionized O/H measurement in-hand to an estimate of total O/H commensurate with the stellar metallicity constraint.
Correspondingly, Figure~\ref{fig:abundances_abs} also includes arrows representing gas-phase corrections and lines corresponding to solar abundance patterns with varying dust which we discuss in turn in this subsection.

First, in order to place our oxygen abundances confidently on an absolute scale, we consider the abundance discrepancy problem associated with ionized nebulae \citep[e.g.][]{tsamisHeavyElementsGalactic2003}.
The vast majority of integrated galaxy studies rely on gas-phase oxygen abundances derived from the strong collisionally-excited lines (CELs) of [\ion{O}{2}] and [\ion{O}{3}], as derived for our systems with the direct-$T_e$ method in Section~\ref{sec:gasmet}.
However, far weaker optical recombination lines (RLs) of oxygen detected in bright nearby \hii{}-regions enable an alternative measurement of O/H, which is found to be systematically larger by a fairly constant factor of order 2 than that derived from the collisionally-excited lines \citep[e.g.][]{peimbertAbundanceRatioGaseous1993,garcia-rojasAbundanceDiscrepancyProblem2007,estebanCarbonOxygenAbundances2014}.
The origin of this offset and the correct metallicity scale remain unclear \citep[see][for more detailed discussion and references]{maiolinoReMetallicaCosmic2019}, especially at low metallicity where recombination line measurements are scarce to non-existent \citep[see e.g.][for an argument for smaller corrections at low metallicity]{bresolinYoungStarsIonized2016}.
To test this in metal-poor dwarf galaxies, we have obtained deep echelle spectra in a complementary program which yield oxygen RL detections for two of the targets in this sample: SB~2 and 82, from which we obtain preliminary RL metallicities of $12+\log\mathrm{O/H}=8.03_{-0.14}^{+0.06}$ and $8.31_{-0.07}^{+0.06}$, respectively (Sanders et al., in-prep).
These measurements are 0.22 and 0.38 dex higher than the direct-$T_e$ measurements we derived for these systems in Table~\ref{tab:basicprop}; consistent with previous \hii{}-region samples presented above.
Motivated by these local findings and for consistency with similar analyses of gas and stellar abundances at $z\sim 2$ \citep[e.g.][]{steidelReconcilingStellarNebular2016,sandersMOSDEFSurveyDirectmethod2020}, we adopt the recombination line scale.
For simplicity, since we do not have RL measurements for the full sample, we apply a constant $+0.24$ dex offset (factor of 1.7) to the direct-$T_e$ O/H abundances from Table~\ref{tab:basicprop} for each galaxy.\footnote{Note also that this offset is slightly larger but comparable to the difference observed between metallicities derived with the direct-$T_e$ method and those found from the same emission line measurements using photoionization modeling analysis with an earlier version of the \cb{} models and a sample of higher-metallicity star-forming regions in \citet{senchynaUltravioletSpectraExtreme2021}. Such analyses provide additional evidence that the direct-$T_e$ method may systematically underestimate gas-phase abundances in similar galaxies.}

Second, we must account for the fact that not all of the oxygen in the ISM is in the ionized gas phase.
In particular, oxygen is depleted from the gas when it is locked into dust grains, depressing the O/H inferred from ionized oxygen relative to the total true ISM O/H.
To account for this, we utilize the results of the detailed abundance analysis presented in \citet{gutkinModellingNebularEmission2016} which forms the basis of the \cb{} models.
The scale of this depletion correction depends upon the assumed dust-to-metal mass ratio \xid{} (see solar abundance lines in Figure~\ref{fig:abundances_abs}).
We note that there is some empirical evidence for evolution in this ratio as a function of metallicity, with metal-poor nearby galaxies showing lower \xid{} on average but with substantial scatter \citep[e.g.][]{devisSystematicMetallicityStudy2019}.
For simplicity and with a lack of direct constraints on this quantity for these systems, we assume in this analysis a solar \xid{}$=0.3$, which leads to a small correction upwards in O/H of $\sim 0.11$ dex (corresponding to $\sim$30\% depletion factor for oxygen).
Assuming a lower $\xid{}=0.1$ would lower this correction to $\sim 0.04$, but we note that the adopted correction is close to that derived for similarly metal-poor galaxies from refractory depletion patterns \citep[e.g.][]{peimbertMgSiFe2010}.

After these corrections, the resulting ISM O/H is readily-converted to a total metallicity $Z$ assuming solar abundances \citep{gutkinModellingNebularEmission2016}.
Since as we have motivated our photospheric metallicity is driven by iron transitions, the stellar $Z$ derived here corresponds to an estimate of Fe/H assuming the same abundance scale.
We thus interpret the residual difference between this stellar $Z$ and gas-phase oxygen derived $Z$ as an inference of $[\mathrm{O/Fe}] \equiv \log \left((\mathrm{O/Fe})/(\mathrm{O/Fe})_\odot\right)$.
The corresponding measurements (accounting for the statistical uncertainties in our direct-$T_e$ O/H measurement and the full posterior in stellar $Z$ estimation) are presented in Table~\ref{tab:ofeh}.

Comparison of the stellar photospheric metallicities with the corrected gas-phase oxygen measurements provides strong evidence for systematically super-solar O/Fe ratios.
This is immediately visible as a bias away from the solar abundance tracks in Figure~\ref{fig:abundances_abs}, enhanced by the RL-CEL correction but present even in comparison with the direct-$T_e$ metallicities.
All six systems are inconsistent with solar O/Fe at at least the 1$\sigma$ level (at over 5 for the well-constrained case of SB~82), with median ratios of 2--4 and a combined constraint over all six systems of $3.0_{-1.2}^{+1.9}$.
While the empirically and theoretically motivated corrections just described both act to enhance this inferred ratio, they do not entirely explain it; conservatively ignoring the larger $0.24$~dex correction to the recombination line scale would still leave the combined sample at $1.7_{-0.7}^{+1.1}$, inconsistent with solar at 1$\sigma$; and even ignoring both the RL and depletion corrections, SB~82 remains above solar at over 3$\sigma$.
We examine these abundance constraints and their implications in Section~\ref{sec:disc_abund}.

\begin{table}
\centering
\caption{
    Comparison of the stellar metallicities to the gas-phase oxygen abundance inferred from the nebular gas.
    We present the total metallicity $Z$ corresponding to the gas-phase oxygen abundance measured using the direct-$T_e$ method (Section~\ref{sec:gasmet}) derived assuming the \citet{gutkinModellingNebularEmission2016} solar abundances with \xid{}$=0.3$ and further adjusted to account for the CEL-RL offset; and the (O/Fe) ratio relative to solar corresponding to the resulting offset from the photospheric-derived $Z$.
    \label{tab:ofeh}
}
\begin{tabular}{lcc}
\hline
Target & $\log Z (\mathrm{gas\;O/H})$ & Inferred \\
Name & (\xid{}$=0.3$, RL-corrected) & $\mathrm{(O/Fe)/(O/Fe)_\odot}$  \\

\hline
\hline

J082555 & -2.85 & $3.8^{+7.7}_{-2.2}$\\
SB2 & -2.48 & $2.8^{+1.2}_{-1.0}$\\
J104457 & -2.81 & $2.2^{+0.7}_{-0.6}$\\
SB82 & -2.36 & $4.3^{+0.7}_{-0.6}$\\
J120202 & -2.79 & $3.7^{+6.0}_{-1.5}$\\
HS1442+4250 & -2.63 & $2.1^{+0.9}_{-0.5}$\\
\hline
combined & -- & $3.0^{+1.9}_{-1.2}$\\
\hline

\end{tabular}
\end{table}

\subsection{Comparison to the stellar wind lines}
\label{sec:windcomp}

In addition to the photospheric transitions focused on in the previous section, the UV continuum region probed by our deep \hstcos{} spectra also contain several lines formed in stellar winds which are strongly constrained by our data (Section~\ref{sec:uvobs}).
In contrast to the photospheric iron lines, the wind lines are highly sensitive to a number of other uncertain assumptions made in the stellar population synthesis modeling.
Most directly, the strength and morphology of the wind lines are dependent on the assumed mass loss rates of luminous and metal-poor massive stars and the detailed modeling of momentum and radiative transfer in these stellar winds including often neglected factors such as wind clumping and shocks: all of which remains uncalibrated by resolved star observations at these low $\lesssim 10\%$ solar metallicities \citep[e.g.][]{bouretQuantitativeSpectroscopyStars2003,hillierTaleTwoStars2003,bouretNoBreakdownRadiatively2015,sundqvistNewPredictionsRadiationdriven2019,pulsAtmosphericNLTEModels2020}.
Our constraints on the stellar photospheric lines provide a unique opportunity to test state-of-the-art wind prescriptions in these models at unprecedentedly low stellar metallicities.
If the atmospheres of the luminous hot stars in these galaxies are treated correctly, the posterior for our fits to the photospheric lines should span the observed wind line strengths.
Issues in matching these lines in turn can be leveraged to provide insight onto the massive stars in these galaxies and the assumptions made in their modeling.

As foreshadowed above, examining the \ion{O}{5} $\lambda 1371$ line provides a demonstrative illustration of the valuable information encoded in these features.
This line is uniformly predicted by the models fit to the photospheric continuum to display a prominent P-Cygni morphology; yet it is in every case observed to instead display narrow absorption near line-center with minimal emission or higher-velocity absorption (Fig.~\ref{fig:otherphotocomp}, Fig.~\ref{fig:otherwindcomp}).
One possibility for this discrepancy is that these galaxies are missing the very luminous and hot stars which dominate this feature in the models, or that these stars are driving far weaker winds than predicted.
But another modeling complication is likely at work in the case of this line: detailed atmospheric modeling of individual O-stars has revealed that \ion{O}{5} $\lambda 1371$ is strongly affected by the inclusion of clumping in the modeling of stellar winds, with unclumped winds producing the strongest emission and high-velocity absorption \citep{bouretQuantitativeSpectroscopyStars2003,bouretLowerMassLoss2005}.
A homogeneous wind structure is generally assumed in the models for high-mass stars employed in population synthesis codes, so this difference strongly suggests that the winds driven by the metal-poor stars in these galaxies are significantly clumped.

\begin{figure*}
\plotone{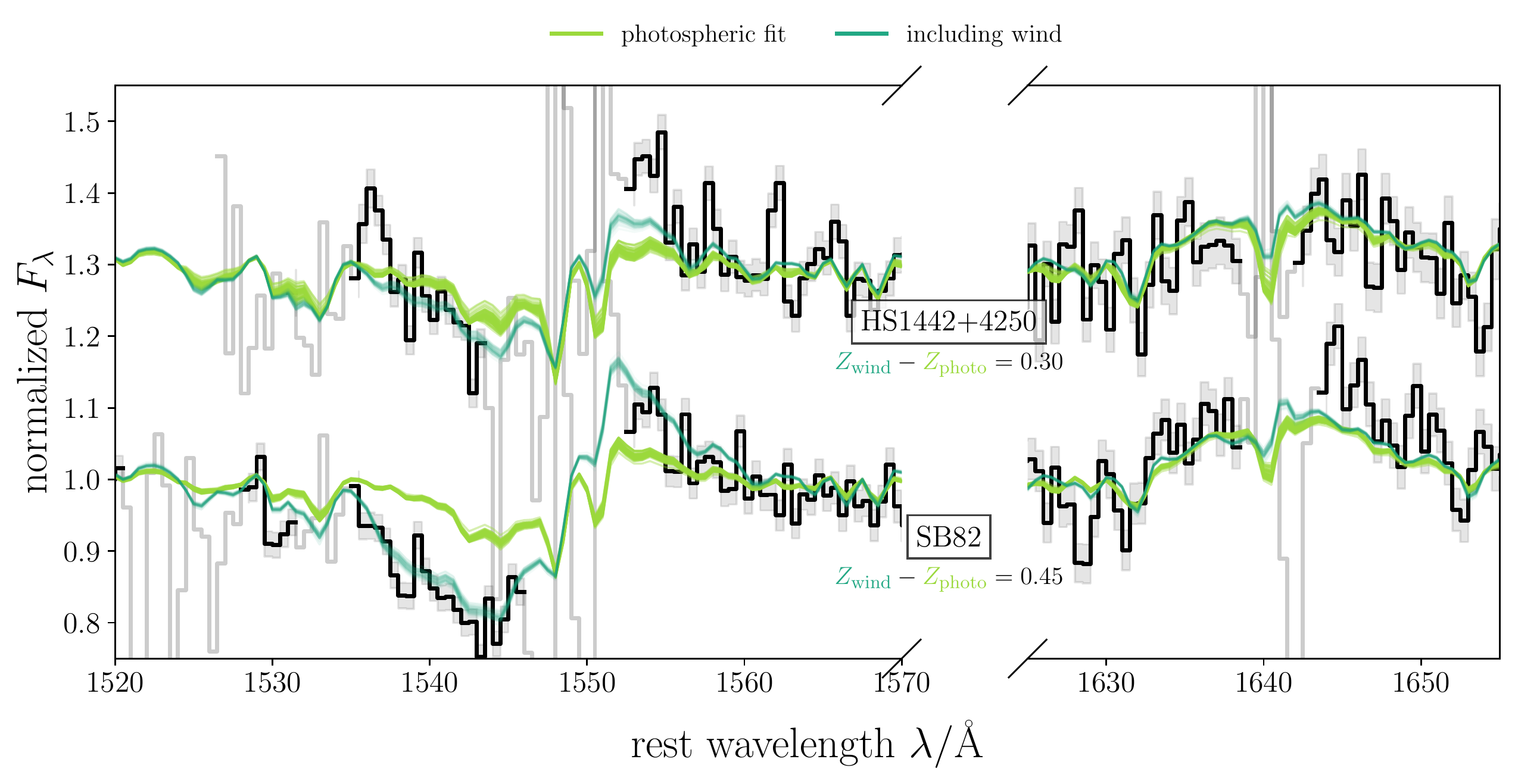}
\caption{
    Same as Fig~\ref{fig:photo1450comp}, but here highlighting the broad stellar wind features at \civ{} and \heii{}.
    While the stellar \heii{} profile is remarkably well-reproduced in both cases, intriguingly the models for SB~82 and (to a lesser extent) HS~1442+4250 struggle to match the prominent stellar \ion{C}{4} profile at the metallicity preferred by the photospheric lines.
    \label{fig:windcomp}
}
\end{figure*}

Comparison of our photospheric fits with the strongest wind lines in our spectral range, \civ{} and \heii{}, represents a crucial test for the stellar population models.
The \ion{C}{4} $\lambda\lambda 1548,1551$ resonant doublet is observed in P-Cygni for normal OB star winds, while the \ion{He}{2} $\lambda 1640$ recombination lines is observed in broad emission in the dense winds driven by Wolf Rayet (WR) and other helium stars as well as very massive hydrogen-burning stars near the Eddington limit.
We focus our analysis in this section on the two highest-S/N spectra in our sample: SB~82 and HS~1442+4250.
In Fig.~\ref{fig:windcomp} we plot their \civ{} and \heii{} profiles with non-stellar contamination masked; the wind comparison for the other spectra and for the bluer wind lines are included in Fig.~\ref{fig:cbs99_wind} and \ref{fig:otherwindcomp}.
First, consider HS~1442+4250, which has a very well detected \civ{} P-Cygni and broad \heii{} emission profile.
The photospheric-only fit does remarkably well at reproducing the absorption trough of \civ{} and the broad \heii{} profile; the latter a testament to the inclusion of new Wolf-Rayet atmospheres in the latest \cb{} models \citep{senchynaUltravioletSpectraExtreme2021}.
However, both the \cb{} and \snn{} models notably underestimate the broad redshifted stellar emission in \civ{}.

The highest-S/N spectrum for SB~82 provides an even greater challenge to the models.
Despite a preferred photospheric metallicity similar to (and slightly below that) of SB~2 and HS~1442+4250, this system presents extremely prominent stellar \civ{} (with an absorption equivalent width of $-2.4$~\AA{}) and strong \heii{}.
While \heii{} is in reasonable agreement with the data, the fit to the photospheric continuum in this galaxy is in substantial tension with the \civ{} wind profile.
Both the observed absorption trough and emission lobe are entirely missed by the model posterior constrained by the photospheric forest, which predicts a significantly weaker feature.
This holds also for the \snn{} models, which prefer a similarly weak \civ{} profile when the photospheric continuum is fit in addition to missing \heii{} (Fig.~\ref{fig:cbs99_wind}).

The \civ{} wind profile is in clear and significant tension with the photospheric continuum for SB~82 (and to a lesser extent HS~1442+4250), in the context of both the \cb{} and \snn{} models.
Analyzed in-isolation and at fixed population age, a stronger wind profile in an integrated galaxy spectrum would generally imply higher metallicities, as these winds are driven by metal line absorption (predominantly iron-peak elements) and the observed lines directly formed by ionic metal species.
To explore this, we consider a second set of fits to the spectrum including the stellar \civ{} profile --- the included wavelength range is indicated in Figure~\ref{fig:windcomp}, and the resulting measurements are presented in Table~\ref{tab:metresults}.
For both SB~82 and HS~1442+4250, better agreement with the \civ{} profile (Fig.~\ref{fig:windcomp}) is only found at substantially higher metallicity than implied by the photospheric fits; specifically, at $+0.45$ and $+0.30$ dex higher.
This improvement in the wind line match comes at the cost of agreement with the photospheric lines; the reduced-$\chi^2$ for the fit for SB~82 including the \civ{} profile is 69\% larger than that for the photospheric-only fit, with the disagreement spread broadly throughout the continuum.
The S/N and low metallicity of the other targets conspire to make detailed analysis of their wind lines more challenging (Fig.~\ref{fig:cbs99_wind}); but repeating this experiment and directly fitting the \civ{} wind profile results in a higher metallicity in four of the six cases, and overall a median change of $+0.2$ dex across the sample.

Another possible explanation for shifts in the strength of stellar wind lines is the effective age or recent star formation history of the galaxy.
Since \civ{} and \heii{} are predominantly powered by hot and generally very luminous stars which power the most significant radiation-driven stellar winds while the UV continuum is produced by a broader range of OB stars, their strength in an integrated galaxy population is dependent on the mixture of these most luminous stars relative to their somewhat older brethren, and thereby on the precise shape of the recent star formation history.
To explore whether this has a significant impact on the mismatch with \civ{}, we conduct an additional set of fits with a star formation history parameterized by two SSPs with independent age and relative UV contribution instead of the variable-age constant SFH (Section~\ref{sec:fitting}), to reflect a star formation mode composed of several entirely discrete cluster bursts rather than a more protracted recent star formation episode.
The results are presented in Table~\ref{tab:metresults} and Figure~\ref{fig:windcomp_sfh}, and discussed in Appendix~\ref{app:sfh}.
We find that while some improvement is made in matching the strong emission lobe of \civ{} in HS~1442+4250 (at the cost of good agreement with stellar \heii{}), adopting a bursty SFH has negligible impact on the strength of the \civ{} absorption trough in SB~82.
As for the assumed CSFH, we find that matching the strength of the \civ{} P-Cygni profile in this system by explicitly fitting it with a 2-SSP model still requires invoking a metallicity significantly higher than that found for the photospheric lines (by $0.44$ dex; Table~\ref{tab:metresults}).
In short, uncertainties in the star formation history do not appear to resolve the difficulty we find in matching the stellar wind lines in these systems.

In summary, we find tension between the stellar photospheric and wind lines in these ultra-deep spectra.
In particular, we find that \civ{} is notably underpredicted by the \cb{} and \snn{} models fit to the photospheric continuum alone.
These differences are most prominent in SB~82 and HS~1442+4250, but we find a suggestion of a similar preference for higher stellar metallicities to match the strength of this wind line in several of the lower-S/N spectra as well.
This is suggestive of tension in the stellar physics underlying these model predictions at these very low metallicities.
We discuss this in more detail in Section~\ref{sec:disc_stellar}.

\section{Discussion}
\label{sec:discuss}

The massive star abundance constraints obtained here provide powerful and otherwise inaccessible insight onto the star-forming regions under study.
We divide our discussion of these results and their implications into three parts; first, we reevaluate the intense nebular emission observed in these systems and their relationship with high-redshift galaxies; next, we discuss lessons learned regarding stellar modeling for metal-poor massive star populations; and finally, we consider their consequences for our understanding of abundances in high-redshift galaxy analogs.

\subsection{Implications for Intense Nebular Emission both Locally and at High-Redshift}
\label{sec:disc_neb}

Local metal-poor dwarf galaxies promise an opportunity to study the constituent ingredients of high-redshift star-forming systems in far greater detail than otherwise possible, especially their low-metallicity young stellar populations and highly-ionized gas.
Local galaxies have already proven instrumental in contextualizing the high-ionization \ion{C}{3}] emission detected at the tip of the reionization-era population that \jwst{} will study \citep[e.g.][]{rigbyIIIEmissionStarforming2015,senchynaUltravioletSpectraExtreme2017}; yet the strong nebular \civ{} at $\sim 20$--$40$~\AA{} rest-frame equivalent width glimpsed in several lensed galaxies at $z>6$ remains mysterious \citep{starkSpectroscopicDetectionIV2015,mainaliEvidenceHardIonizing2017,schmidtGrismLensAmplifiedSurvey2017}.
While nebular \civ{} is essentially entirely absent from canonical UV atlases of star-forming galaxies \citep[e.g.][]{leithererUltravioletLineSpectra2001}, \hstcos{} has in recent years unveiled a small population of tens of low-redshift star-forming galaxies powering nebular \civ{} alongside prominent \heii{} with no clear indications of non-stellar ionizing sources \citep{bergCarbonOxygenAbundances2016,senchynaUltravioletSpectraExtreme2017,senchynaPhotometricIdentificationMMT2019,bergChemicalEvolutionCarbon2019,bergIntenseIVHe2019}.
However, several key questions remain unanswered by this body of line detections.
First, it remains unclear why local systems with otherwise very similar nebular properties display such a broad range of nebular \civ{} equivalent widths \citep[e.g.][]{senchynaExtremelyMetalpoorGalaxies2019}.
And second, the highest equivalent width emission in this doublet yet detected locally is in J104457 at $11$ \AA{} \citep[][and see Table~\ref{tab:uvmeas}]{bergIntenseIVHe2019}, which falls short of that encountered at $z>6$ by a factor of 2--4.
Without actual signatures of the massive stars underlying this emission or a clear view of the \civ{} profiles, attempts to answer these questions have thus far been restricted to correlations with gas-phase metallicities and other nebular properties which provide only a partial view of the underlying physics.

In this work we have targeted six of these local nebular \civ{} emitters with spectra deep enough to place strong constraints on stellar photospheric and wind signatures.
These targets were selected primarily on this high-ionization emission, and span a significant range of inferred ages (2--20 Myr assuming a constant SFH; Table~\ref{tab:sedfits}) gas-phase metallicities from $12+\log\mathrm{O/H}=7.5$ (5\% solar; J082555, J104457, J120202) up to the relatively oxygen-rich SB~82 at $7.93$ (15\% solar; Table~\ref{tab:basicprop}).
Surprisingly, even for SB~82 and 2 which lie above 10\% solar in O/H (the traditional formal limit defining XMPs), the photospheric continuum features in all six targets prefer stellar metallicities Z (driven by Fe/H) below 10\% solar ($\log Z<-2.8$).
That this sample of nebular \civ{} emitters with diverse nebular properties including SB~82 at nearly SMC metallicity in gas-phase O/H is uniformly populated by extremely metal-poor stars suggests that this emission indeed may require the hard ionizing radiation field these sub-10\% stellar populations are capable of powering.
This finding lends further support to the notion that nebular \civ{} may be a remarkably useful tracer of these extremely metal-poor stellar populations at high redshift where it is detected alongside other UV line ratios consistent with star formation and when [\ion{O}{3}] $\lambda 4363$ will be intractable.

While this new dimension of stellar metallicity appears to help explain how SB~82 and 2 power nebular \civ{} at relatively high metallicity, it may not fully illuminate the scatter found in the strength of \civ{} among the other objects.
In particular, consider the strongest nebular \civ{} emitter in our sample and known locally among star-forming galaxies at all, J104457.
This galaxy is clearly more metal-poor in both stars (at $\log Z \simeq -3.1$) and gas (at $12+\log\mathrm{O/H}=7.5$) than the next-strongest \civ{} emitter in our sample, HS~1442+4250 ($\log Z \simeq -2.9$, $12+\log\mathrm{O/H}=7.7$).
This is consistent with this picture in which a harder stellar ionizing radiation field (and potentially hotter gas temperature) serve to boost this emission at lower metallicities.
However, the other two lowest-O/H galaxies in our sample J082555 and J120202 present nearly identical O/H and H$\beta$ equivalent widths to J104457 (Table~\ref{tab:basicprop}), yet power \civ{} at only 2.1 and 1.5~\AA{}; 5 times smaller than J104457.
If stellar metallicity were the only other variable at play, one might expect J082555 and J120202 to harbor more metal-rich stellar populations with correspondingly softer radiation fields.
But while the broad constraints on their metallicity from the UV data do leave open the narrow possibility for slightly more metal-rich stars than J104457, both of these galaxies are likely at a similar $\log Z\lesssim -3.1$ (Table~\ref{tab:metresults}).
Some other factor is likely at play then to explain the wide range in \civ{} strengths between these three systems.

A clue as to another potentially crucial variable in determining the strength of nebular \civ{} resides directly in the high resolution spectral profiles obtained for \civ{} itself (Section~\ref{sec:uvmeas}).
The profiles of both components of \civ{} in the three enigmatic lowest-O/H galaxies just discussed (J082555, J104457, J120202) fall into a clear sequence (Figure~\ref{fig:civ_overview}).
The lowest EW \civ{} emitter of these three, J120202, shows a clear P-Cygni profile on 100~km/s scales with no emission blueward of line center.
The next most prominent, J082555, presents a broad, purely emission profile with a strongly asymmetric redshifted peak; and the strongest \civ{} emission (in J104457) is broad and clearly double-peaked in both \civ{} components with no sign of underlying absorption.
We argue that this pattern, very similar to a Ly$\alpha$ profile sequence and followed by the entire sample presented here, is strong evidence that \civ{} emission is subjected to resonant scattering before it escapes these dwarf galaxies, as argued by \citet{bergIntenseIVHe2019}.
In other words, these profiles suggest that the emergent equivalent width of nebular \civ{} may be lower in J120202 and J082555 than in J104457 in part due to a significantly higher column density of outflowing triply-ionized carbon along the line of sight into the former galaxies.

\begin{figure}
\plotone{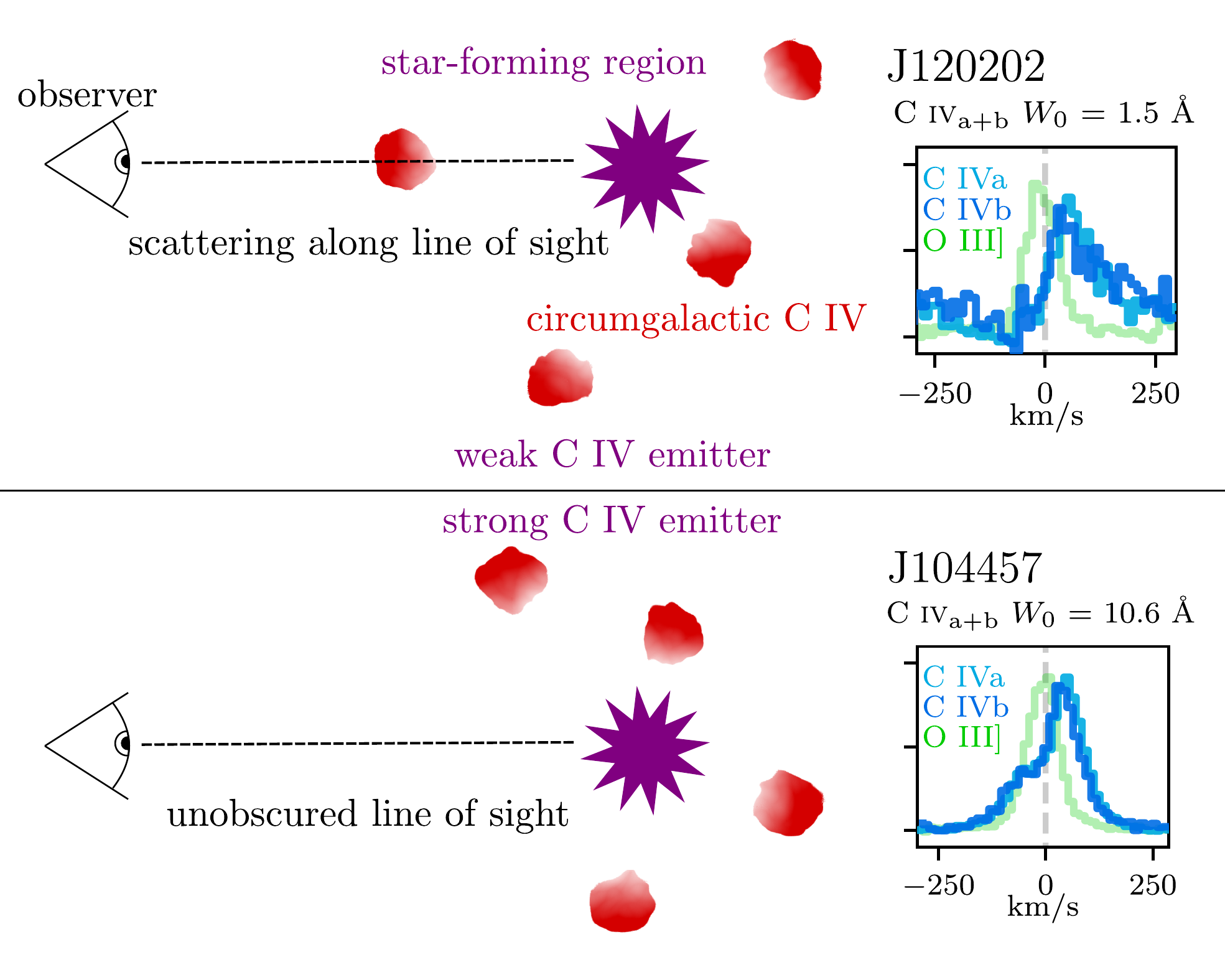}
\caption{
    Schematic illustration of the \civ{} scattering picture we propose.
    Scattering in spatially-compact (of-order 100~pc) circumgalactic \civ{} `clouds' which in some cases intercept the line-of-sight to the target star-forming regions would naturally explain why some targets with otherwise very similar metallicities and inferred ages show a broad range of emergent nebular \civ{} equivalent widths.
    The \civ{} emission profile of the relatively-weak emitter J120202 shows a P-Cygni profile with a strongly redshifted emission peak indicative of a high column density of interstitial \civ{}, whereas the prominent emission in J104457 is double-peaked and suggestive of a relatively unobscured pathway out of the galaxy for \civ{} photons.
    \label{fig:civ_schematic}
}
\end{figure}

Our observations suggest that resonant scattering may regulate the escape of stellar-photoionized \civ{} emission produced in these local galaxies.
While the column densities of \civ{} implied by such a scenario are difficult to produce in the \hii{} regions themselves or in a collapsed superwind \citep{grayCatastrophicCoolingSuperwinds2019}, substantial quantities of \civ{}-absorbing gas associated with $z>1$ galaxies have long been noted in quasar absorption line spectra \citep[e.g.][]{steidelHighRedshiftExtensionSurvey1990,chenOriginIVAbsorption2001,adelbergerConnectionGalaxiesIntergalactic2005,rudieColumnDensityKinematics2019} and \hstcos{} has revealed that this holds true for the circumgalactic medium (CGM) of a wide range of galaxies in the nearby Universe as well \citep[e.g.][]{chenOriginIVAbsorption2001,borthakurImpactStarburstsCircumgalactic2013,bordoloiCOSDwarfsSurveyCarbon2014}.
In particular, the detection statistics of \civ{} absorption along quasar sightlines around low-$z$ sub-$L_\star$ galaxies suggest a patchy distribution of \civ{} consistent with being virially-bound, but with a particularly high covering fraction of absorption close-in to the star-forming subset \citep[][]{bordoloiCOSDwarfsSurveyCarbon2014}.
This is suggestive of a picture similar to that encountered at high redshift, in which small-scale ($\sim 100$~pc) clumps of \civ{} gas sustained by the extragalactic UV background are present in the metal-enriched outflows or `fountains' powered by supernovae energy input in these star-forming galaxies \citep[e.g.][]{rauchSmallScaleStructureHigh2001,rudieColumnDensityKinematics2019}.
In this view, the strong blueshifted absorption in the \civ{} profiles of J120202 and SB~82 suggests that our line-of-sight to these star-forming regions is intercepted by a high column density \civ{} cloud in the target galaxy CGM, which suppresses the measured \civ{} flux.
Conversely, the double-peaked emission profile in J104457 suggests relatively minimal \civ{} scattering, suggesting a comparatively clear line-of-sight in this case and potentially explaining in part its dramatically higher measured equivalent width of \civ{}.
We illustrate this proposed situation schematically in Figure~\ref{fig:civ_schematic}.

A significant role for CGM scattering in \civ{} escape in these systems has important consequences for our comparison to high-redshift star-forming galaxies.
One might expect that a CGM primarily enriched by stellar yields might be significantly lower-metallicity in the early Universe, and thus potentially less likely to reach high column densities in \civ{}.
Blind absorption line system surveys conducted with very high-redshift quasars provide an opportunity to test this experimentally by constraining the incidence of \civ{} absorbers as a function of redshift.
Interestingly, such surveys have revealed a dramatic decline in the incidence of \civ{} absorption systems above redshift $z>5$, likely due to some combination of lower carbon abundances and a softer metagalactic ionizing background \citep[e.g.][]{beckerHighRedshiftMetalsDecline2009,cooperHeavyElementAbsorption2019}.
The extremely low rate of \civ{} absorption detection at these high redshifts strongly suggests that the covering fraction of \civ{} in the CGM of high-redshift star-forming halos is substantially lower than in the local Universe.
If this holds, resonant scattering in the CGM while still possible is much less likely to substantially affect \civ{} emission powered by metal-poor stars in these early systems.
This is consistent with the high detection rate of very strong \civ{} in early reionization-era spectra, and would mean that \civ{} emission at these redshifts is likely to be an even more complete tracer of highly-ionized gas around metal-poor massive stars than locally.

While variable optical depths to scattering plausibly explains some of the scatter in \civ{} strength among local systems and differences with the $z>6$ Universe, it does not resolve all of this tension.
In particular, the narrowly double-peaked emission profile found for the strongest local emitter, J104457, indicates relatively minimal scattering: not enough to suggest an intrinsic unabsorbed equivalent width approaching the 20--40~\AA{} encountered at $z>6$.
In other words, we have still not found any star-forming galaxies in the nearby universe with intrinsic \civ{} emission comparable to reionization era detections --- empirically, it remains unclear whether massive stars could power such prominent \civ{}.

However, our data strengthens the case that local UV sample simply do not yet reach the metallicities and ages where stars are able to power \civ{} most efficiently.
The local record holder J104457 stands out as one of the most metal-poor galaxies in gas-phase O/H with UV coverage \citep{senchynaExtremelyMetalpoorGalaxies2019}.
However, our stellar continuum fits indicate that the stars in that system are not dramatically more metal-poor than the gas; we infer an O/Fe ratio relative to solar of $2.2_{-0.6}^{+0.7}$ (Table~\ref{tab:ofeh}), whereas assembling systems at very high redshift might be expected to harbor significantly more alpha-enriched gas \citep[near the pure Type~II SNe limit of $\mathrm{(O/Fe)/(O/Fe)_\odot}\simeq 5$; e.g.][]{nomotoNucleosynthesisYieldsCorecollapse2006} and correspondingly lower-metallicity, hotter massive stars.
The extremely young age of J104457 implied by its very high EW of 264~\AA{} in H$\beta$ is also noteworthy.
While lower-O/H galaxies have been studied by \hstcos{} like I~Zw~18 and SBS~0335-052E, the EWs of the optical nebular lines in the star-forming regions targeted are actually uniformly below 200~\AA{} in H$\beta$ and thus suggestive of less extremely young ages than in J104457 \citep{senchynaExtremelyMetalpoorGalaxies2019,woffordStarsGasMost2021}.
It is very possible that a combination of the hard ionizing radiation field from an extremely iron-poor $<5\%$ solar and very young stellar population\footnote{As expected from photoionization modeling of high-redshift \civ{} emitters; e.g.\ \citet{starkSpectroscopicDetectionIV2015}.} combined with relatively high oxygen and carbon abundances\footnote{Potentially influenced by a combination of both Type II SNe yields and the carbon-rich output of Population~III supernovae \citep[e.g.][]{frebelSegueUnevolvedFossil2014,cookeBigBangNucleosynthesis2015,frebelNearFieldCosmologyExtremely2015}.} might lead to significantly stronger \civ{} equivalent widths than encountered in J104457.
The lack of UV observations of extremely metal-poor dwarf galaxies dominated by recent star formation is largely a product of their typical faintness which has contributed also to their delayed discovery \citep[e.g.][]{hsyuSearchingLowestmetallicityGalaxies2018,izotovLowredshiftLowestmetallicityStarforming2019,senchynaPhotometricIdentificationMMT2019,kojimaSubaruHSCMachine2019}, but measurement of nebular \civ{} and other high-ionization UV lines in exemplar systems in this regime would be of significant value in the interpretation of \jwst{} spectra at the highest redshifts.

Irrespective of the results of the search for the most prominent \civ{} in the local Universe, extremely metal-poor galaxies nearby still provide our \textit{only} opportunity to directly calibrate the ionizing spectra produced by low-metallicity stellar populations.
These spectra will allow us to ask directly whether stellar population synthesis models are able to reproduce the high-ionization nebular emission lines including \civ{} at the age and \citep[crucially, non-zero; e.g.][]{kehrigExtendedHeII2018} metallicity inferred from the photospheric lines.
We will reserve detailed modeling of the high-ionization nebular lines alongside the stellar continuum to a later paper, but note here that our results already provide several important lessons along these lines.
First, we have argued that the CGM can significantly scatter resonant \civ{} emission; since this depresses the emergent flux in \civ{} captured in a spectroscopic aperture, the measured flux in nebular \civ{} must be interpreted as a lower limit on the intrinsic emission from the \hii{} region.
While this may require more work to incorporate into photoionization modeling schema, these \civ{} lower limits still provide unique leverage on the $\sim 50$ eV ionizing radiation field \citep[e.g.][]{platConstraintsProductionEscape2019}.
In addition, our results suggest that conducting such an analysis with the stellar metallicity set to that implied by gas-phase O/H and solar abundances with no additional correction may lead to substantial errors (up to nearly an order of magnitude at maximal O/Fe) in the inferred stellar metallicity,\footnote{It is interesting to note however that this is not always the case. In particular, an enhancement in O/Fe of 2.2 over solar will effectively cancel-out the RL-CEL and depletion corrections we have adopted in this work. In this context, previous work where these factors are neglected but agreement between stellar and gas-phase metallicities are found \citep[e.g.][]{chisholmConstrainingMetallicitiesAges2019} might instead be reinterpreted as evidence for modest $\alpha$/Fe-enhancement.} leaving any comparison to stellar models subject to significant systematic errors.
Direct assessments of the stellar metallicity such as that conducted here bring this sort of analysis further beyond model testing and towards prescriptive calibration of the stellar ionizing spectrum as a function of fundamental stellar properties; a keystone for an era in which we hope to extract meaningful physical constraints from multiplexed high-redshift nebular spectra.

\subsection{New Challenges to State-of-the-Art Stellar Models}
\label{sec:disc_stellar}

The interpretation of star-forming galaxy spectra in the era of \jwst{} and the ELTs will increasingly depend on the accuracy of models for the evolution and atmospheres of massive stars at very low metallicity; and stellar winds remain a particularly crucial and uncertain component of these models.
Stellar winds both directly modulate the hard ionizing spectrum so crucial to accurately interpreting nebular line emission, and via mass and angular momentum loss they shape the evolutionary channels that both isolated and interacting massive stars follow.
In addition, the prominent resonant UV stellar wind lines themselves like \civ{} provide some of the most direct and accessible insight into the massive stars inhabiting distant faint galaxies; but any insight derived from these lines is only as valuable as their modeling is correct.
Unfortunately, with only a handful of resolved OB stars accessible in the Local Group outside the Magellanic Clouds, and detections of weak stellar features exceedingly rare in unresolved metal-poor galaxies, testing models for these winds below 20\% solar metallicity remains challenging.

The spectra presented in this paper provide some of the deepest constraints yet obtained on the UV continua of extremely metal-poor $Z/Z_\odot\lesssim 10\%$ massive stars.
Notably, these data provide access to both the integrated stellar wind signatures as well as the metallicity-sensitive forest of iron photospheric absorption features, presenting a powerful challenge to stellar population synthesis models.
The wind lines are subject to particularly uncertain physics adopted in the underlying models, especially the scaling of mass loss rates with stellar metallicity and luminosity and the NLTE modeling of ionized gas in expanding atmospheres.
If this physics is treated successfully, we should be able to find consistency with these wind line profiles when the photospheric continuum is fit alone.

But as explored in detail in Section~\ref{sec:windcomp}, our experiment reveals signs of significant tension between the wind and photospheric line detections in our spectra.
First, we find that the models preferred by the photospheric lines underestimate the strength of the \civ{} P-Cygni profile in the two highest-S/N spectra of our sample (SB~82 and HS~1442+4250).
Direct fits to the wind lines require systematically higher stellar metallicities generally independent of the assumed star formation history, by up to $0.4$ dex in the most extreme case of SB~82.
While stellar \heii{}~$\lambda 1640$ is remarkably well matched to the data in most cases, the higher-ionization wind-sensitive line \ion{O}{5}~$\lambda 1371$ is predicted by the models to display a strong P-Cygni signature which is not detected in our data.
And finally, the models best matched to the iron forest also appear to underestimate the strength of several individual strong photospheric lines of $\alpha$ elements, particularly \ion{S}{5} (Section~\ref{sec:photcomp}).

This difficulty we have found in simultaneously matching the wind and photospheric lines is suggestive of discord in the modeling of the strong integrated stellar wind features in these young star-forming galaxies.
In particular, we highlight that fits to the \civ{} P-Cygni profile can result in metallicities up to 3 times larger than those derived from the stellar photospheric lines alone.
This echoes several other recent analyses where tension has been found between the UV stellar wind lines and other galaxy properties in extreme star-forming galaxies.
First, \citet{senchynaUltravioletSpectraExtreme2021} found that the \cb{} models struggled to match the very strong stellar \heii{} wind line and to a lesser extent \civ{} when fit alongside the strong optical nebular lines in a sample of moderately metal-poor star-forming regions (near LMC metallicity, $12+\log\mathrm{O/H}\simeq 8.2$).
In this case, the strongest wind profiles were only reproduced by invoking significantly higher stellar metallicities in clear tension with the gas-phase metallicity and electron temperature.
In another recent study focused on the canonical extremely metal-deficient galaxy SBS~0335-052E (near the low-end of the gas-phase metallicity range spanned by the targets in this work, at $12+\log\mathrm{O/H}=7.3$-$7.5$ depending on method), \citet{woffordStarsGasMost2021} found that direct fits to the weak \civ{} P-Cygni profile detected in their deep stacked G160M spectrum resulted in inferred metallicities substantially larger (by $\sim 0.5$--$0.7$ dex) than inferred from the nebular gas lines.
These results differ from our current analysis in several important ways, most crucially in their use of nebular gas lines as a metallicity reference rather than the stellar photospheric forest.
However, they all concur that fits to the strong wind lines in young star-forming systems commonly result in metallicities systematically larger than those derived from other methods.

These findings have important implications for modeling metal-poor star-forming galaxies at all redshifts.
By virtue of its prominence, the strong \civ{} P-Cygni feature is the most likely massive star signature to be directly detected in star-forming galaxies at high-redshift.
The resonant wind line complexes especially \civ{} have long been recognized as some of the most accessible UV abundance indicators for unresolved star-forming galaxies \citep[e.g.][]{heckmanUltravioletSpectroscopicProperties1998,keelFarUltravioletSpectroscopyStarforming2004,crowtherReliabilityCIVl1549Abundance2006}, and are likely to provide our most direct leverage on the metallicity of the massive stars in the brightest $z>6$ galaxies with \jwst{} and the ELTs.
However, our results and other recent results in the literature suggest that at-present, direct fits with state-of-the-art models to these features especially in metal-poor and high-sSFR systems (as expected at high redshift) can yield metallicity estimates that are biased significantly (as much as $\gtrsim 0.4$~dex) towards higher values.
Until the origin of this discrepancy is addressed, metallicity estimates driven by the UV wind lines should be treated with some caution.

While troubling from the perspective of galaxy abundance measurement, the tension we have found in modeling the wind lines may actually contain significant information about the metal-poor massive stars inhabiting these nearby galaxies.
The discrepancy between the photospheric fit and the \civ{} profile especially evident in SB~82 is suggestive of a potential deficiency in the treatment of luminous stars driving strong winds in this system.
This is less likely to be due to higher-order complexities in photospheric or wind modeling; the predominant effect of unaccounted for microturbulence would be to further depress the inferred photospheric metallicities (Appendix~\ref{app:otherdetails}), and the strength of the stellar \civ{} wind line is particularly robust to the inclusion of effects like clumping and X-ray emission from shocked gas in the wind which would naturally explain the difficulty we find in reproducing \ion{O}{5}~$\lambda 1371$ \citep[see Section~\ref{sec:windcomp}; e.g.][]{bouretLowerMassLoss2005}.
Instead, either an underestimation of mass loss rates for the very luminous stars in that galaxy or an overabundance of the most massive wind-driving stars relative to the IMF we have assumed \citep[possibly produced by binary mass transfer; e.g.][]{senchynaUltravioletSpectraExtreme2021} could potentially enhance the strength of \civ{} without substantially changing the effective equivalent width of the photospheric iron lines.
Clues such as this revealed in the stellar features imprinted on integrated light spectra of star-forming regions may be one of our most direct opportunities to confront detailed massive star models in these particulars at very low metallicities.

However, another possibility which might also explain why SB~82 is much more discrepant with the models than the otherwise broadly similar SB~2 and HS~1442+4250 is that the highly $\alpha$-enhanced abundances we have inferred for this system might significantly change the appearance of the massive stars in this system.
While iron dominates the EUV opacities which control wind launching in the normal optically-thin winds that P-Cygni \civ{} is produced in, an increase in the abundance of $\alpha$ elements by 0.6 dex can impact on the strength of lines produced by $\alpha$ element species.
The variation found in C/O abundance in addition to $\alpha/$Fe further complicates this picture; but the UV \ion{O}{3}] and \ion{C}{3}] nebular lines in SB~82 suggest only a moderately low value of $[\mathrm{C/O}]=-0.26$ \citep[typical for similar metal-poor dwarf galaxies;][]{senchynaUltravioletSpectraExtreme2017} which combined with our inferred $[\mathrm{O/Fe}]=0.64$ results in a significant residual enhancement in carbon of 0.4 dex over the abundances assumed in the underlying stellar models.
Intriguingly, \citet{eldridgeEffectStellarEvolution2012} performed an experiment varying the carbon abundance in \textsc{WM-BASIC} NLTE atmosphere models for O stars, and found that the most dramatic changes to the emergent \civ{} P-Cygni profile with varied C/O occurred near a moderately-low metallicity of $Z=0.004$; quite close to the metallicity we infer for SB~82 from the fits incorporating the \civ{} profile.

Our results underscore both the power and the challenges inherent in studying the UV stellar continuum.
While expensive to produce, a next-generation set of stellar population synthesis models with varied elemental abundance patterns in the massive star atmospheres themselves would allow us to disentangle the impact of these patterns on the emergent spectrum and strong wind lines.
Developing such a grid would take us substantially closer to both confident interpretation of the most expensive rest-UV spectroscopy that \jwst{} will provide and robust constraints on the mass function and evolution of extremely metal-poor massive stars.

\subsection{Abundances in Extreme Nearby Star-Forming Galaxies and the Search for `Analogues'}
\label{sec:disc_abund}

\begin{figure*}
\plotone{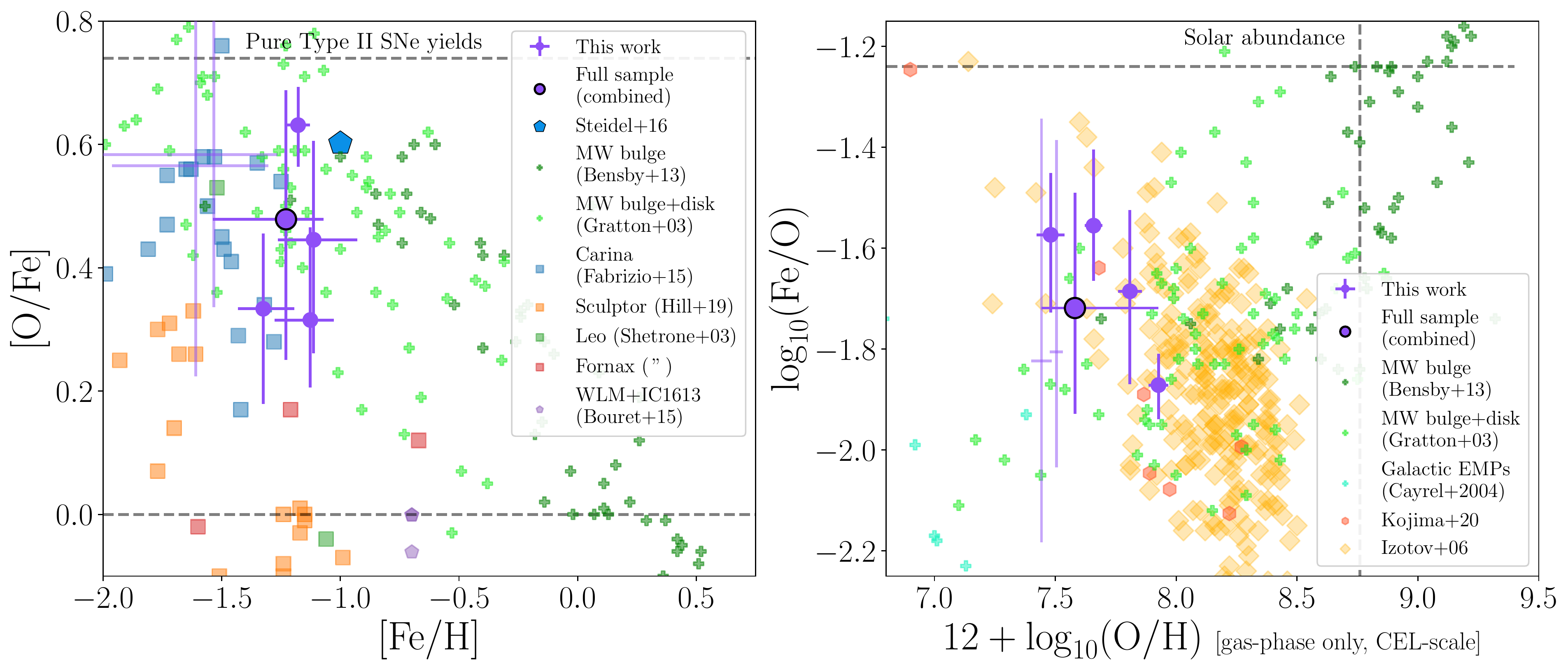}
\caption{Iron abundances derived from the stellar photospheric lines in our ultra-deep \hstcos{} UV spectra and oxygen abundances from the nebular O,H lines in the optical, in the context of other complementary abundance samples in the local Universe.
In particular, we plot Milky Way bulge and disk stars \citep{bensbyChemicalEvolutionGalactic2013,grattonAbundancesMetalpoorStars2003,cayrelFirstStarsAbundance2004}; Local Group dSph stars from Carina \citep{fabrizioCarinaProjectVIII2015}, Sculptor \citep{hillVLTFLAMESHighresolution2019}, Leo, and Fornax \citep{shetroneVLTUVESAbundances2003}; O stars in the dIrrs WLM and IC~1613 \citep{bouretNoBreakdownRadiatively2015}; and on the right, blue compact dwarf galaxies with gas-phase iron abundances \citep[plotted in absolute Fe/O space for consistency with previous nebular studies;][]{izotovChemicalCompositionMetalpoor2006,kojimaEMPRESSIIHighly2020}.
We plot the two low-S/N systems (J0825555 and J120202) as transparent crosses, and display an outlined point and errorbar representing the average measurement integrated over our sample posterior.
Note also that for ease of comparison with archival blue compact dwarf galaxy measurements, the gas-phase O/H measurement on the right panel x-axis is the direct-$T_e$ measurement without any correction for depletion or the RL-CEL scale difference.
\label{fig:abundances}
}
\end{figure*}

One of the key questions facing those studying local star-forming galaxies as reference systems is how and to what degree these systems differ from their high-redshift counterparts.
Abundance ratios and the $\alpha/\mathrm{Fe}$ ratio in particular have been highlighted as a critical point of potential difference between star-forming systems in the early and present-day Universe, with many authors emphasizing the idea that offsets between local and high-redshift galaxies might be explained by a near or complete lack of significantly $\alpha$-enriched star-forming systems in the nearby Universe \citep[e.g.][]{steidelReconcilingStellarNebular2016,stromNebularEmissionLine2017,sandersMOSDEFSurveyDirectmethod2020,toppingMOSDEFLRISSurveyInterplay2020}.
But despite substantial interest, the $\alpha/\mathrm{Fe}$ ratios and the corresponding shape of the long-term star formation histories of local unresolved metal-poor galaxies in general remain remarkably poorly-constrained.
Attempts to divine the long-term star formation history through direct detection of old stellar populations have proven extremely challenging even in the brightest and nearest metal-poor galaxies outside the Local Group \citep[see for instance the long debate as to the age of I~Zw~18:][]{searleInferencesCompositionTwo1972,hunterMassiveStarsZW1995,izotovDeepHubbleSpace2004,annibaliStarFormationHistory2013}.
The stellar metallicities necessary to constrain the relationship between massive star iron abundances and gas-phase oxygen require expensive space telescope UV observations or deep near-infrared data if red supergiants are present, and thus only a handful of quantitative measurements independent of the gas-phase exist in unresolved local galaxies \citep[e.g.][]{larsenAnatomyYoungMassive2008,lardoRedSupergiantsCosmic2015,hernandezChemicalAbundancesTwo2017,hernandezFirstMetallicityStudy2019,chisholmConstrainingMetallicitiesAges2019}.

Our constraints on the photospheric iron in the massive stars inhabiting local \civ{}-emitters provide our most direct view yet on the stellar abundances in extreme reionization-era analogue galaxies.
In Figure~\ref{fig:abundances} we plot the [O/Fe] and Fe/H measurements for our systems alongside measurements for Galactic stars, Local Group dSph and dIrr stars, and previous gas-phase only estimates for blue compact dwarfs (see caption for references).
Our galaxies are uniformly metal poor, all residing at $[\mathrm{Fe/H}]<-1.0$; old stars in the Local Group at these metallicities span from uniformly $\alpha$-enhanced in the Milky Way bulge and disk up to the Type~II SNe limit at $\sim 5$ times solar \citep[e.g.][]{kobayashiGalacticChemicalEvolution2006,nomotoNucleosynthesisYieldsCorecollapse2006} down to near-solar in some of the dwarf spheroidal galaxies (dSphs) with more protracted star formation histories.
If our target dwarf galaxies had substantial stellar mass buildup at early times $\gtrsim 1$~Gyr ago and were strongly enriched by Type~Ia SNe yields like these dSphs, we would expect to encounter similarly near-solar values in our comparison of gas-phase oxygen and the iron in recently-formed massive stars.
However, instead we find evidence for uniform $\alpha$-enhancement in the targeted \civ{}-emitters, ranging from 2.1--4.4 times solar with a median of 3.1 and $>1.5$ solar at at least 1$\sigma$ significance in all targets -- physically consistent with slightly diluted Type~II SNe yields and well within the broad locus of values measured in the dSphs and Milky Way.

Our measurement of uniform O/Fe-enrichment in the targeted \civ{}-emitters confirms that galaxies with at least some of the crucial chemical imprints of legitimately `young' high-redshift galaxies can indeed be found nearby.
The targeted systems appear to host a combination of iron-poor but oxygen-rich stars and gas comparable to the conditions encountered in the Milky Way bulge or Carina in their early assembly (Fig.~\ref{fig:abundances}).
This high detection rate of $\alpha$-enrichment is likely directly linked to the fact that we have selected our sample on the presence of high equivalent-width nebular \civ{} emission, immediately ensuring they are especially high in sSFR and potentially preferentially identifying galaxies with particularly iron-poor massive stars \citep[e.g.][]{senchynaExtremelyMetalpoorGalaxies2019}.
While the true fraction of such chemically-young galaxies among local star-forming dwarfs is unclear, our results suggest that selections focused on high equivalent width and high-ionization nebular emission are likely one of the best ways to find nearby galaxies with stars and gas abundances comparable to those encountered in the early Universe.

While having photospheric stellar abundances is the ideal case, the bulk of studies especially in a post-\hst{} landscape will be forced to rely on optical spectroscopy.
Gas-phase lines of ionized iron \citep[primarily only of \ion{Fe}{3}, but in high-quality spectra spanning \ion{Fe}{2}--\ion{Fe}{5} in the range 4200--5300; e.g.][]{izotovChemicalCompositionMetalpoor2006,bergCharacterizingExtremeEmission2021} provide an alternative way to estimate iron abundances in similar systems especially at moderately low metallicities, albeit subject to potentially large systematic uncertainties in the ionization correction, depletion scale, and best-available atomic data \citep[e.g.][]{rodriguezFeIVDiscrepancy2005,izotovChemicalCompositionMetalpoor2006}.
In the right panel of Fig.~\ref{fig:abundances} we compare our measurements to those of other star-forming dwarf galaxies with oxygen and iron abundances from gas-phase emission line constraints \citep[from][]{izotovChemicalCompositionMetalpoor2006,kojimaEMPRESSIIHighly2020}.
The targets analyzed here are among the lowest metallicity in gas-phase O/H with measurements of iron abundances from any source.
Recent analyses have found a broad range of Fe/O in the ionized gas, including measurements as high as the solar abundance of $\log(\mathrm{Fe/O})\simeq -1.2$ at very low metallicity \citep{kojimaEMPRESSIIHighly2020,isobeEMPRESSIVExtremely2021}.
In contrast, we find a systematically subsolar Fe/O across our sample (persistent even if we were to ignore the depletion and RL correction; Section~\ref{sec:abundcomp}), in closer agreement with the locus of the \citet{izotovChemicalCompositionMetalpoor2006} sample at low O/H (where depletion effects should be smallest, though not necessarily negligible).
This is further reinforced by the detailed study of \citet{bergCharacterizingExtremeEmission2021}, where the gas-phase iron abundance in J104457 is estimated using four-zone photoionization modeling with multiple iron species resulting in Fe/O (though dependent on ICF) in broad agreement with our estimates.
While the wide range of local results suggests some caution must still be applied, the agreement we find with \citet{bergCharacterizingExtremeEmission2021} suggests that at very low metallicity and with minimal dust the careful estimation of the gas-phase iron abundances may yield a complementary indirect path towards estimating stellar iron abundances locally.

The ratio of oxygen to iron we have inferred potentially contains significant information about the chemical enrichment history of these galaxies.
Their placement in $\alpha$/Fe-Fe/H space is a step towards connecting these unresolved field star-forming dwarf galaxies to the much better-studied resolved dwarf irregular and spheroidal systems in the Local Group \citep{tolstoyStarFormationHistoriesAbundances2009}.
In a traditional closed-box view, our measurement of moderate to high $\alpha$-enrichment would suggest that these galaxies are dominated to varying degrees by the yields from recent star formation and are in this sense broadly `young', with a significant fraction of their total stellar mass built-up recently.
However, several factors neglected in this view are crucial to robustly quantifying this statement: notably pristine gas inflows and SNe-driven outflows \citep[e.g.][]{dalcantonMetallicityGalaxyDisks2007,chisholmMetalenrichedGalacticOutflows2018} and the potential contribution of other exotic stellar enrichment events like hypernovae and pair-instability supernovae with different Fe/O yields than canonical core-collapse SNe \citep[e.g.][]{nomotoNucleosynthesisStarsChemical2013,isobeEMPRESSIVExtremely2021}.
While the targeted galaxies do appear to be relatively chemically-young according to the canonical interpretation of the $\alpha/\mathrm{Fe}$ clock, robust constraints on the possible star formation and chemical enrichment histories leading to their present state must await detailed modeling.

\section{Summary}
\label{sec:summary}

In this work we presented new ultra-deep \hstcos{} spectroscopy for six nebular \civ{} emitters in the nearby Universe.
These spectra provide a unique window onto the massive star populations underlying their high-ionization emission.
Our data place strong constraints on both the photospheric and wind features of these massive stars, as well as on the detailed profiles of the nebular emission that they power.
We directly model the normalized stellar continuum and derive metallicity constraints driven by stellar iron absorption, which we then compare both to the gas-phase abundances and to the stellar wind profiles in the highest-S/N spectra in our sample.
Our primary conclusions are summarized as-follows:
\begin{itemize}
    \item The strength and profile of nebular \civ{} emission in these high-S/N spectra closely follow a sequence from weak emission with prominent blueshifted absorption to very high equivalent width emission with a broad double-peaked profile (Section~\ref{sec:uvmeas}).
    We suggest that this is strong evidence that resonant scattering is taking place in the escape of this \civ{} emission, and that this is most likely occurring in the circumgalactic medium of the target galaxies where \civ{} absorbers are frequently encountered (Section~\ref{sec:disc_neb}).
    This scattering may help explain some of the otherwise puzzling variation found in \civ{} strength observed locally, but there is good evidence to suggest that this is less likely to suppress \civ{} in $z>6$ galaxies.
    \item Despite the fact that our sample probes a fairly wide range of gas-phase metallicities $12+\log\mathrm{O/H}\simeq 7.4$--$7.9$, the photospheric metallicities we derive from the stellar continuum fits are uniformly low, at 2--8\% solar (Section~\ref{sec:results}).
    This further solidifies the notion that nebular \civ{} in star-forming galaxies is associated with the hard ionizing radiation fields produced in young stellar populations below $10\%$ solar.
    Detailed comparison with the gas-phase oxygen abundances reveals evidence for $\alpha$-enhancement, with [O/Fe] ranging from 2--5 across the sample (median $3.0_{-1.2}^{+1.9}$; Section~\ref{sec:abundcomp}).
    This enhancement persists at 1-$\sigma$ even in the conservative case of no correction to the gas-phase CEL abundance scale.
    This reveals that these galaxies are dominated by stars and gas with chemistry closer to that of their primordial high-redshift counterparts than to the bulk of the evolved low-redshift Universe.
    \item The strong constraints placed simultaneously on stellar photospheric lines and prominent stellar wind complexes represents a unique challenge for stellar population synthesis models.
    We find intriguing evidence for a mismatch between the prominent stellar \civ{} P-Cygni profiles and that predicted by the stellar models constrained by the photospheric lines in the highest-S/N spectra in our sample.
    In the context of the models, these wind lines in isolation require metallicities substantially higher (by $0.2$--$0.4$ dex) than the photospheric lines prefer across our sample.
    Caution should thus be applied in interpreting metallicities derived from stellar wind lines directly as is routine at the highest redshifts.
    We suggest that this tension could reflect deficiencies in the treatment of winds for luminous metal-poor stars; but we cannot rule-out the possibility that accounting for significant $\alpha$-enhancement relative to solar abundances in the stellar atmosphere may also directly reproduce the anomalously-strong \civ{} wind signature alongside weak photospheric iron absorption.
\end{itemize}

Our results forecast a promising future for the continued use of star-forming metal-poor dwarf galaxies as fundamental calibrators for both galactic and stellar astrophysics.
Their importance in both roles will only increase as the next generation of large ground-based and space telescopes arrive on-sky, revealing orders of magnitude larger samples of galaxies dominated by low-metallicity stellar populations.
However, several key areas of attention must remain in focus as this work proceeds.
First, the impending loss of ultraviolet spectroscopic capabilities from \hst{} will severely hamper this work; observations of metal-poor star-forming galaxies are of especially high priority in the coming years, both as high-redshift templates and extremely low-metallicity stellar calibrators.
A high signal-to-noise reference library of ultraviolet spectra for low-metallicity star-forming regions spanning as many of the photospheric and wind complexes in the FUV as possible will provide the empirical laboratory against which metal-poor stellar models can be usefully compared: for which this sample and the COS Legacy Archive Spectroscopic SurveY (CLASSY, Berg et al.\ submitted) represent a significant start.
In addition, alternative pathways in the optical and infrared towards constraining the fundamental properties of the massive stars in these systems should be thoroughly explored.
And second, the latest stellar population synthesis prescriptions are already being pushed to their limits in reproducing the detailed UV spectra of star-forming regions including those presented here.
A next-generation sequence of stellar population synthesis models with underlying atmosphere grids where important properties including abundances and wind modeling parameters are varied in physically-motivated ways will be necessary to fully explore the information content of these stellar continua, as well as to enable a full assessment of the quantitative systematic uncertainty these variables introduce to the modeling of integrated spectra at high-redshift.

\begin{acknowledgments}

This research is based on observations made with the NASA/ESA Hubble Space Telescope and supported by a grant from the Space Telescope Science Institute, which is operated by the Association of Universities for Research in Astronomy, Inc., under NASA contract NAS 5-26555. 
These observations are associated with programs GO:15646 and GO:15881.
Observations reported here were obtained at the MMT Observatory, a joint facility of the University of Arizona and the Smithsonian Institution.
Some of the computing for this project was performed on the Memex cluster.
We would like to thank Carnegie Institution for Science and the Carnegie Sci-Comp Committee for providing computational resources and support that contributed to these research results.

P.~S.\ was generously supported by a Carnegie Fellowship through the Carnegie Observatories during the completion of this manuscript.
D.P.S.\ acknowledges support from the National Science Foundation through the grant AST-1410155.
JC and SC acknowledge support from the ERC via an Advanced Grant under grant agreement no. 321323-NEOGAL.

This research made use of Astropy, a community-developed core python package for Astronomy \citep{astropycollaborationAstropyCommunityPython2013}; Matplotlib \citep{hunterMatplotlib2DGraphics2007}; Numpy and SciPy \citep{jonesSciPyOpenSource2001}; the SIMBAD database, operated at CDS, Strasbourg, France; and NASA's Astrophysics Data System.
   
\end{acknowledgments}

\bibliography{zoterolib}{}

\begin{thebibliography}{}
\expandafter\ifx\csname natexlab\endcsname\relax\def\natexlab#1{#1}\fi
\providecommand{\url}[1]{\href{#1}{#1}}
\providecommand{\dodoi}[1]{doi:~\href{http://doi.org/#1}{\nolinkurl{#1}}}
\providecommand{\doeprint}[1]{\href{http://ascl.net/#1}{\nolinkurl{http://ascl.net/#1}}}
\providecommand{\doarXiv}[1]{\href{https://arxiv.org/abs/#1}{\nolinkurl{https://arxiv.org/abs/#1}}}

\bibitem[{Adelberger {et~al.}(2005)Adelberger, Shapley, Steidel, Pettini, Erb,
  \& Reddy}]{adelbergerConnectionGalaxiesIntergalactic2005}
Adelberger, K.~L., Shapley, A.~E., Steidel, C.~C., {et~al.} 2005, \apj, 629,
  636, \dodoi{10.1086/431753}

\bibitem[{Annibali {et~al.}(2013)Annibali, Cignoni, Tosi, van~der Marel,
  Aloisi, Clementini, Contreras~Ramos, Fiorentino, Marconi, \&
  Musella}]{annibaliStarFormationHistory2013}
Annibali, F., Cignoni, M., Tosi, M., {et~al.} 2013, \aj, 146, 144,
  \dodoi{10.1088/0004-6256/146/6/144}

\bibitem[{{Astropy Collaboration} {et~al.}(2013){Astropy Collaboration},
  Robitaille, Tollerud, Greenfield, Droettboom, Bray, Aldcroft, Davis,
  Ginsburg, Price-Whelan, Kerzendorf, Conley, Crighton, Barbary, Muna,
  Ferguson, Grollier, Parikh, Nair, Unther, Deil, Woillez, Conseil, Kramer,
  Turner, Singer, Fox, Weaver, Zabalza, Edwards, Azalee~Bostroem, Burke, Casey,
  Crawford, Dencheva, Ely, Jenness, Labrie, Lim, Pierfederici, Pontzen, Ptak,
  Refsdal, Servillat, \&
  Streicher}]{astropycollaborationAstropyCommunityPython2013}
{Astropy Collaboration}, Robitaille, T.~P., Tollerud, E.~J., {et~al.} 2013,
  \aap, 558, A33, \dodoi{10.1051/0004-6361/201322068}

\bibitem[{Becker {et~al.}(2009)Becker, Rauch, \&
  Sargent}]{beckerHighRedshiftMetalsDecline2009}
Becker, G.~D., Rauch, M., \& Sargent, W. L.~W. 2009, \apj, 698, 1010,
  \dodoi{10.1088/0004-637X/698/2/1010}

\bibitem[{Bensby {et~al.}(2014)Bensby, Feltzing, \&
  Oey}]{bensbyExploringMilkyWay2014}
Bensby, T., Feltzing, S., \& Oey, M.~S. 2014, \aap, 562, A71,
  \dodoi{10.1051/0004-6361/201322631}

\bibitem[{Bensby {et~al.}(2013)Bensby, Yee, Feltzing, Johnson, Gould, Cohen,
  Asplund, Mel{\'e}ndez, Lucatello, Han, Thompson, {Gal-Yam}, Udalski, Bennett,
  Bond, Kohei, Sumi, Suzuki, Suzuki, Takino, Tristram, Yamai, \&
  Yonehara}]{bensbyChemicalEvolutionGalactic2013}
Bensby, T., Yee, J.~C., Feltzing, S., {et~al.} 2013, \aap, 549, A147,
  \dodoi{10.1051/0004-6361/201220678}

\bibitem[{Berg {et~al.}(2019{\natexlab{a}})Berg, Chisholm, Erb, Pogge, Henry,
  \& Olivier}]{bergIntenseIVHe2019}
Berg, D.~A., Chisholm, J., Erb, D.~K., {et~al.} 2019{\natexlab{a}}, \apj, 878,
  L3, \dodoi{10.3847/2041-8213/ab21dc}

\bibitem[{Berg {et~al.}(2021)Berg, Chisholm, Erb, Skillman, Pogge, \&
  Olivier}]{bergCharacterizingExtremeEmission2021}
---. 2021, arXiv:2105.12765 [astro-ph].
\newblock \doarXiv{2105.12765}

\bibitem[{Berg {et~al.}(2019{\natexlab{b}})Berg, Erb, Henry, Skillman, \&
  McQuinn}]{bergChemicalEvolutionCarbon2019}
Berg, D.~A., Erb, D.~K., Henry, R. B.~C., Skillman, E.~D., \& McQuinn, K. B.~W.
  2019{\natexlab{b}}, \apj, 874, 93, \dodoi{10.3847/1538-4357/ab020a}

\bibitem[{Berg {et~al.}(2016)Berg, Skillman, Henry, Erb, \&
  Carigi}]{bergCarbonOxygenAbundances2016}
Berg, D.~A., Skillman, E.~D., Henry, R. B.~C., Erb, D.~K., \& Carigi, L. 2016,
  \apj, 827, 126, \dodoi{10.3847/0004-637X/827/2/126}

\bibitem[{Bordoloi {et~al.}(2014)Bordoloi, Tumlinson, Werk, Oppenheimer,
  Peeples, Prochaska, Tripp, Katz, Dav{\'e}, Fox, Thom, Ford, Weinberg,
  Burchett, \& Kollmeier}]{bordoloiCOSDwarfsSurveyCarbon2014}
Bordoloi, R., Tumlinson, J., Werk, J.~K., {et~al.} 2014, \apj, 796, 136,
  \dodoi{10.1088/0004-637X/796/2/136}

\bibitem[{Borthakur {et~al.}(2013)Borthakur, Heckman, Strickland, Wild, \&
  Schiminovich}]{borthakurImpactStarburstsCircumgalactic2013}
Borthakur, S., Heckman, T., Strickland, D., Wild, V., \& Schiminovich, D. 2013,
  \apj, 768, 18, \dodoi{10.1088/0004-637X/768/1/18}

\bibitem[{Bouret {et~al.}(2005)Bouret, Lanz, \&
  Hillier}]{bouretLowerMassLoss2005}
Bouret, J.-C., Lanz, T., \& Hillier, D.~J. 2005, \aap, 438, 301,
  \dodoi{10.1051/0004-6361:20042531}

\bibitem[{Bouret {et~al.}(2003)Bouret, Lanz, Hillier, Heap, Hubeny, Lennon,
  Smith, \& Evans}]{bouretQuantitativeSpectroscopyStars2003}
Bouret, J.-C., Lanz, T., Hillier, D.~J., {et~al.} 2003, \apj, 595, 1182,
  \dodoi{10.1086/377368}

\bibitem[{Bouret {et~al.}(2015)Bouret, Lanz, Hillier, Martins, Marcolino, \&
  Depagne}]{bouretNoBreakdownRadiatively2015}
---. 2015, \mnras, 449, 1545, \dodoi{10.1093/mnras/stv379}

\bibitem[{Bouret {et~al.}(2013)Bouret, Lanz, Martins, Marcolino, Hillier,
  Depagne, \& Hubeny}]{bouretMassiveStarsLow2013}
Bouret, J.-C., Lanz, T., Martins, F., {et~al.} 2013, \aap, 555, A1,
  \dodoi{10.1051/0004-6361/201220798}

\bibitem[{Bresolin {et~al.}(2016)Bresolin, Kudritzki, Urbaneja, Gieren, Ho, \&
  Pietrzy{\'n}ski}]{bresolinYoungStarsIonized2016}
Bresolin, F., Kudritzki, R.-P., Urbaneja, M.~A., {et~al.} 2016, \apj, 830, 64,
  \dodoi{10.3847/0004-637X/830/2/64}

\bibitem[{Bressan {et~al.}(2012)Bressan, Marigo, Girardi, Salasnich, Dal~Cero,
  Rubele, \& Nanni}]{bressanPARSECStellarTracks2012}
Bressan, A., Marigo, P., Girardi, L., {et~al.} 2012, \mnras, 427, 127,
  \dodoi{10.1111/j.1365-2966.2012.21948.x}

\bibitem[{Bruzual \& Charlot(2003)}]{bruzualStellarPopulationSynthesis2003}
Bruzual, G., \& Charlot, S. 2003, \mnras, 344, 1000,
  \dodoi{10.1046/j.1365-8711.2003.06897.x}

\bibitem[{Caffau {et~al.}(2008)Caffau, Ludwig, Steffen, Ayres, Bonifacio,
  Cayrel, Freytag, \& Plez}]{caffauPhotosphericSolarOxygen2008}
Caffau, E., Ludwig, H.-G., Steffen, M., {et~al.} 2008, \aap, 488, 1031,
  \dodoi{10.1051/0004-6361:200809885}

\bibitem[{Caffau {et~al.}(2011)Caffau, Ludwig, Steffen, Freytag, \&
  Bonifacio}]{caffauSolarChemicalAbundances2011}
Caffau, E., Ludwig, H.-G., Steffen, M., Freytag, B., \& Bonifacio, P. 2011,
  Solar Physics, 268, 255, \dodoi{10.1007/s11207-010-9541-4}

\bibitem[{Cayrel {et~al.}(2004)Cayrel, Depagne, Spite, Hill, Spite, François,
  Plez, Beers, Primas, Andersen, Barbuy, Bonifacio, Molaro, \&
  Nordstr\"{o}m}]{cayrelFirstStarsAbundance2004}
Cayrel, R., Depagne, E., Spite, M., {et~al.} 2004, \aap, 416, 1117,
  \dodoi{10.1051/0004-6361:20034074}

\bibitem[{Chabrier(2003)}]{chabrierGalacticStellarSubstellar2003}
Chabrier, G. 2003, \pasp, 115, 763, \dodoi{10.1086/376392}

\bibitem[{Charlot \& Longhetti(2001)}]{charlotNebularEmissionStarforming2001}
Charlot, S., \& Longhetti, M. 2001, \mnras, 323, 887,
  \dodoi{10.1046/j.1365-8711.2001.04260.x}

\bibitem[{Chen {et~al.}(2001)Chen, Lanzetta, \&
  Webb}]{chenOriginIVAbsorption2001}
Chen, H.-W., Lanzetta, K.~M., \& Webb, J.~K. 2001, \apj, 556, 158,
  \dodoi{10.1086/321537}

\bibitem[{Chen {et~al.}(2015)Chen, Bressan, Girardi, Marigo, Kong, \&
  Lanza}]{chenPARSECEvolutionaryTracks2015}
Chen, Y., Bressan, A., Girardi, L., {et~al.} 2015, \mnras, 452, 1068,
  \dodoi{10.1093/mnras/stv1281}

\bibitem[{Chevallard \&
  Charlot(2016)}]{chevallardModellingInterpretingSpectral2016}
Chevallard, J., \& Charlot, S. 2016, \mnras, 462, 1415,
  \dodoi{10.1093/mnras/stw1756}

\bibitem[{Chisholm {et~al.}(2019)Chisholm, Rigby, Bayliss, Berg, Dahle,
  Gladders, \& Sharon}]{chisholmConstrainingMetallicitiesAges2019}
Chisholm, J., Rigby, J.~R., Bayliss, M., {et~al.} 2019, \apj, 882, 182,
  \dodoi{10.3847/1538-4357/ab3104}

\bibitem[{Chisholm {et~al.}(2018)Chisholm, Tremonti, \&
  Leitherer}]{chisholmMetalenrichedGalacticOutflows2018}
Chisholm, J., Tremonti, C., \& Leitherer, C. 2018, \mnras, 481, 1690,
  \dodoi{10.1093/mnras/sty2380}

\bibitem[{Cooke(2015)}]{cookeBigBangNucleosynthesis2015}
Cooke, R.~J. 2015, \apjl, 812, L12, \dodoi{10.1088/2041-8205/812/1/L12}

\bibitem[{Cooper {et~al.}(2019)Cooper, Simcoe, Cooksey, Bordoloi, Miller,
  Furesz, Turner, \& Ba{\~n}ados}]{cooperHeavyElementAbsorption2019}
Cooper, T.~J., Simcoe, R.~A., Cooksey, K.~L., {et~al.} 2019, \apj, 882, 77,
  \dodoi{10.3847/1538-4357/ab3402}

\bibitem[{Crowther {et~al.}(2006)Crowther, Prinja, Pettini, \&
  Steidel}]{crowtherReliabilityCIVl1549Abundance2006}
Crowther, P.~A., Prinja, R.~K., Pettini, M., \& Steidel, C.~C. 2006, \mnras,
  368, 895, \dodoi{10.1111/j.1365-2966.2006.10164.x}

\bibitem[{Cullen {et~al.}(2021)Cullen, Shapley, McLure, Dunlop, Sanders,
  Topping, Reddy, Amorin, Begley, Bolzonella, Calabro, Carnall, Castellano,
  Cimatti, Cirasuolo, Cresci, Fontana, Fontanot, Garilli, Guaita, Hamadouche,
  Hathi, Mannucci, McLeod, Pentericci, Saxena, Talia, \&
  Zamorani}]{cullenNIRVANDELSSurveyRobust2021}
Cullen, F., Shapley, A.~E., McLure, R.~J., {et~al.} 2021, arXiv e-prints, 2103,
  arXiv:2103.06300

\bibitem[{Dalcanton(2007)}]{dalcantonMetallicityGalaxyDisks2007}
Dalcanton, J.~J. 2007, \apj, 658, 941, \dodoi{10.1086/508913}

\bibitem[{De~Vis {et~al.}(2019)De~Vis, Jones, Viaene, Casasola, Clark, Baes,
  Bianchi, Cassara, Davies, De~Looze, Galametz, Galliano, Lianou, Madden,
  {Manilla-Robles}, Mosenkov, Nersesian, Roychowdhury, Xilouris, \&
  Ysard}]{devisSystematicMetallicityStudy2019}
De~Vis, P., Jones, A., Viaene, S., {et~al.} 2019, \aap, 623, A5,
  \dodoi{10.1051/0004-6361/201834444}

\bibitem[{Draine(2011)}]{drainePhysicsInterstellarIntergalactic2011}
Draine, B.~T. 2011, Physics of the Interstellar and Intergalactic Medium

\bibitem[{Eldridge \& Stanway(2012)}]{eldridgeEffectStellarEvolution2012}
Eldridge, J.~J., \& Stanway, E.~R. 2012, \mnras, 419, 479,
  \dodoi{10.1111/j.1365-2966.2011.19713.x}

\bibitem[{Esteban {et~al.}(2014)Esteban, {Garc{\'i}a-Rojas}, Carigi, Peimbert,
  Bresolin, {L{\'o}pez-S{\'a}nchez}, \&
  {Mesa-Delgado}}]{estebanCarbonOxygenAbundances2014}
Esteban, C., {Garc{\'i}a-Rojas}, J., Carigi, L., {et~al.} 2014, \mnras, 443,
  624, \dodoi{10.1093/mnras/stu1177}

\bibitem[{Fabrizio {et~al.}(2015)Fabrizio, Nonino, Bono, Primas, Th{\'e}venin,
  Stetson, Cassisi, Buonanno, Coppola, {da Silva}, Dall'Ora, Ferraro, Genovali,
  Gilmozzi, Iannicola, Marconi, Monelli, Romaniello, \&
  Walker}]{fabrizioCarinaProjectVIII2015}
Fabrizio, M., Nonino, M., Bono, G., {et~al.} 2015, \aap, 580, A18,
  \dodoi{10.1051/0004-6361/201525753}

\bibitem[{Fanelli {et~al.}(1992)Fanelli, O'Connell, Burstein, \&
  Wu}]{fanelliSpectralSynthesisUltraviolet1992}
Fanelli, M.~N., O'Connell, R.~W., Burstein, D., \& Wu, C.-C. 1992, \apjs, 82,
  197, \dodoi{10.1086/191714}

\bibitem[{Ferland {et~al.}(2017)Ferland, Chatzikos, Guzm\'{a}n, Lykins, van
  Hoof, Williams, Abel, Badnell, Keenan, Porter, \&
  Stancil}]{ferland2017ReleaseCloudy2017}
Ferland, G.~J., Chatzikos, M., Guzm\'{a}n, F., {et~al.} 2017, \rmxaa, 53, 385

\bibitem[{Feroz {et~al.}(2009)Feroz, Hobson, \&
  Bridges}]{ferozMULTINESTEfficientRobust2009}
Feroz, F., Hobson, M.~P., \& Bridges, M. 2009, \mnras, 398, 1601,
  \dodoi{10.1111/j.1365-2966.2009.14548.x}

\bibitem[{Fitzpatrick(1999)}]{fitzpatrickCorrectingEffectsInterstellar1999}
Fitzpatrick, E.~L. 1999, \pasp, 111, 63, \dodoi{10.1086/316293}

\bibitem[{Foreman-Mackey {et~al.}(2013)Foreman-Mackey, Hogg, Lang, \&
  Goodman}]{foreman-mackeyEmceeMCMCHammer2013}
Foreman-Mackey, D., Hogg, D.~W., Lang, D., \& Goodman, J. 2013, \pasp, 125,
  306, \dodoi{10.1086/670067}

\bibitem[{Fox {et~al.}(2019)Fox, James, Frazer, \&
  Fischer}]{foxFluxCalibrationNew2019}
Fox, A., James, B.~L., Frazer, E.~M., \& Fischer, W.~J. 2019, Instrument
  Science Report COS 2019-9, 9 pages

\bibitem[{Frebel \& Norris(2015)}]{frebelNearFieldCosmologyExtremely2015}
Frebel, A., \& Norris, J.~E. 2015, \araa, 53, 631,
  \dodoi{10.1146/annurev-astro-082214-122423}

\bibitem[{Frebel {et~al.}(2014)Frebel, Simon, \&
  Kirby}]{frebelSegueUnevolvedFossil2014}
Frebel, A., Simon, J.~D., \& Kirby, E.~N. 2014, \apj, 786, 74,
  \dodoi{10.1088/0004-637X/786/1/74}

\bibitem[{Froese~Fischer \&
  Tachiev(2004)}]{froesefischerBreitPauliEnergyLevels2004}
Froese~Fischer, C., \& Tachiev, G. 2004, Atomic Data and Nuclear Data Tables,
  87, 1, \dodoi{10.1016/j.adt.2004.02.001}

\bibitem[{Gallazzi {et~al.}(2005)Gallazzi, Charlot, Brinchmann, White, \&
  Tremonti}]{gallazziAgesMetallicitiesGalaxies2005}
Gallazzi, A., Charlot, S., Brinchmann, J., White, S. D.~M., \& Tremonti, C.~A.
  2005, \mnras, 362, 41, \dodoi{10.1111/j.1365-2966.2005.09321.x}

\bibitem[{{Garc{\'i}a-Rojas} \&
  Esteban(2007)}]{garcia-rojasAbundanceDiscrepancyProblem2007}
{Garc{\'i}a-Rojas}, J., \& Esteban, C. 2007, \apj, 670, 457,
  \dodoi{10.1086/521871}

\bibitem[{Gordon {et~al.}(2003)Gordon, Clayton, Misselt, Landolt, \&
  Wolff}]{gordonQuantitativeComparisonSmall2003}
Gordon, K.~D., Clayton, G.~C., Misselt, K.~A., Landolt, A.~U., \& Wolff, M.~J.
  2003, \apj, 594, 279, \dodoi{10.1086/376774}

\bibitem[{Gratton {et~al.}(2003)Gratton, Carretta, Desidera, Lucatello, Mazzei,
  \& Barbieri}]{grattonAbundancesMetalpoorStars2003}
Gratton, R.~G., Carretta, E., Desidera, S., {et~al.} 2003, \aap, 406, 131,
  \dodoi{10.1051/0004-6361:20030754}

\bibitem[{Gray {et~al.}(2019)Gray, Oey, Silich, \&
  Scannapieco}]{grayCatastrophicCoolingSuperwinds2019}
Gray, W.~J., Oey, M.~S., Silich, S., \& Scannapieco, E. 2019, \apj, 887, 161,
  \dodoi{10.3847/1538-4357/ab510d}

\bibitem[{Graziani {et~al.}(2019)Graziani, Courtois, Lavaux, Hoffman, Tully,
  Copin, \& Pomar{\`e}de}]{grazianiPeculiarVelocityField2019}
Graziani, R., Courtois, H.~M., Lavaux, G., {et~al.} 2019, \mnras, 488, 5438,
  \dodoi{10.1093/mnras/stz078}

\bibitem[{Gutkin {et~al.}(2016)Gutkin, Charlot, \&
  Bruzual}]{gutkinModellingNebularEmission2016}
Gutkin, J., Charlot, S., \& Bruzual, G. 2016, \mnras, 462, 1757,
  \dodoi{10.1093/mnras/stw1716}

\bibitem[{Heap {et~al.}(2006)Heap, Lanz, \&
  Hubeny}]{heapFundamentalPropertiesOType2006}
Heap, S.~R., Lanz, T., \& Hubeny, I. 2006, \apj, 638, 409,
  \dodoi{10.1086/498635}

\bibitem[{Heckman {et~al.}(1998)Heckman, Robert, Leitherer, Garnett, \& van~der
  Rydt}]{heckmanUltravioletSpectroscopicProperties1998}
Heckman, T.~M., Robert, C., Leitherer, C., Garnett, D.~R., \& van~der Rydt, F.
  1998, \apj, 503, 646, \dodoi{10.1086/306035}

\bibitem[{Hernandez {et~al.}(2017)Hernandez, Larsen, Trager, Groot, \&
  Kaper}]{hernandezChemicalAbundancesTwo2017}
Hernandez, S., Larsen, S., Trager, S., Groot, P., \& Kaper, L. 2017, Astronomy
  \&amp; Astrophysics, Volume 603, id.A119,
  {$<$}NUMPAGES{$>$}24{$<$}/NUMPAGES{$>$} pp., 603, A119,
  \dodoi{10.1051/0004-6361/201730550}

\bibitem[{Hernandez {et~al.}(2019)Hernandez, Larsen, Aloisi, Berg, Blair, Fox,
  Heckman, James, Long, Skillman, \&
  Whitmore}]{hernandezFirstMetallicityStudy2019}
Hernandez, S., Larsen, S., Aloisi, A., {et~al.} 2019, \apj, 872, 116,
  \dodoi{10.3847/1538-4357/ab017a}

\bibitem[{Higson {et~al.}(2019)Higson, Handley, Hobson, \&
  Lasenby}]{higsonDynamicNestedSampling2019}
Higson, E., Handley, W., Hobson, M., \& Lasenby, A. 2019, Statistics and
  Computing, 29, 891, \dodoi{10.1007/s11222-018-9844-0}

\bibitem[{Hill {et~al.}(2019)Hill, Sk{\'u}lad{\'o}ttir, Tolstoy, Venn,
  Shetrone, Jablonka, Primas, Battaglia, {de Boer}, Fran{\c c}ois, Helmi,
  Kaufer, Letarte, Starkenburg, \& Spite}]{hillVLTFLAMESHighresolution2019}
Hill, V., Sk{\'u}lad{\'o}ttir, {\'A}., Tolstoy, E., {et~al.} 2019, \aap, 626,
  A15, \dodoi{10.1051/0004-6361/201833950}

\bibitem[{Hillier {et~al.}(2003)Hillier, Lanz, Heap, Hubeny, Smith, Evans,
  Lennon, \& Bouret}]{hillierTaleTwoStars2003}
Hillier, D.~J., Lanz, T., Heap, S.~R., {et~al.} 2003, \apj, 588, 1039,
  \dodoi{10.1086/374329}

\bibitem[{Hirschauer \& {et
  al.}(2021)}]{hirschauerCosmicOriginsSpectrograph2021}
Hirschauer, A.~S., \& {et al.} 2021, Cosmic {{Origins Spectrograph Instrument
  Handbook}}, {{Version}} 13.0 ({Baltimore}: {STScI})

\bibitem[{Hol {et~al.}(2006)Hol, Schon, \&
  Gustafsson}]{holResamplingAlgorithmsParticle2006}
Hol, J.~D., Schon, T.~B., \& Gustafsson, F. 2006, in 2006 {{IEEE Nonlinear
  Statistical Signal Processing Workshop}}, 79--82,
  \dodoi{10.1109/NSSPW.2006.4378824}

\bibitem[{Hsyu {et~al.}(2018)Hsyu, Cooke, Prochaska, \&
  Bolte}]{hsyuSearchingLowestmetallicityGalaxies2018}
Hsyu, T., Cooke, R.~J., Prochaska, J.~X., \& Bolte, M. 2018, \apj, 863, 134,
  \dodoi{10.3847/1538-4357/aad18a}

\bibitem[{Hunter \& Thronson(1995)}]{hunterMassiveStarsZW1995}
Hunter, D.~A., \& Thronson, Jr., H.~A. 1995, \apj, 452, 238,
  \dodoi{10.1086/176295}

\bibitem[{Hunter(2007)}]{hunterMatplotlib2DGraphics2007}
Hunter, J.~D. 2007, Computing In Science \& Engineering, 9, 90–95,
  \dodoi{10.1109/MCSE.2007.55}

\bibitem[{Irimia \& Froese~Fischer(2005)}]{irimiaBreitPauliOscillator2005}
Irimia, A., \& Froese~Fischer, C. 2005, Physica Scripta, 71, 172,
  \dodoi{10.1238/Physica.Regular.071a00172}

\bibitem[{Isobe {et~al.}(2021)Isobe, Ouchi, Suzuki, Moriya, Nakajima, Nomoto,
  Rauch, Harikane, Kojima, Ono, Fujimoto, Inoue, Kim, Komiyama, Kusakabe, Lee,
  Maseda, Matthee, {Michel-Dansac}, Nagao, Nanayakkara, Nishigaki, Onodera,
  Sugahara, \& Xu}]{isobeEMPRESSIVExtremely2021}
Isobe, Y., Ouchi, M., Suzuki, A., {et~al.} 2021, arXiv:2108.03850 [astro-ph].
\newblock \doarXiv{2108.03850}

\bibitem[{Izotov {et~al.}(2019)Izotov, Guseva, Fricke, \&
  Henkel}]{izotovLowredshiftLowestmetallicityStarforming2019}
Izotov, Y.~I., Guseva, N.~G., Fricke, K.~J., \& Henkel, C. 2019, \aap, 623,
  A40, \dodoi{10.1051/0004-6361/201834768}

\bibitem[{Izotov {et~al.}(2006)Izotov, Stasińska, Meynet, Guseva, \&
  Thuan}]{izotovChemicalCompositionMetalpoor2006}
Izotov, Y.~I., Stasińska, G., Meynet, G., Guseva, N.~G., \& Thuan, T.~X. 2006,
  \aap, 448, 955, \dodoi{10.1051/0004-6361:20053763}

\bibitem[{Izotov \& Thuan(2004)}]{izotovDeepHubbleSpace2004}
Izotov, Y.~I., \& Thuan, T.~X. 2004, \apj, 616, 768, \dodoi{10.1086/424990}

\bibitem[{Jones {et~al.}(2001)Jones, Oliphant, Peterson,
  {et~al.}}]{jonesSciPyOpenSource2001}
Jones, E., Oliphant, T., Peterson, P., {et~al.} 2001, SciPy: Open source
  scientific tools for Python

\bibitem[{Keel {et~al.}(2004)Keel, Holberg, \&
  Treuthardt}]{keelFarUltravioletSpectroscopyStarforming2004}
Keel, W.~C., Holberg, J.~B., \& Treuthardt, P.~M. 2004, \aj, 128, 211,
  \dodoi{10.1086/421367}

\bibitem[{Kehrig {et~al.}(2018)Kehrig, V\'{i}lchez, Guerrero,
  Iglesias-P\'{a}ramo, Hunt, Duarte-Puertas, \&
  Ramos-Larios}]{kehrigExtendedHeII2018}
Kehrig, C., V\'{i}lchez, J.~M., Guerrero, M.~A., {et~al.} 2018, \mnras, 480,
  1081, \dodoi{10.1093/mnras/sty1920}

\bibitem[{Kobayashi {et~al.}(2006)Kobayashi, Umeda, Nomoto, Tominaga, \&
  Ohkubo}]{kobayashiGalacticChemicalEvolution2006}
Kobayashi, C., Umeda, H., Nomoto, K., Tominaga, N., \& Ohkubo, T. 2006, \apj,
  653, 1145, \dodoi{10.1086/508914}

\bibitem[{Kojima {et~al.}(2019)Kojima, Ouchi, Rauch, Ono, Isobe, Fujimoto,
  Harikane, Hashimoto, Hayashi, Komiyama, Kusakabe, Kim, Lee, Mukae, Nagao,
  Onodera, Shibuya, Sugahara, Umemura, \& Yabe}]{kojimaSubaruHSCMachine2019}
Kojima, T., Ouchi, M., Rauch, M., {et~al.} 2019, arXiv:1910.08559 [astro-ph].
\newblock \doarXiv{1910.08559}

\bibitem[{Kojima {et~al.}(2020)Kojima, Ouchi, Rauch, Ono, Nakajima, Isobe,
  Fujimoto, Harikane, Hashimoto, Hayashi, Komiyama, Kusakabe, Kim, Lee, Mukae,
  Nagao, Onodera, Shibuya, Sugahara, Umemura, \&
  Yabe}]{kojimaEMPRESSIIHighly2020}
---. 2020, arXiv:2006.03831 [astro-ph].
\newblock \doarXiv{2006.03831}

\bibitem[{Kourkchi {et~al.}(2020)Kourkchi, Courtois, Graziani, Hoffman,
  Pomar{\`e}de, Shaya, \& Tully}]{kourkchiCosmicflows3TwoDistanceVelocity2020}
Kourkchi, E., Courtois, H.~M., Graziani, R., {et~al.} 2020, \aj, 159, 67,
  \dodoi{10.3847/1538-3881/ab620e}

\bibitem[{Kroupa {et~al.}(1993)Kroupa, Tout, \&
  Gilmore}]{kroupaDistributionLowMassStars1993}
Kroupa, P., Tout, C.~A., \& Gilmore, G. 1993, \mnras, 262, 545,
  \dodoi{10.1093/mnras/262.3.545}

\bibitem[{Kurucz(2006)}]{kuruczRobertKuruczOnline2006}
Kurucz, R.~L. 2006, Robert {{L}}. {{Kurucz}} on-Line Database of Observed and
  Predicted Atomic Transitions

\bibitem[{Lanz \& Hubeny(2003)}]{lanzGridNonLTELineblanketed2003}
Lanz, T., \& Hubeny, I. 2003, \apjs, 146, 417, \dodoi{10.1086/374373}

\bibitem[{Lardo {et~al.}(2015)Lardo, Davies, Kudritzki, Gazak, Evans, Patrick,
  Bergemann, \& Plez}]{lardoRedSupergiantsCosmic2015}
Lardo, C., Davies, B., Kudritzki, R.~P., {et~al.} 2015, \apj, 812, 160,
  \dodoi{10.1088/0004-637X/812/2/160}

\bibitem[{Larsen {et~al.}(2008)Larsen, Origlia, Brodie, \&
  Gallagher}]{larsenAnatomyYoungMassive2008}
Larsen, S.~S., Origlia, L., Brodie, J., \& Gallagher, J.~S. 2008, \mnras, 383,
  263, \dodoi{10.1111/j.1365-2966.2007.12528.x}

\bibitem[{Leitherer {et~al.}(2014)Leitherer, Ekstr\"{o}m, Meynet, Schaerer,
  Agienko, \& Levesque}]{leithererEffectsStellarRotation2014}
Leitherer, C., Ekstr\"{o}m, S., Meynet, G., {et~al.} 2014, \apjs, 212, 14,
  \dodoi{10.1088/0067-0049/212/1/14}

\bibitem[{Leitherer {et~al.}(2001)Leitherer, Le{\~a}o, Heckman, Lennon,
  Pettini, \& Robert}]{leithererUltravioletLineSpectra2001}
Leitherer, C., Le{\~a}o, J. R.~S., Heckman, T.~M., {et~al.} 2001, \apj, 550,
  724, \dodoi{10.1086/319814}

\bibitem[{Leitherer {et~al.}(2010)Leitherer, Ortiz~Ot\'{a}lvaro, Bresolin,
  Kudritzki, Lo~Faro, Pauldrach, Pettini, \&
  Rix}]{leithererLibraryTheoreticalUltraviolet2010}
Leitherer, C., Ortiz~Ot\'{a}lvaro, P.~A., Bresolin, F., {et~al.} 2010, \apjs,
  189, 309, \dodoi{10.1088/0067-0049/189/2/309}

\bibitem[{Leitherer {et~al.}(2011)Leitherer, Tremonti, Heckman, \&
  Calzetti}]{leithererUltravioletSpectroscopicAtlas2011}
Leitherer, C., Tremonti, C.~A., Heckman, T.~M., \& Calzetti, D. 2011, \aj, 141,
  37, \dodoi{10.1088/0004-6256/141/2/37}

\bibitem[{Mainali {et~al.}(2017)Mainali, Kollmeier, Stark, Simcoe, Walth,
  Newman, \& Miller}]{mainaliEvidenceHardIonizing2017}
Mainali, R., Kollmeier, J.~A., Stark, D.~P., {et~al.} 2017, \apjl, 836, L14,
  \dodoi{10.3847/2041-8213/836/1/L14}

\bibitem[{Maiolino \& Mannucci(2019)}]{maiolinoReMetallicaCosmic2019}
Maiolino, R., \& Mannucci, F. 2019, \aap Review, 27, 3,
  \dodoi{10.1007/s00159-018-0112-2}

\bibitem[{Nomoto {et~al.}(2013)Nomoto, Kobayashi, \&
  Tominaga}]{nomotoNucleosynthesisStarsChemical2013}
Nomoto, K., Kobayashi, C., \& Tominaga, N. 2013, \araa, 51, 457,
  \dodoi{10.1146/annurev-astro-082812-140956}

\bibitem[{Nomoto {et~al.}(2006)Nomoto, Tominaga, Umeda, Kobayashi, \&
  Maeda}]{nomotoNucleosynthesisYieldsCorecollapse2006}
Nomoto, K., Tominaga, N., Umeda, H., Kobayashi, C., \& Maeda, K. 2006, Nuclear
  Physics A, 777, 424, \dodoi{10.1016/j.nuclphysa.2006.05.008}

\bibitem[{Peimbert \& Peimbert(2010)}]{peimbertMgSiFe2010}
Peimbert, A., \& Peimbert, M. 2010, \apj, 724, 791,
  \dodoi{10.1088/0004-637X/724/1/791}

\bibitem[{Peimbert {et~al.}(1993)Peimbert, Storey, \&
  {Torres-Peimbert}}]{peimbertAbundanceRatioGaseous1993}
Peimbert, M., Storey, P.~J., \& {Torres-Peimbert}, S. 1993, \apj, 414, 626,
  \dodoi{10.1086/173108}

\bibitem[{Plat {et~al.}(2019)Plat, Charlot, Bruzual, Feltre,
  {Vidal-Garc{\'i}a}, Morisset, Chevallard, \&
  Todt}]{platConstraintsProductionEscape2019}
Plat, A., Charlot, S., Bruzual, G., {et~al.} 2019, \mnras, 490, 978,
  \dodoi{10.1093/mnras/stz2616}

\bibitem[{Prinja \& Crowther(1998)}]{prinjaHSTUVMeasurementsWind1998}
Prinja, R.~K., \& Crowther, P.~A. 1998, \mnras, 300, 828,
  \dodoi{10.1046/j.1365-8711.1998.01963.x}

\bibitem[{Puls {et~al.}(2020)Puls, Najarro, Sundqvist, \&
  Sen}]{pulsAtmosphericNLTEModels2020}
Puls, J., Najarro, F., Sundqvist, J.~O., \& Sen, K. 2020, Astronomy \&
  Astrophysics, 642, A172, \dodoi{10.1051/0004-6361/202038464}

\bibitem[{Rafelski(2018)}]{rafelskiCOSDataHandbook2018}
Rafelski, M. 2018, {{COS Data Handbook}}, {{Version}} 4.0 ({Baltimore: STScI})

\bibitem[{Rauch {et~al.}(2001)Rauch, Sargent, \&
  Barlow}]{rauchSmallScaleStructureHigh2001}
Rauch, M., Sargent, W. L.~W., \& Barlow, T.~A. 2001, \apj, 554, 823,
  \dodoi{10.1086/321402}

\bibitem[{Rigby {et~al.}(2015)Rigby, Bayliss, Gladders, Sharon, Wuyts, Dahle,
  Johnson, \& Peña-Guerrero}]{rigbyIIIEmissionStarforming2015}
Rigby, J.~R., Bayliss, M.~B., Gladders, M.~D., {et~al.} 2015, \apjl, 814, L6,
  \dodoi{10.1088/2041-8205/814/1/L6}

\bibitem[{Rigby {et~al.}(2018)Rigby, Bayliss, Chisholm, Bordoloi, Sharon,
  Gladders, Johnson, {Paterno-Mahler}, Wuyts, Dahle, \&
  Acharyya}]{rigbyMagellanEvolutionGalaxies2018}
Rigby, J.~R., Bayliss, M.~B., Chisholm, J., {et~al.} 2018, \apj, 853, 87,
  \dodoi{10.3847/1538-4357/aaa2fc}

\bibitem[{Rix {et~al.}(2004)Rix, Pettini, Leitherer, Bresolin, Kudritzki, \&
  Steidel}]{rixSpectralModelingStarforming2004}
Rix, S.~A., Pettini, M., Leitherer, C., {et~al.} 2004, \apj, 615, 98,
  \dodoi{10.1086/424031}

\bibitem[{Rodr\'{i}guez \& Rubin(2005)}]{rodriguezFeIVDiscrepancy2005}
Rodr\'{i}guez, M., \& Rubin, R.~H. 2005, \apj, 626, 900, \dodoi{10.1086/429958}

\bibitem[{Rudie {et~al.}(2019)Rudie, Steidel, Pettini, Trainor, Strom, Hummels,
  Reddy, \& Shapley}]{rudieColumnDensityKinematics2019}
Rudie, G.~C., Steidel, C.~C., Pettini, M., {et~al.} 2019, \apj, 885, 61,
  \dodoi{10.3847/1538-4357/ab4255}

\bibitem[{Ryabchikova {et~al.}(2015)Ryabchikova, Piskunov, Kurucz, Stempels,
  Heiter, Pakhomov, \& Barklem}]{ryabchikovaMajorUpgradeVALD2015}
Ryabchikova, T., Piskunov, N., Kurucz, R.~L., {et~al.} 2015, Physica Scripta,
  90, 054005, \dodoi{10.1088/0031-8949/90/5/054005}

\bibitem[{Sanders {et~al.}(2020)Sanders, Shapley, Reddy, Kriek, Siana, Coil,
  Mobasher, Shivaei, Freeman, Azadi, Price, Leung, Fetherolf, {de Groot}, Zick,
  Fornasini, \& Barro}]{sandersMOSDEFSurveyDirectmethod2020}
Sanders, R.~L., Shapley, A.~E., Reddy, N.~A., {et~al.} 2020, \mnras, 491, 1427,
  \dodoi{10.1093/mnras/stz3032}

\bibitem[{Sargent \& Searle(1970)}]{sargentIsolatedExtragalacticII1970}
Sargent, W. L.~W., \& Searle, L. 1970, \apj, 162, L155, \dodoi{10.1086/180644}

\bibitem[{Schlafly \& Finkbeiner(2011)}]{schlaflyMeasuringReddeningSloan2011}
Schlafly, E.~F., \& Finkbeiner, D.~P. 2011, \apj, 737, 103,
  \dodoi{10.1088/0004-637X/737/2/103}

\bibitem[{Schmidt {et~al.}(2017)Schmidt, Huang, Treu, Hoag, Brada{\v c}, Henry,
  Jones, Mason, Malkan, Morishita, Pentericci, Trenti, Vulcani, \&
  Wang}]{schmidtGrismLensAmplifiedSurvey2017}
Schmidt, K.~B., Huang, K.-H., Treu, T., {et~al.} 2017, \apj, 839, 17,
  \dodoi{10.3847/1538-4357/aa68a3}

\bibitem[{Searle \& Sargent(1972)}]{searleInferencesCompositionTwo1972}
Searle, L., \& Sargent, W. L.~W. 1972, \apj, 173, 25, \dodoi{10.1086/151398}

\bibitem[{Searle {et~al.}(1973)Searle, Sargent, \&
  Bagnuolo}]{searleHistoryStarFormation1973}
Searle, L., Sargent, W. L.~W., \& Bagnuolo, W.~G. 1973, \apj, 179, 427,
  \dodoi{10.1086/151882}

\bibitem[{Senchyna \& Stark(2019)}]{senchynaPhotometricIdentificationMMT2019}
Senchyna, P., \& Stark, D.~P. 2019, \mnras, 484, 1270,
  \dodoi{10.1093/mnras/stz058}

\bibitem[{Senchyna {et~al.}(2021)Senchyna, Stark, Charlot, Chevallard, Bruzual,
  \& {Vidal-Garc{\'i}a}}]{senchynaUltravioletSpectraExtreme2021}
Senchyna, P., Stark, D.~P., Charlot, S., {et~al.} 2021, \mnras, 503, 6112,
  \dodoi{10.1093/mnras/stab884}

\bibitem[{Senchyna {et~al.}(2019)Senchyna, Stark, Chevallard, Charlot, Jones,
  \& {Vidal-Garc{\'i}a}}]{senchynaExtremelyMetalpoorGalaxies2019}
Senchyna, P., Stark, D.~P., Chevallard, J., {et~al.} 2019, \mnras, 488, 3492,
  \dodoi{10.1093/mnras/stz1907}

\bibitem[{Senchyna {et~al.}(2017)Senchyna, Stark, Vidal-Garc\'{i}a, Chevallard,
  Charlot, Mainali, Jones, Wofford, Feltre, \&
  Gutkin}]{senchynaUltravioletSpectraExtreme2017}
Senchyna, P., Stark, D.~P., Vidal-Garc\'{i}a, A., {et~al.} 2017, \mnras, 472,
  2608, \dodoi{10.1093/mnras/stx2059}

\bibitem[{Shapley {et~al.}(2003)Shapley, Steidel, Pettini, \&
  Adelberger}]{shapleyRestFrameUltravioletSpectra2003}
Shapley, A.~E., Steidel, C.~C., Pettini, M., \& Adelberger, K.~L. 2003, \apj,
  588, 65, \dodoi{10.1086/373922}

\bibitem[{Shetrone {et~al.}(2003)Shetrone, Venn, Tolstoy, Primas, Hill, \&
  Kaufer}]{shetroneVLTUVESAbundances2003}
Shetrone, M., Venn, K.~A., Tolstoy, E., {et~al.} 2003, \aj, 125, 684,
  \dodoi{10.1086/345966}

\bibitem[{Skilling(2004)}]{skillingNestedSampling2004}
Skilling, J. 2004, 735, 395, \dodoi{10.1063/1.1835238}

\bibitem[{Skilling(2006)}]{skillingNestedSamplingGeneral2006}
---. 2006, Bayesian Analysis, 1, 833, \dodoi{10.1214/06-BA127}

\bibitem[{Speagle(2020)}]{speagleDYNESTYDynamicNested2020}
Speagle, J.~S. 2020, \mnras, 493, 3132, \dodoi{10.1093/mnras/staa278}

\bibitem[{Stark {et~al.}(2015)Stark, Walth, Charlot, Cl\'{e}ment, Feltre,
  Gutkin, Richard, Mainali, Robertson, Siana, Tang, \&
  Schenker}]{starkSpectroscopicDetectionIV2015}
Stark, D.~P., Walth, G., Charlot, S., {et~al.} 2015, \mnras, 454, 1393,
  \dodoi{10.1093/mnras/stv1907}

\bibitem[{Steidel(1990)}]{steidelHighRedshiftExtensionSurvey1990}
Steidel, C.~C. 1990, \apjs, 72, 1, \dodoi{10.1086/191407}

\bibitem[{Steidel {et~al.}(2016)Steidel, Strom, Pettini, Rudie, Reddy, \&
  Trainor}]{steidelReconcilingStellarNebular2016}
Steidel, C.~C., Strom, A.~L., Pettini, M., {et~al.} 2016, \apj, 826, 159,
  \dodoi{10.3847/0004-637X/826/2/159}

\bibitem[{Storey {et~al.}(2014)Storey, Sochi, \&
  Badnell}]{storeyCollisionStrengthsNebular2014}
Storey, P.~J., Sochi, T., \& Badnell, N.~R. 2014, \mnras, 441, 3028,
  \dodoi{10.1093/mnras/stu777}

\bibitem[{Storey \& Zeippen(2000)}]{storeyTheoreticalValuesOIII2000}
Storey, P.~J., \& Zeippen, C.~J. 2000, \mnras, 312, 813,
  \dodoi{10.1046/j.1365-8711.2000.03184.x}

\bibitem[{Strom {et~al.}(2021)Strom, Rudie, Steidel, \&
  Trainor}]{stromChemicalAbundanceScaling2021}
Strom, A.~L., Rudie, G.~C., Steidel, C.~C., \& Trainor, R.~F. 2021,
  arXiv:2111.06410 [astro-ph].
\newblock \doarXiv{2111.06410}

\bibitem[{Strom {et~al.}(2017)Strom, Steidel, Rudie, Trainor, Pettini, \&
  Reddy}]{stromNebularEmissionLine2017}
Strom, A.~L., Steidel, C.~C., Rudie, G.~C., {et~al.} 2017, \apj, 836, 164,
  \dodoi{10.3847/1538-4357/836/2/164}

\bibitem[{Sundqvist {et~al.}(2019)Sundqvist, Bj{\"o}rklund, Puls, \&
  Najarro}]{sundqvistNewPredictionsRadiationdriven2019}
Sundqvist, J.~O., Bj{\"o}rklund, R., Puls, J., \& Najarro, F. 2019,
  arXiv:1910.06586 [astro-ph].
\newblock \doarXiv{1910.06586}

\bibitem[{Tayal(2007)}]{tayalOscillatorStrengthsElectron2007}
Tayal, S.~S. 2007, \apjs, 171, 331, \dodoi{10.1086/513107}

\bibitem[{Tayal \&
  Zatsarinny(2010)}]{tayalBreitPauliTransitionProbabilities2010}
Tayal, S.~S., \& Zatsarinny, O. 2010, \apjs, 188, 32,
  \dodoi{10.1088/0067-0049/188/1/32}

\bibitem[{Todt {et~al.}(2015)Todt, Sander, Hainich, Hamann, Quade, \&
  Shenar}]{todtPotsdamWolfRayetModel2015}
Todt, H., Sander, A., Hainich, R., {et~al.} 2015, \aap, 579, A75,
  \dodoi{10.1051/0004-6361/201526253}

\bibitem[{Tolstoy {et~al.}(2009)Tolstoy, Hill, \&
  Tosi}]{tolstoyStarFormationHistoriesAbundances2009}
Tolstoy, E., Hill, V., \& Tosi, M. 2009, \araa, 47, 371,
  \dodoi{10.1146/annurev-astro-082708-101650}

\bibitem[{Topping {et~al.}(2020{\natexlab{a}})Topping, Shapley, Reddy, Sanders,
  Coil, Kriek, Mobasher, \& Siana}]{toppingMOSDEFLRISSurveyConnection2020}
Topping, M.~W., Shapley, A.~E., Reddy, N.~A., {et~al.} 2020{\natexlab{a}},
  \mnras, 499, 1652, \dodoi{10.1093/mnras/staa2941}

\bibitem[{Topping {et~al.}(2020{\natexlab{b}})Topping, Shapley, Reddy, Sanders,
  Coil, Kriek, Mobasher, \& Siana}]{toppingMOSDEFLRISSurveyInterplay2020}
---. 2020{\natexlab{b}}, \mnras, 495, 4430, \dodoi{10.1093/mnras/staa1410}

\bibitem[{Tsamis {et~al.}(2003)Tsamis, Barlow, Liu, Danziger, \&
  Storey}]{tsamisHeavyElementsGalactic2003}
Tsamis, Y.~G., Barlow, M.~J., Liu, X.-W., Danziger, I.~J., \& Storey, P.~J.
  2003, \mnras, 338, 687, \dodoi{10.1046/j.1365-8711.2003.06081.x}

\bibitem[{Tully {et~al.}(2008)Tully, Shaya, Karachentsev, Courtois, Kocevski,
  Rizzi, \& Peel}]{tullyOurPeculiarMotion2008}
Tully, R.~B., Shaya, E.~J., Karachentsev, I.~D., {et~al.} 2008, \apj, 676, 184,
  \dodoi{10.1086/527428}

\bibitem[{Verhamme {et~al.}(2006)Verhamme, Schaerer, \&
  Maselli}]{verhamme3DLyaRadiation2006}
Verhamme, A., Schaerer, D., \& Maselli, A. 2006, \aap, 460, 397,
  \dodoi{10.1051/0004-6361:20065554}

\bibitem[{Vidal-Garc\'{i}a {et~al.}(2017)Vidal-Garc\'{i}a, Charlot, Bruzual, \&
  Hubeny}]{vidal-garciaModellingUltravioletlineDiagnostics2017}
Vidal-Garc\'{i}a, A., Charlot, S., Bruzual, G., \& Hubeny, I. 2017, \mnras,
  470, 3532, \dodoi{10.1093/mnras/stx1324}

\bibitem[{Wiese {et~al.}(1996)Wiese, Fuhr, \&
  Deters}]{wieseAtomicTransitionProbabilities1996}
Wiese, W.~L., Fuhr, J.~R., \& Deters, T.~M. 1996, Atomic transition
  probabilities of carbon, nitrogen, and oxygen : a critical data compilation.
  Edited by W.L. Wiese, J.R. Fuhr, and T.M. Deters. Washington, DC : American
  Chemical Society ... for the National Institute of Standards and Technology
  (NIST) c1996. QC 453 .W53 1996. Also Journal of Physical and Chemical
  Reference Data, Monograph 7. Melville, NY: AIP Press

\bibitem[{Wofford {et~al.}(2021)Wofford, {Vidal-Garc{\'i}a}, Feltre,
  Chevallard, Charlot, Stark, Herenz, \& Hayes}]{woffordStarsGasMost2021}
Wofford, A., {Vidal-Garc{\'i}a}, A., Feltre, A., {et~al.} 2021, \mnras, 500,
  2908, \dodoi{10.1093/mnras/staa3365}

\end{thebibliography}
\bibliographystyle{aasjournal}

\appendix

\section{On Other Stellar Population Synthesis Assumptions}
\label{app:othermods}

\subsection{Comparison to Starburst99}

Differing assumptions in both stellar evolution and crucially atmosphere predictions for hot stars could lead to variation between assumed model frameworks in the stellar metallicities derived from the continuum fits described herein.
To test this, we generate a grid of CSFH models from the latest public version of \snn{} \citep{leithererEffectsStellarRotation2014} and repeat the same analysis as described in Section~\ref{sec:fittingdetails} with a few small differences.
We focus on the fully-theoretical UV spectral predictions output by \snn{} \citep{leithererLibraryTheoreticalUltraviolet2010}, which provide spectra at a resolution of $0.4$~\AA{} over the targeted wavelength regime.
The sharp photospheric transitions evident in these spectra relative to \cb{} evince the higher base resolution of the internal stellar atmosphere grids used in these models; we find similar agreement between models and data when the flexible Gaussian smoothing performed before model comparison is applied to the model spectra rather than the data as in the case of \cb{}.
We generate a grid of CSFH models with a \citet{kroupaDistributionLowMassStars1993} IMF and upper mass cutoff of 100~$M_\odot$ at ages from $0$--$100$~Myr and at the full set of five metallicities provided from $Z=0.0004$--$0.05$.

The resulting metallicity estimates from the \snn{} CSFH fits with constrained age prior are displayed alongside our fiducial metallicities from \cb{} in Table~\ref{tab:modcomp}.
Morphologically, only small differences are visible in the key photospheric line complexes between these models.
In general the results from the two models also agree remarkably well, with inferred metallicities in agreement within $1.5~\sigma$ with only one exception: HS~1442+4250, for which \snn{} prefers a $0.36$ dex lower metallicity ($\sim 2\sigma$).
Much more significant disagreement is visible in the strength of the stellar wind lines especially \heii{} (Figures~\ref{fig:cbs99_wind} and \ref{fig:otherwindcomp}).
This is attributable to the significant changes made to the treatment of luminous massive stars with Wolf Rayet-like atmospheres in the latest \cb{} models \citep[see also][]{senchynaUltravioletSpectraExtreme2021}, and underscores the importance of treating inferences drawn from the wind lines with additional caution.

\begin{table}
\centering
\caption{Stellar metallicities derived from photospheric line fits for our fiducial \cb{} stellar population synthesis results compared to those from \snn{}.
\label{tab:modcomp}}
\begin{tabular}{lcc}

Target & \multicolumn{2}{c}{$\log Z$ CSFH, with age prior} \\
Name & \cb{} & \snn{}  \\

\hline
\hline

J082555 & $-3.43^{+0.34}_{-0.49}$ & $-3.19^{+0.20}_{-0.14}$ \\
SB2 & $-2.93^{+0.19}_{-0.15}$ & $-3.11^{+0.21}_{-0.19}$ \\
J104457 & $-3.14^{+0.13}_{-0.11}$ & $-3.16^{+0.19}_{-0.16}$ \\
SB82 & $-3.00^{+0.05}_{-0.04}$ & $-3.09^{+0.15}_{-0.17}$ \\
J120202 & $-3.35^{+0.23}_{-0.43}$ & $-3.26^{+0.16}_{-0.10}$ \\
HS1442+4250 & $-2.94^{+0.10}_{-0.15}$ & $-3.32^{+0.09}_{-0.06}$ \\

\hline

\end{tabular}
\end{table}

\begin{figure}
\plotone{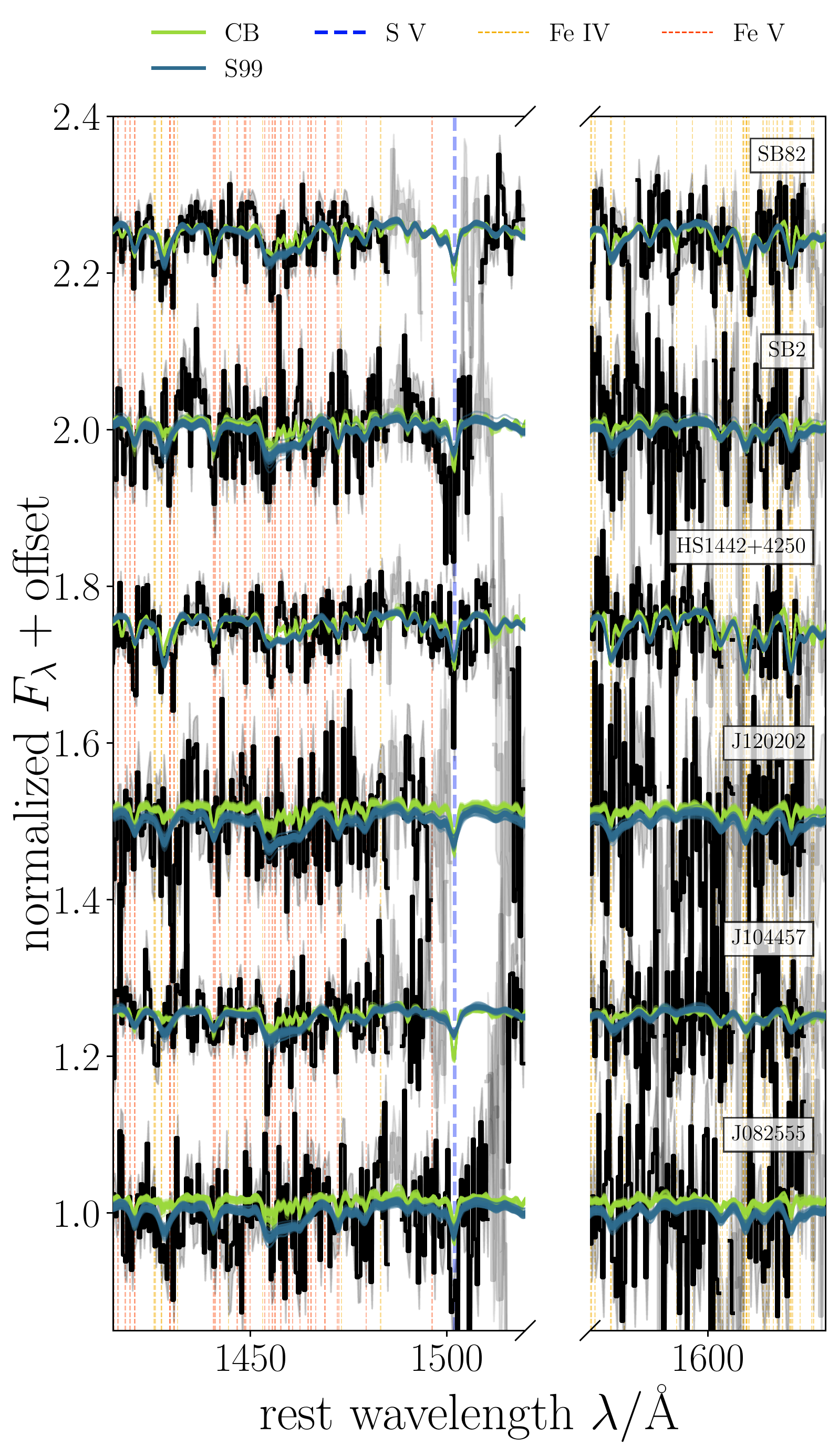}
\caption{
    Fits to the photospheric continuum with both \cb{} and \snn{}, focused on regions dominated by \ion{Fe}{4} and \ion{Fe}{5} absorption.
    \label{fig:cbs99_photo}
    }
\end{figure}

\begin{figure}
\plotone{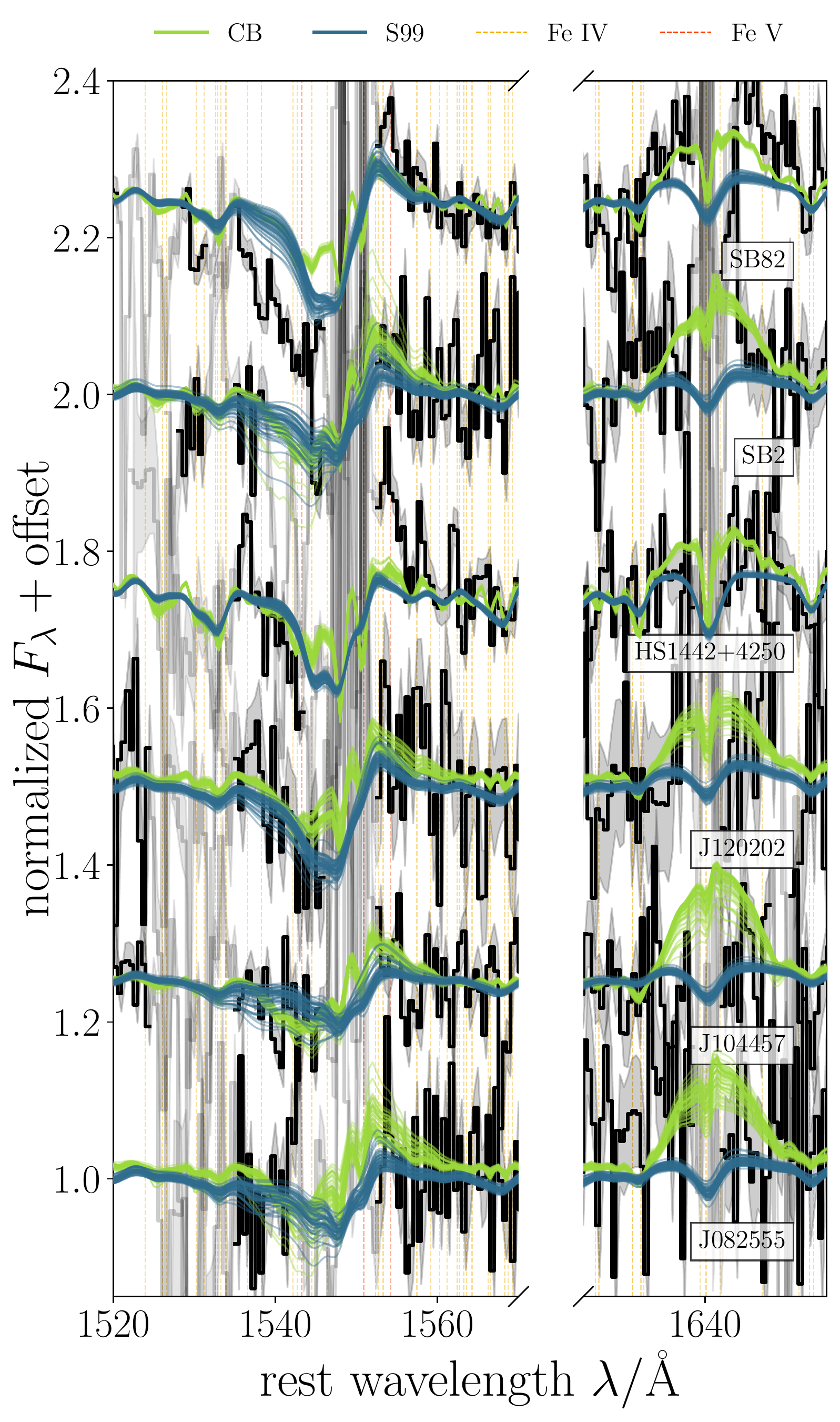}
\caption{
    Same as Fig.~\ref{fig:cbs99_photo}, but focused on the stellar wind line profiles of \civ{} and \heii{}.
    \label{fig:cbs99_wind}
    }
\end{figure}

\begin{figure}
\plotone{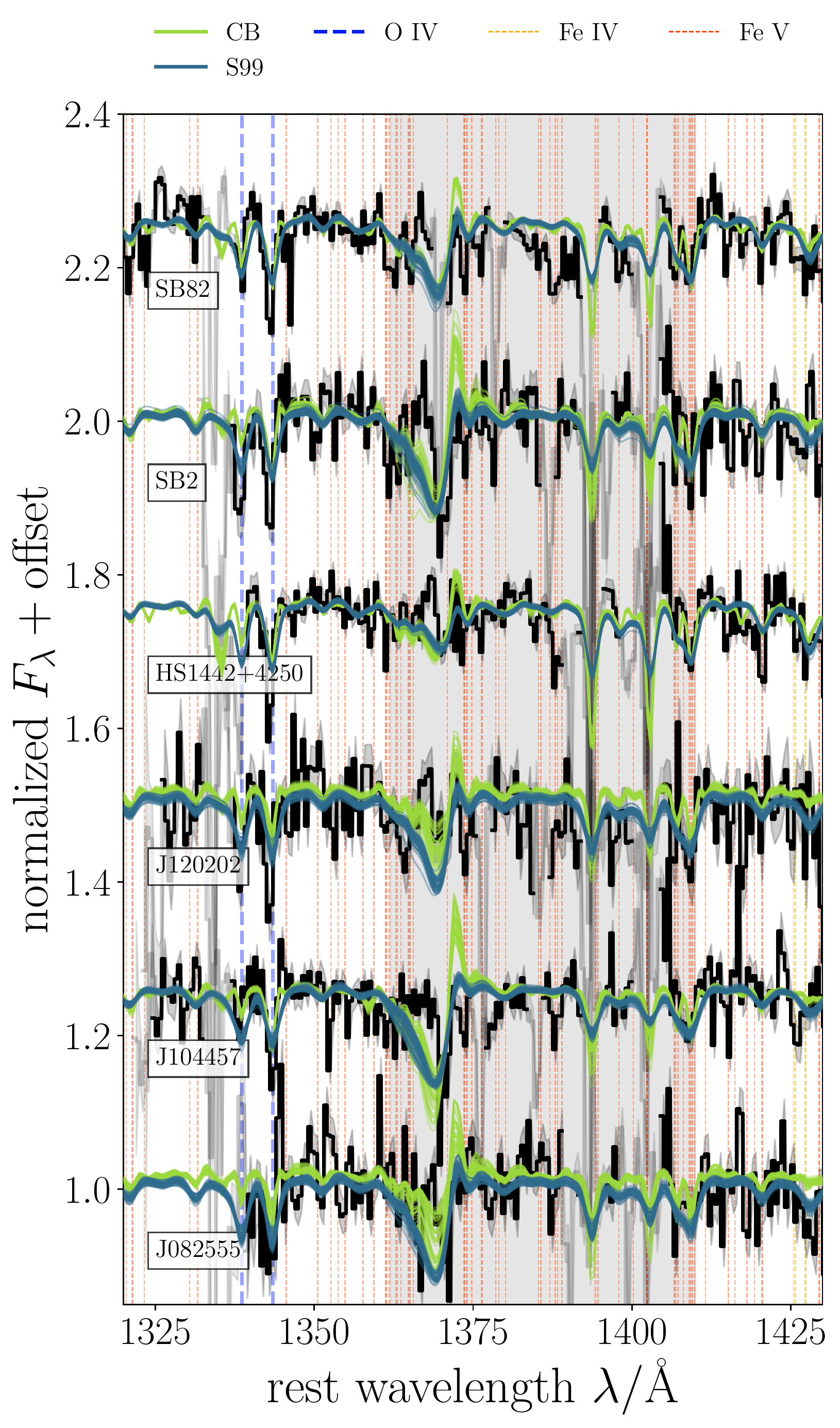}
\caption{
    Same as Fig.~\ref{fig:cbs99_photo}, but here focusing on the blue end of our G160M spectra, including the \ion{O}{5} $\lambda 1371$ and \ion{S}{4} $\lambda\lambda 1393,1402$ resonant doublet.
    The greyed-out wavelength region is never included in our fits to avoid issues in matching \ion{O}{5} $\lambda 1371$, \ion{Si}{4} $\lambda 1393,1402$, and the ISM-contaminated \ion{C}{2} lines present there.
    \label{fig:otherwindcomp}
}
\end{figure}

\subsection{On the star formation history}
\label{app:sfh}

As discussed in Sections~\ref{sec:fitting} and \ref{sec:results}, a bursty star formation history as parameterized by a discrete combination of SSPs rather than a CSFH can change the relative impact of the most luminous stars on the UV continuum.
The character of these effects are often degenerate with other potential deficiencies in the models including most obviously the high-mass IMF shape and cutoff.
Here we briefly discuss the results of fitting the photospheric continuum with a combination of two SSPs rather than a CSFH, which are presented in Table~\ref{tab:metresults}.
We also considered a combination of a CSFH model with a superimposed SSP both with variable age and UV contribution, with broadly qualitatively similar results.
In general, we find that the impact on the photospheric metallicities is to add scatter at a level of $\sim 0.1$--$0.2$ dex rather than introduce any systematic offsets, with a median change in $\log Z$ of $-0.02$ dex when the SFH is shifted from continuous to a two SSP formulation.

\begin{figure}
\plotone{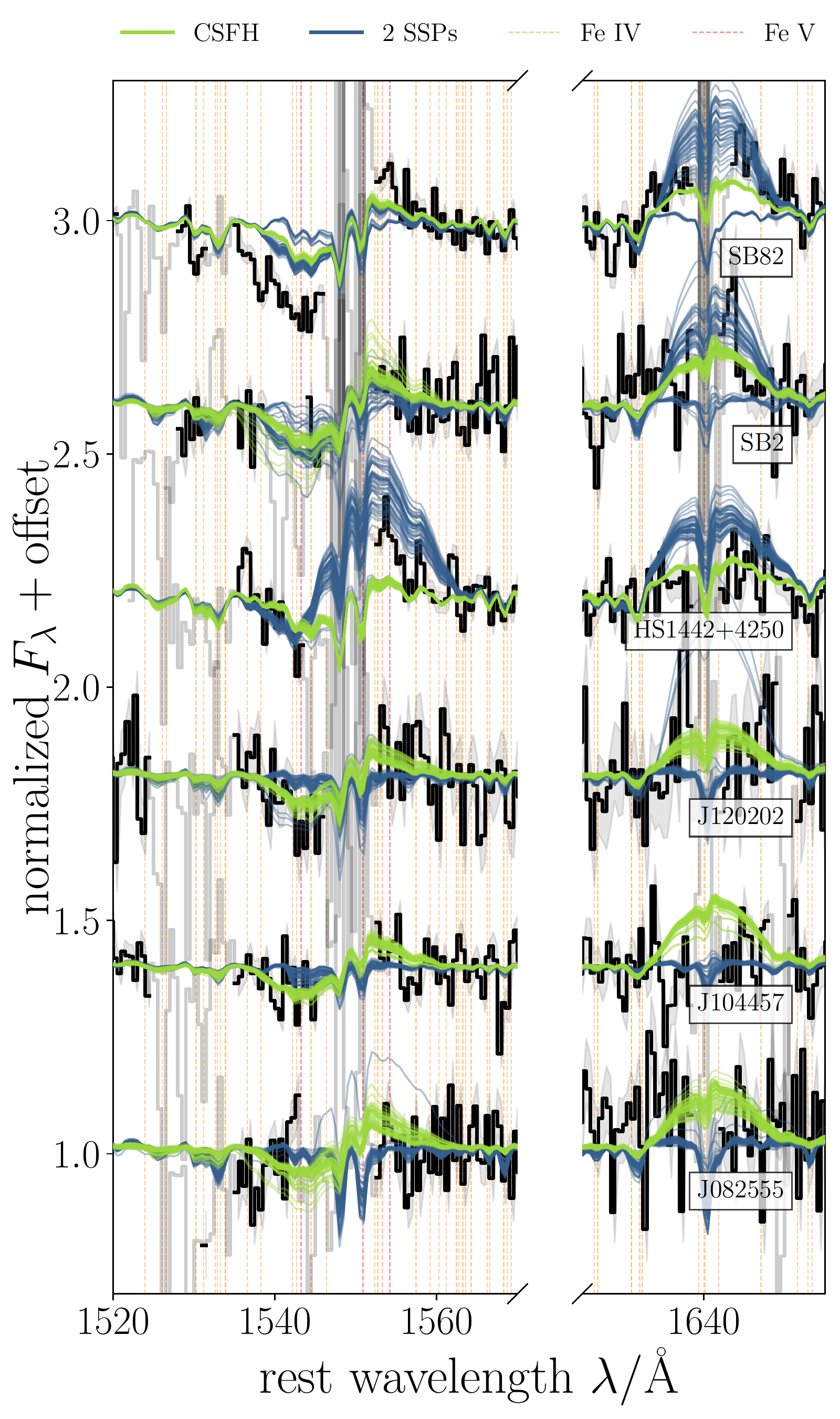}
\caption{
    Comparison of fits to the photospheric continuum in predictions for the stellar wind lines \civ{} (left) and \heii{} (right) with the \cb{} models assuming a constant star formation history with variable age (CSFH, light green) and assuming a combination of two SSPs (dark blue).
    The primary effect of adopting a bursty SFH as in the latter case is an increase in scatter for the stellar \heii{} line, and in some cases (notably HS~1442+4250) additional flexibility in the emission lobe of the \civ{} profile --- both of which are likely driven by Wolf-Rayet spectra assigned to very luminous stars.
    Minimal impact is observed for the depth of the absorption trough of \civ{} which remains highly discrepant for SB~82.
    \label{fig:windcomp_sfh}
}
\end{figure}

As visible in Figure~\ref{fig:windcomp_sfh}, the effects on the wind lines are similarly scattered.
The strength of the stellar \heii{} line shows the most variation, with the posteriors of the 2-SSP fits generally bracketing the CSFH fits but sometimes extending to significantly stronger or weaker signatures.
The effect on the \civ{} P-Cygni profile is generally smaller.
Most notably, the emission lobe of \civ{} in the best-fit models for HS~1442+4250 is significantly stronger and closer to the data in the 2-SSP case than in the CSFH model.
We interpret this as a consequence of an increased proportion of luminous stars with Wolf Rayet-like spectra in the models.
Though while the \civ{} emission lobe is closer to that observed for HS~1442+4250, both it and the \heii{} are notably over-predicted by the posterior spread in this case.
Crucially, essentially no impact is observed in the absorption trough for \civ{}; the observed profile for SB~82 is still substantially stronger than predicted by the models fit to the photospheric lines.

In general, the wind morphological agreement is worse for the 2-SSP models than for the CSFH fits; and of the two high-S/N and discrepant models, only the \civ{} emission mismatch in HS~1442+4250 is aided by adopting a bursty SFH.

\section{On the robustness of metallicities derived from the UV continuum}
\label{app:otherdetails}

Here we explore several potential sources of systematic uncertainty which may generally impact on the derivation of stellar metallicities from UV photospheric lines in unresolved star-forming galaxy spectra.
As discussed, our main conclusions are unlikely to be impact significantly by these considerations, but they should be considered in more detail moving forward as results at low and high redshift begin to rely on the increasingly precise inference of $Z_\star$ from such observations.

\subsection{Metallicity retrieval at low S/N}
\label{app:snretrieval}

One important concern in metallicity retrieval from stellar continuum fitting is the impact of signal-to-noise in the observed spectra.
At low-S/N, noise in the continuum could act to artificially mimic or enhance real variation due to the forest of photospheric absorption lines, potentially increasing the inferred metallicity \citep[e.g.][]{toppingMOSDEFLRISSurveyInterplay2020}.
In principle, our Bayesian approach to modeling the continuum accounting directly for the spectral uncertainties (including a model for potentially underestimated errors) should produce results consistent with the true metallicity even at relatively low S/N.
To test this, we perform a series of simulations using our fiducial \cb{} models.
For S/N per model wavelength pixel ranging from 5--50, we resample the model spectrum with Gaussian uncertainties added and run the resulting simulated spectrum through the same machinery described in Section~\ref{sec:fittingdetails} which has been applied to extract metallicities from our observed spectra.
For the purposes of this experiment we focus on a CSFH model with `true' metallicity $Z=0.0002$ and age 10~Myr; we find qualitatively similar (and generally more favorable) results at higher metallicities where the added noise less-readily affects the stronger photospheric lines.
We extract metallicity constraints from the fits to the same wind-masked regions probed by the observed spectra, specifically focusing on 1420--1510 and 1565--1630~\AA{}.

\begin{figure}
\plotone{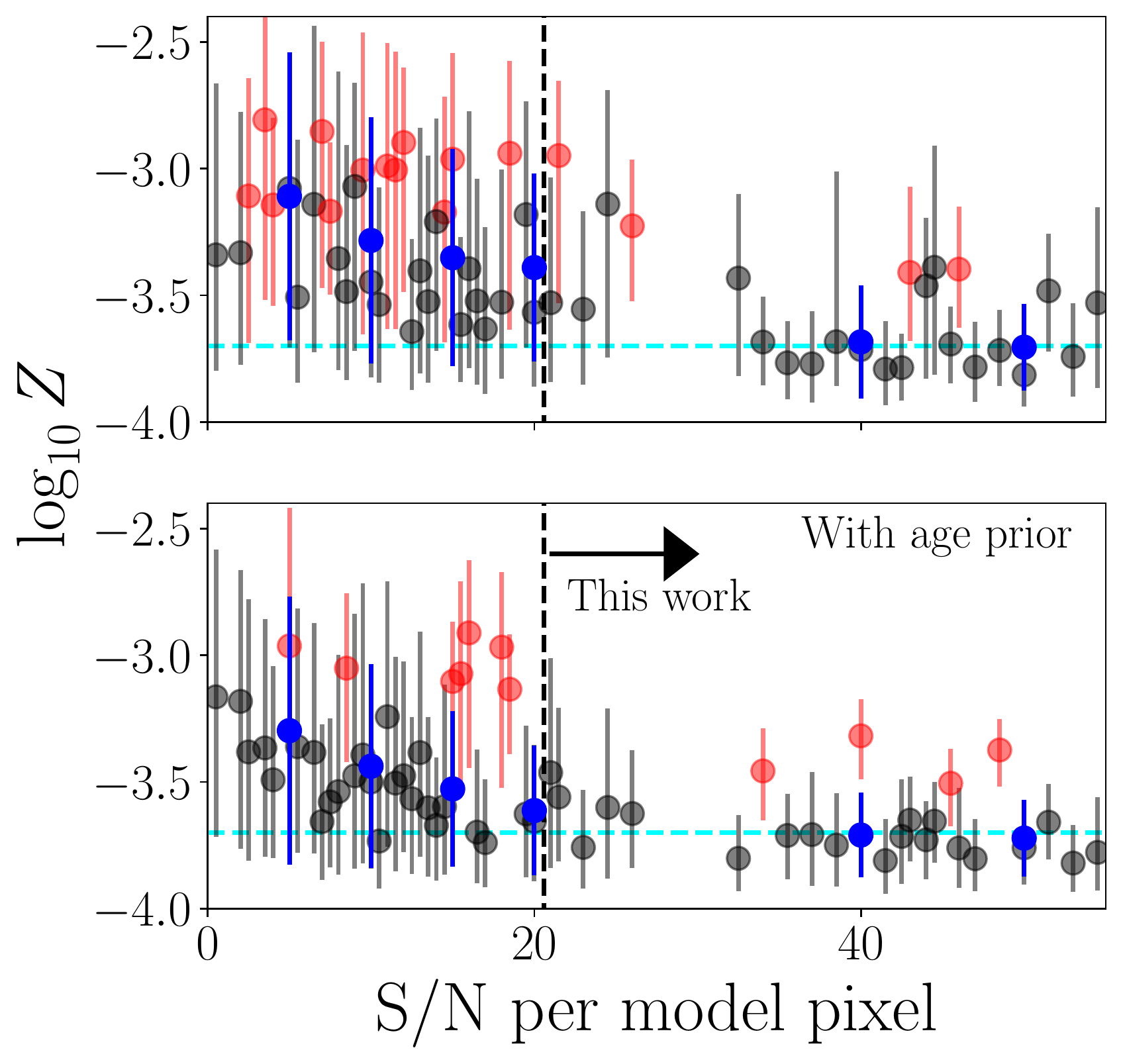}
\caption{
    Demonstration of metallicity retrieval applying our photospheric continuum fitting code to mock spectra.
    Both panels display the derived 68\% confidence interval on $\log Z$ from fits to different realizations of a mock spectrum (CSFH, 10~Myr, $Z=0.0002$) resampled at S/N per 0.5~\AA{} model pixel of $[5,10,15,20,40,50]$.
    Each fit is a black point, offset on the x-axis for clarity; except for cases where the $1\sigma$ confidence range does not overlap with the true metallicity (indicated as a cyan dashed line), in which case the point is colored red.
    The median recovered metallicity and average uncertainty in each S/N bin is indicated by a blue point.
    The top panel displays fits with a CSFH with flat priors in $\log t$ and $\log Z$; while the bottom shows results with a 0.05-dex Gaussian prior centered on the true age.
    For S/N$\gtrsim 20$, the vast majority of individual realizations produce results consistent with the true metallicity within $1\sigma$, and results with a prior on age show negligible bias towards higher metallicities.
    \label{fig:metretrieval}
}
\end{figure}

We perform several sets of simulations at each S/N and plot the results as a function of this ratio in Figure~\ref{fig:metretrieval}.
As expected, this experiment shows a bias towards higher inferred metallicities at low S/N ($<20$), likely a combination of both noise imitating photospheric features and a regression of the posterior distribution towards the prior (flat in $-4\leq \log Z \leq -1.4$) as the data loses informative power.
However, incorporating a Gaussian prior on the age centered on the true value significantly improves this outlook at low S/N, reducing the rate at which trials result in a posterior metallicity distribution inconsistent at $1\sigma$ with the true metallicity from $\sim 40-50\%$ to $\sim 20\%$ in this low-S/N regime (broadly consistent with expectations for the 68\% confidence interval utilized).
Additionally, note that our deep COS spectra all reside at S/N$>20$ per 0.5~\AA{} model pixel (even assuming the conservatively large formal COS uncertainties: Table~\ref{tab:cosobs}).
In this regime, both sets of simulations suggest that bias in the inferred metallicities should be minimal and that such noise is unlikely to introduce error beyond that already accounted for by our uncertainty analysis.

\subsection{Other considerations}

A crucial parameter which has not generally been taken into consideration in the derivation of metallicities from integrated galaxy spectra is microturbulence.
Detailed NLTE atmosphere modeling of individual OB star spectra in the Local Group has revealed the importance of including significant of-order $\sim 10$~km/s microturbulent velocities in order for the models to reach a consistent match for the important photospheric line strengths and profiles \citep[e.g.][]{hillierTaleTwoStars2003,heapFundamentalPropertiesOType2006,bouretMassiveStarsLow2013}.
Whether this finding corresponds to real physical microscopic motions or other physics remains unclear; but crucially, the modeled strength of the UV iron lines are found to increase with increasing microturbulent velocities, complicating iron abundance derivations \citep[e.g.][]{bouretNoBreakdownRadiatively2015}.
The TLUSTY atmosphere grids undergirding the OB star predictions for the \cb{} models we have adopted assume a typical microturbulent velocity of 10~km/s \citep[][and Y.\ Chen, private communication]{lanzGridNonLTELineblanketed2003}.
It is important to note that if a larger microturbulent velocity were adopted, we would expect to find stronger iron lines at fixed $Z$ in the models and thus infer smaller stellar metallicities from our fitting; increasing the tension with the gas-phase abundances.
Also, while complicated by the likelihood of subsolar $\alpha$/Fe ratios relative to the models, the fact that the microturbulence-sensitive \ion{S}{5}~$\lambda 1502$ line is not overestimated by the models lends further support to the notion that the assumed microturbulence is not overestimated (and thus that our iron abundances are not correspondingly underestimated).

\label{lastpage}

\end{document}